\documentclass[a4paper,11pt]{article}
\pdfoutput=1 

\usepackage{jcappub} 

\usepackage[T1]{fontenc} 
\usepackage{physics}
\usepackage{feynmp-auto}
\usepackage{simpler-wick}
\usepackage{bm}
\usepackage{float}
\usepackage{xcolor}    
\usepackage{hyperref}  
\usepackage{booktabs}  
\usepackage{array}     

\newcommand{\Dird}{\delta_{\rm D}^{[3]}}
\newcommand{\kron}{\delta^{\rm K}}

\newcommand*\Bell{\ensuremath{\boldsymbol\ell}}

\title{\boldmath Full Parity-Violating Trispectrum in Axion Inflation: Reduction to Low-D Integrals}

\author[a]{Matthew A. Reinhard,}
\author[b,c]{Zachary Slepian,}
\author[b,d]{Jiamin Hou,}
\author[b]{\& Alessandro Greco}

\affiliation[a]{Department of Physics, University of Florida, Gainesville, FL 32611, USA}
\affiliation[b]{Department of Astronomy, University of Florida, Gainesville, FL 32611, USA}
\affiliation[c]{Lawrence Berkeley National Laboratory, 1 Cyclotron Road, Berkeley CA 94720, USA}
\affiliation[d]{Max-Planck-Institut f\"ur Extraterrestrische Physik, Postfach 1312, Giessenbachstrasse 1, 85748 Garching bei M\"unchen, Germany}

\emailAdd{matt.reinhard@ufl.edu}
\emailAdd{zslepian@ufl.edu}
\emailAdd{jiamin.hou@mpe.mpg.de}
\emailAdd{alessandro.greco@ufl.edu}

\abstract{Recent measurements of the galaxy 4-Point Correlation Function (4PCF) have seemingly detected non-zero parity-odd modes at high significance. Since gravity, the primary driver of galaxy formation and evolution is parity-even, any parity violation, if genuine, is likely to have been produced by some new parity-violating mechanism in the early Universe. Here we investigate an inflationary model with a Chern-Simons interaction between an axion and a $U(1)$ gauge field, where the axion itself is the inflaton field. Evaluating the trispectrum (Fourier-space analogue of the 4PCF) of the primordial curvature perturbations is an involved calculation with very high-dimensional loop integrals. We demonstrate how to simplify these integrals and perform all angular integrations analytically by reducing the integrals to convolutions and exploiting the Convolution Theorem. This leaves us with low-dimensional radial integrals that are much more amenable to efficient numerical evaluation. This paper is the first in a series in which we will use these results to compute the full late-time 4PCF for axion inflation, thence enabling constraints from upcoming 3D spectroscopic surveys such as Dark Energy Spectroscopic Instrument (DESI), Euclid, or Roman.}

\begin{document}
\maketitle
\flushbottom
\allowdisplaybreaks

\section{Introduction}
The first measurements of the Cosmic Microwave Background (CMB) anisotropies introduced two major problems for Big Bang cosmology \cite{COBE}. The uniformity in temperature of the CMB was unexpected for regions that were thought to have never been in causal contact. This is known as the horizon problem. Additionally, the temperature fluctuation power spectrum of the CMB corresponds to a Universe which has zero spatial curvature. Why the curvature takes on this very special value (it would have been zero to 59 decimal places at the Planck time) is known as the flatness problem \cite{Guth}.  

These two problems may be resolved by assuming a period of exponential, superluminal expansion of the early Universe, known as inflation \cite{Guth}. Inflation generally posits the existence of a scalar field, known as the inflaton, whose energy density dominates the primordial Universe, driving the expansion of space \cite{Inflation_Review}. Cosmic inflation also gives a mechanism for the production of the small relative temperature fluctuations in the CMB \cite{2020}. Quantum fluctuations of the inflaton field produce primordial curvature perturbations. These pockets of spacetime curvature accumulate matter and radiation, the imprint of which can be seen in the CMB anisotropies. These over-dense regions were also the seeds from which galaxies formed. Therefore, inflationary models provide a direct link between the behavior of quantum fields in the early Universe and the distribution of galaxies at late times \cite{Structure_Formation}.

The N-Point Correlation Functions (NPCF) are a statistical tool that has proven to be of great utility in cosmology and beyond \cite{Slepian_3PCF,Portillo_2018,peebles1980, arkanihamed2015cosmologicalcolliderphysics, Bartolo_2004}. The NPCF measures the correlation of some quantity at $N$ different spatial locations. The Fourier Transform (FT) of the 2-Point Correlation Function is known as the power spectrum. CMB measurements have shown that the scalar power spectrum is nearly scale-invariant, matching the predictions of single field slow-roll inflation. Both the galaxy 2PCF and 3PCF have been used to make complementary measurements of Baryon Acoustic Oscillation features in the distribution of galaxies, allowing us to better constrain the expansion history of the Universe \cite{Alam_2017, Alam_2021, Eisenstein_2005, Percival_2007}.

The galaxy 4PCF measures the excess clustering of quartets of galaxies over and above that which a spatially random distribution would possess. The 4PCF is the simplest statistic sensitive to parity violation in the 3D distribution of matter \cite{Cahn_PRL}. \cite{Jeong_2012} earlier pointed out that correlating two position-dependent power spectra also probes parity, and \cite{Shiraishi_2016} proposed the use of the 4PCF of the CMB anisotropies. A 3D parity transformation is simply taking $(x,y,z) \to (-x,-y,-z)$, i.e. spatial inversion. Geometrically, it corresponds to a mirror reflection followed by a $180^\circ$ rotation. Recently, fast algorithms have been developed to measure the 4PCF \cite{sunseri_22, Philcox_2021}. The search proposed by \cite{Cahn_PRL} was carried out using the 4PCF of Sloan Digital Sky Survey (SDSS) Baryon Oscillation Spectroscopic Survey (BOSS) galaxies, finding $7.1 \sigma$ evidence for parity violation in the CMASS sample of roughly $800,000$ Luminous Red Galaxies (LRGs) at $0.43 < z < 0.7$ and roughly $3\sigma$ evidence in the LOWZ (280,000 galaxies) lower-redshift sample of LRGs  ($0.16 < z < 0.36$) \cite{4PCF}; see also \cite{Philcox_2022}. These measurements both used tetrahedra, formed of sets of four galaxies at a time, that had sides ranging up to 328 Mpc/$h$ (\S5.1.4 of \cite{4PCF}); hence, on cosmological scales, where baryonic physics should not be important due to simple causality arguments. In particular, if we assume only standard model interactions, effects in galaxy formation such as feedback, supernovae, jets, and accretion, e.g.~\cite{nelson2021illustristngsimulationspublicdata, Peroux_2020, Benson_2010}, cannot contribute to parity violation. These processes involve the electromagnetic force which is parity even.

Despite these intriguing measurements, questions remain regarding the robustness of the results. Improvements can be made through various approaches, including the development of new statistical methods~\cite{Jamieson2024:POP}, the identification of distinctive features in parity-odd signals~\cite{Hou2024:BAOodd}, and the explicit fitting of theoretical models, which serves as the primary motivation for this paper.

On large scales, galaxy formation is primarily governed by gravity, which is 
 also parity even. Perturbation theory has been used to compute the gravitationally induced even-parity galaxy 4PCF \cite{williams_paper} and trispectrum (FT of the 4PCF) \cite{Bertolini_2016}. Any parity violation in large-scale structure (LSS) on these scales most likely came from very early in the Universe, during inflation. This is the only period during which the energy scales involved and physical interactions present are not sufficiently well-constrained by the Standard Model that parity violation due to them is ruled out.  

In this work, we focus on a parity-violating inflationary model with the interaction $\phi F \tilde{F}$ between the axion field $\phi$ and a $U(1)$ gauge field (the analog of the photon). $F$ is the analog of the electromagnetic tensor and $\tilde{F}$ is its dual. We will refer to this model as ``axion inflation''. The first model utilizing an axion-like field as the inflaton was known as natural inflation \cite{Freese}, and several variations of this model have been studied since \cite{Kim_2005, Dimopoulos_2008, Kaloper_2009, Anber_2010, Bartolo_2015, Bartolo_2015_2, Adshead_2018, Liu_2020}. 
 
 For the model considered in this paper, the axion-gauge field interaction produces disparate amounts of positive and negative helicity gauge quanta. This essentially means one will have a preference for one handedness of the gauge quanta over the other---exactly what parity violation is. Since these gauge quanta have energy density, General Relativity implies that they then produce curvature perturbations; these latter seed the galaxies that we observe at late times. This model is of particular interest for interpreting the parity-odd trispectrum observed in the BOSS analysis, since it predicts that the largest parity-odd signal comes from an equilateral configuration of wave vectors~\cite{Caravano_2023, Niu_2023}, also see~\citep[e.g.][]{Fujita_2024}. In contrast,~\cite{Cabass_2023, Cabass_2023_2} proposed various models for parity violation, but the largest parity-odd contribution in those models come from the collapsed or squeezed limits.

 To produce a prediction for this late-time signal, one must first compute the primordial curvature trispectrum. One then would apply linear transfer functions, possibly non-linear evolution (which should not be significant on the scales mostly relevant for the BOSS result), redshift-space distortions \cite{Hamilton_1998}, and finally a galaxy biasing scheme \cite{Desjacques_2018}. Previous works have calculated the tree-level trispectrum for parity violating inflationary models, such as axion inflation  \cite{Shiraishi_2016, Liu_2020} and modified gravity \cite{Cyril_23,moretti2024breakingparitycasetrispectrum}.

 However, in practice the calculation of the primordial curvature trispectrum is already greatly complicated by the fact that the parity-odd component is only present at the one-loop level (or higher), if statistical isotropy is preserved. The presence of loop integrals on each of the internal momenta lead to a 16-dimensional integral. \cite{Niu_2023} was the first to compute the one-loop correction to the trispectrum for axion inflation, the model considered in this work, and \cite{Fujita_2024} calculated the trispectrum for a similar inflationary model, but with the axion as a spectator field. These works numerically evaluated the loop integrals by brute force for a few different configurations of the four external wave-vectors; however, the bulk of the parameter space remains unexplored due to this approach's numerical cost. 
 
 In this paper, we present an analytic method to accelerate dramatically the numerical computation of the primordial trispectrum for axion inflation. This acceleration is achieved by separating the angular and radial components of the trispectrum. We analytically evaluate the angular integrals and perform a change of variables on the radial integrals so as to facilitate efficient numerical computation. 

In \S\ref{Section2} we present background material, including the equation of motion for the gauge field. In \S \ref{sec:prod}, we present the equations that govern the evolution of the inflaton perturbations. We discuss how to compute the primordial trispectrum in \S\ref{sec:corrfun} and present the operator ordering for each of the contributing pieces. In \S\ref{Section5} we demonstrate how the different ways of contracting the operator orderings of \S\ref{sec:corrfun} lead to only four independent kinds of diagrams. Furthermore, we decompose the trispectrum into four pieces, one for each of these diagram types. Each piece contains integrals over conformal time and the loop momenta. In \S\ref{Section6} we show that these integrals are convolutions, and make use of the Convolution Theorem to turn the problem into one of computing FTs. We demonstrate how to analytically compute the angular parts of these FTs, leaving us with only radial integrals. We use a change of variables to further reduce the dimensionality of these radial integrals in \S\ref{Section7}. In \S\ref{Section8}, we show how to put the (greatly simplified) integrals back into our four pieces of the trispectrum, giving us a complete expression for the primordial trispectrum. Additionally, we provide a table with a description of all the important ingredients necessary for constructing the trispectrum, as well as their location and mass dimension. In \S\ref{Section9}, we outline how efficient numerical integration of the low-dimensional radial integrals might proceed. \S\ref{Section10} concludes. 

We include several appendices to clarify mathematical steps and list results that are omitted in the main part of this paper. Appendix \ref{AppendixA} includes the multi-argument delta function expansions that we use extensively in \S\ref{Section5}. In Appendix \ref{AppendixB}, we list all the convolution integrals from \S\ref{Section6} as well as the form they take after applying the Convolution Theorem. In Appendix \ref{AppendixC}, we demonstrate how to perform the angular integration for the Fourier integrals listed in Appendix \ref{AppendixB}. Throughout this work, we use an approximation for the gauge field mode functions given in Eq. (\ref{A:approx_1}). This is a good approximation only in a certain region of the integration space. However, our method for reducing the dimensionality of the loop integrals requires that we integrate over the whole space. In Appendix \ref{AppendixD}, we verify numerically that extending the integration range (outside of the region of good approximation) for integrals containing these mode functions does not significantly change the result. In \S\ref{Section5} we state that the trispectrum cross terms are parity-even; Appendix \ref{AppendixE} provides a proof of this claim.

\section{\textit{U}(1) Axion Inflation}
\label{Section2}
We consider an inflationary model with an axion field $\phi$ coupled to a $U(1)$ gauge field. Here, the axion \textit{itself} is the inflaton, driving the accelerated expansion of space. During inflation, the axion's potential energy will thus dominate the energy density, as is standard in the single-field ``slow-roll'' picture \cite{Baumann_2009}. This inflationary model has been extensively studied in \cite{Anber_2010, Barnaby_2011, Niu_2023}. For a pedagogical review see \cite{Morbelli_23}. For our presentation of the setup here, we follow closely the methods and notation of \citep{Barnaby_2011_2}.  

The action $\mathcal{S}$ is
\begin{align}
\mathcal{S} = \int \dd[4] x \sqrt{-g} \Bigg [ \frac{1}{2}M^2_{\rm p}R -\frac{1}{2} \partial_\mu \phi \, \partial^\mu \phi -\frac{1}{4}F_{\mu \nu} \,F^{\mu \nu} -\frac{1}{4\Lambda} \phi\, \tilde F^{\mu \nu}\, F_{\mu \nu} - V(\phi) \Bigg ]
\label{eq:Ldens}
\end{align}
where $g$ is the determinant of the metric, $M_{\rm p}$ is the reduced Planck mass, $R$ is the Ricci scalar, $\partial_{\mu}$ is a partial derivative with respect to the $\mu^{th}$ dimension, and $\mu \in \{0,1,2,3\}$ runs over time and space indices. $F_{\mu \nu} \equiv \partial_\mu A_\nu - \partial_\nu A_\mu $ is the analog of the Faraday tensor for this field, with $A_{\mu}$ the associated $U(1)$ gauge field. $\Lambda^{-1}$ is the coupling constant, which controls the strength of the interaction. $V(\phi)$ is the inflaton potential. Tilde denotes the dual, and we have 
\begin{align} 
\tilde F^{\mu \nu} \equiv  \epsilon^{\mu \nu \rho \sigma} F_{\rho \sigma}/2 
\end{align}
where $ \epsilon^{\mu \nu \rho \sigma}$ is the Levi-Civita tensor,
\begin{align}
\epsilon^{\mu \nu \rho \sigma} = \frac{1}{\sqrt{-g}}\eta^{\mu \nu \rho \sigma}.
\end{align}
$\eta^{\mu \nu \rho \sigma}$
is $+1$ for even permutations of its indices, $-1$ for odd permutations, and zero if any indices are repeated. 

We assume the potential is sufficiently flat to allow for a slowly rolling inflaton field. The (potential) slow-roll parameters are defined as
\begin{align}
\epsilon \equiv  \frac{M^2_{\rm p}}{2}\bigg(\frac{V'(\phi)}{V(\phi)}\bigg)^2, \hspace{5mm} \eta \equiv M^2_{\rm p}\bigg(\frac{V''(\phi)}{V(\phi)}\bigg).
\end{align}
We will assume that $\epsilon \ll 1$ and $|\eta| \ll 1$.

During inflation, the rapid expansion of space very quickly flattens any initial curvature. Thus we assume a flat, homogeneous, and isotropic spacetime geometry. This geometry is described by the Friedmann-Lemaitre-Robertson-Walker metric:
\begin{align}
ds^2 = -dt^2 + a^2(t)\, \dd \vb{x} \vdot \dd \vb{x} = a^2(\tau)[- \dd \tau^2 + \dd \vb{x} \vdot \dd \vb{x}]. 
\end{align}
$ds^2$ is the line element, and
we have set the speed of light $c = 1$ here and do so throughout this work. $t$ is the cosmic time; $a(t)$ is the scale factor describing the dimensionless size of the Universe, and is conventionally set to unity at present. The dot product  $\dd \vb{x} \vdot \dd \vb{x}$ is Euclidean and $\dd \vb{x}$ is a spatial 3-vector. The conformal time $\tau$ is related to cosmic time $t$ by $\dd t = a \,\dd \tau$.

We have ignored metric perturbations as these would simply add subleading terms to the equations of motion of the inflaton and gauge field. Further details about the inclusion of metric perturbations can be found in \cite{Barnaby_2011_2}. In this section, we are interested in the production of gauge quanta due to the rolling of the homogeneous inflaton background, thus we will also ignore inflaton perturbations. The inflaton perturbations will be reintroduced in \S \ref{sec:prod}.

For the gauge field $A_\mu$, we choose the Coulomb gauge; this means that $\div \vb A = 0$. We also set the time component of $A_{\mu}$ to be zero, i.e. $A_0 = 0$. This choice enables us to use the 3-vector $\vb A$ in what follows. We further assume that the gauge field preserves statistical isotropy, i.e. it has a vanishing vacuum expectation value. We thus have the gauge field equation of motion as:
\begin{equation}
    \label{eq:A_eom}
    \vb A'' - \laplacian \vb A - \frac{\bar{\phi}'(\tau)}{\Lambda}\grad \cross \vb A = 0,
\end{equation}
where prime denotes a partial derivative with respect to the conformal time, i.e. $\vb A' = \partial{\vb A}/\partial{\tau}$.

We may now define fields $\vb E$ and $\vb B$ in analogy with the more familiar electric and magnetic fields. We have
\begin{equation}
\vb E = -{a(\tau)^{-2}} \vb A', \hspace{7mm} \vb B = {a(\tau)^{-2}}\, \curl \vb A.
\label{eqn:EB}
\end{equation}

We write the gauge field $\vb A$ (evaluated at conformal time $\tau$ and 3D position $\vb x$) as an inverse FT:
\begin{equation}
\vb A(\tau, \vb x) = \sum_{\lambda = \pm } \int \frac{\dd[3] \vb k}{(2\pi)^{3}} \left [ \bm{\hat{\epsilon}}_\lambda (\vu k) a_\lambda (\vb k) A_\lambda (\tau, k) e^{-i \vb k \vdot \vb x} + \text{h.c.} \right ]  
\label{eq:A}
\end{equation}
The wave-vector dependence of the Fourier-space field is decomposed into positive and negative helicity states $\bm{\hat{\epsilon}}_\lambda \equiv (\vu \theta + i\lambda \vu \phi)/\sqrt{2}$, hence the sum over $\lambda = \pm$ in front of the integral; h.c. stands for ``Hermitian conjugate'' of the previous term. The helicity eigenstates $\epsilon_{\lambda}$ are also known as circular polarization vectors and satisfy the relations:
\begin{align}
\vb k \cdot \bm{\hat{\epsilon}}_{\pm} (\vu k) = 0, \hspace{2mm} \vb k \times \bm{\hat \epsilon}_{\pm}(\vu k) = \mp ik\bm{\hat \epsilon}_{\pm}(\vu k), \hspace{2mm} \bm{\hat\epsilon}_{\pm}(-\vu k) = \bm{\hat\epsilon}_{\pm}^*(\vu k), 
\label{eq:pol_vec}
\end{align}
with normalization $\bm{\hat \epsilon}_{\lambda}^*(\vu k) \vdot \bm{\hat \epsilon}_{\lambda'}(\vu k) = \delta^{\rm K}_{\lambda \lambda'}$, where $^*$ denotes complex conjugate and $\delta^{\rm K}_{\lambda \lambda'}$ is the Kronecker Delta, unity when its subscripts are equal and zero otherwise.

In Eq. (\ref{eq:A}), $a_{\lambda}$ is the gauge field annihilation operator; its Hermitian conjugate, which we denote $a_{\lambda}^{\dag}$, is the gauge field creation operator. These operators obey the commutation relations
\begin{equation} \label{CCR}
\comm{a_\lambda(\vb k)}{a_{\lambda'}^\dag(\vb k')} = (2\pi)^3 \delta^{\rm K} _{\lambda \lambda'} \, \Dird (\vb k - \vb k'), \hspace{2mm} \comm{a_\lambda(\vb k)}{a_{\lambda'}(\vb k')} = 0,  \end{equation}
with  $\Dird$ the 3D Dirac delta function. In quantum mechanics one promotes the position and momentum of a particle to operators, and then imposes the canonical commutation relations on these operators. In quantum field theory, one replaces classical fields with operators and then canonically quantizes these operators. The commutation relations for the creation and annihilation operators given in Eq. (\ref{CCR}) ensure that our fields have the correct canonical commutation relations \cite{Chen_2010}.

Here and throughout, our FT convention is that an inverse FT has a negative $i$ in the exponential and is normalized as $\dd[3] \vb k/(2\pi)^3$; a forward FT has no factors of $\pi$ and a positive sign in the exponential. We note that both the sign in the exponential and the normalization differ from those used in \citep{Barnaby_2011_2}, which has a positive sign for an inverse FT and normalizes forward and inverse FTs symmetrically, with factors of $(2\pi)^{-3/2}$ each.

Using the Fourier-space representation of the gauge field ${\bf A}$, Eq. (\ref{eq:A}), in the vectorial equation of motion for it, Eq. (\ref{eq:A_eom}), we obtain the equation of motion for the amplitudes of the plus and minus modes of ${\bf A}$ as
\begin{align}
\label{eq: A_mf_eom}
    A_{\pm}^{''}(\tau, k) + k^2A_{\pm}(\tau, k) \mp \frac{2k\xi}{\tau}A_{\pm}(\tau, k) = 0,   
\end{align}
where 
$$\xi \equiv \frac{\dot{\Bar{\phi}}}{2\Lambda H(\tau)}.$$
$\Bar{\phi}$ is the background value of the inflaton field  and is spatially constant. We recall from the action, Eq. (\ref{eq:Ldens}), that $\Lambda^{-1}$ is the axion-gauge field coupling, and $H(\tau)$ is the Hubble parameter. We have used the fact that $\tau \approx -1/(a(\tau)H(\tau))$ for $\epsilon \ll 1$.  During slow-roll inflation, $\xi$ is approximately constant. 

We note that at early times ($\tau \to -\infty$) the last term in Eq. (\ref{eq: A_mf_eom}), which is the only symmetry-breaking term, is negligible, and so both $A_{+}$ and $A_{-}$ will obey the equation of a simple harmonic oscillator, and gauge quanta of either helicity are equally produced. As inflation progresses, one of these modes is enhanced. The frequency of this oscillator is
\begin{align}
\omega_{\pm} = \sqrt{k^2 \mp 2k\xi/\tau}.
\label{eqn:osc_freq}
\end{align}
By analogy to the standard classical simple harmonic oscillator, we may regard the second term in the square-root above, $\mp 2 k \xi/\tau$, as the square of a time-dependent mass. We assume $\dot{ \Bar{\phi}} > 0$ so that the $A_-$ mode has an imaginary frequency (also called a tachyonic instability) for $-k\tau < 2\xi$, leading to exponential production of modes with wave-numbers satisfying this condition.  In the interval $(8\xi)^{-1} \lesssim -k\tau \lesssim 2\xi$, the solution to Eq. (\ref{eq: A_mf_eom}) is well-approximated by \cite{Barnaby_2011}:
\begin{align} \label{A:approx_1}
    A_{-}(\tau, k) \approx \frac{1}{\sqrt{2k}}\left(\frac{-k\tau}{2\xi}\right)^{1/4} e^{\pi \xi - 2\sqrt{-2\xi k\tau}}
\end{align}
Almost all of the gauge field fluctuations produced  are in this region of $k \tau$. From this point on, we thus ignore the $A_{+}$ modes since they experience no tachyonic instability and so their production is negligible.

A parity transformation takes a positive helicity mode function to a negative helicity one and vice versa. \textbf{The imbalance in production of the two helicity modes is thus a violation of parity symmetry, which has arisen due to the $\phi \tilde{F} F$ interaction between the axion and the gauge field}.

Eq. (\ref{eq:A_eom}) describes the production of gauge quanta. These particles may back-react, altering the time evolution of the inflaton background. The backreaction must be controlled so as not to spoil inflation. A detailed analysis of this backreaction in \citep{Barnaby_2011_2}, finds that slow-roll inflation is preserved if two conditions are satisfied:
\begin{align}
    &\frac{H(\tau)^2}{2\pi|\dot{\Bar{\phi}}|} \ll 13\,\xi^{3/2}\,e^{-\pi \xi},\\
    &\frac{H(\tau)}{M_{\rm p}} \ll 146\,\xi^{3/2}\,e^{-\pi \xi}.\nonumber
\end{align}
Towards the end of inflation there may be a significant contribution to the trispectrum (possibly both even and odd) due to the backreaction \cite{Barnaby_2012, Linde_2013}. In this work, we assume that this effect is negligible. The stronger the interaction between the axion and the gauge field  ($\Lambda^{-1}$) the earlier during inflation that backreaction will become prevalent. According to \cite{figueroa2023}, as inflation progresses and the backreaction becomes significant, the production of $A_+$ modes will increase, reducing the imbalance of positive and negative helicity modes, and thus suppressing parity violation towards the end of inflation. Our result for the trispectrum will therefore be most accurate for scales larger than some cutoff corresponding to the transition into the strong backreaction regime. Thus when searching for evidence of axion inflation in the parity-odd galaxy 4PCF, it may be useful to consider only those larger scales which exited the horizon during the earlier part of inflation, when this suppression is not relevant. 

The inverse decay of the gauge quanta into inflaton fluctuations will be the source of parity violation in the primordial trispectrum. We study these inverse decay fluctuations in \S\ref{sec:prod}.

\section{Production of Inflaton Fluctuations}
\label{sec:prod}
In addition to the backreaction, the gauge quanta can also source small fluctuations in the inflaton field through an inverse decay process. Here, we study these inverse decay fluctuations. We consider small fluctuations $\delta \phi$ in the inflaton field, $\phi$, around its spatially uniform (but time-dependent) background $\bar\phi$. We  thus write
\begin{align}
\phi(\tau, \vb x) = \Bar \phi(\tau) + \delta \phi(\tau, \vb x).
\label{eqn:phi-decomp}
\end{align}
From the action, Eq.  (\ref{eq:Ldens}), we have the inflaton field's equation of motion as
\begin{align}
\phi '' + 2\mathcal{H}(\tau)\phi ' - \laplacian \phi + a(\tau)^2 \dv {V}{\phi} = \frac{a(\tau)^2}{\Lambda} \,\vb E \vdot \vb B 
\label{eqn:phi-eom}
\end{align}
where $\mathcal{H}(\tau) = d \ln a/d\tau$ is the conformal Hubble parameter and $\vb E$ and $\vb B$ are defined in Eq. (\ref{eqn:EB}). We recall that $V$ is the inflationary potential. We have again ignored metric perturbations since their inclusion introduces only subleading terms in Eq. (\ref{eqn:phi-eom}). We note that the gauge field $\vb A$ arises in the source term on the right-hand side of Eq. (\ref{eqn:phi-eom}) through $\vb E$ and $\vb B$, leading to production of inflaton particles via inverse decays.
 
Substituting Eq. (\ref{eqn:phi-decomp}) into Eq. (\ref{eqn:phi-eom}), we obtain the equation of motion for the perturbations as
\begin{equation}
\delta \phi '' + 2\mathcal{H}(\tau)\delta \phi ' - \laplacian \delta \phi + a(\tau)^2 \dv [2] {V}{\bar\phi} \delta \phi  = \frac{a(\tau)^2}{\Lambda} \left ( \vb E \vdot \vb B - \expval{\vb E \vdot \vb B} \right ),
\label{eq:fluc_eom}
\end{equation}
where $\expval{\vb E \vdot \vb B}$ denotes a spatial average of the dot product of the electric and magnetic field vectors. 
We may now further separate the fluctuations $\delta \phi$ in the inflaton field into: 
{\bf i)} vacuum fluctuations and 
{\bf ii)} fluctuations produced by inverse decay of the gauge quanta. 

These correspond respectively to the {\bf i)} homogeneous and {\bf ii)} particular solutions of the differential equation (\ref{eq:fluc_eom}). With this in mind, we write the full solution as
\begin{equation}
\delta \phi (\tau, \vb x) = \delta \phi^\text{vac} (\tau, \vb x) + \delta \phi^\text{inv} (\tau, \vb x).    
\end{equation}
These two types of fluctuations are statistically independent of each other because they arise via two independent mechanisms. 

We now write the fluctuations as an inverse FT:
\begin{align} \label{Q_MF}
\delta \phi (\tau, \vb x) = \int \frac{\dd [3] \vb k}{(2\pi)^{3}}\,\frac{Q(\tau, \vb k)}{a(\tau)} e^{-i\vb k \vdot \vb x};
\end{align}
$Q$ is conventionally called the ``mode function''. Now inserting this decomposition into  the perturbations' equation of motion, Eq. (\ref{eq:fluc_eom}), we obtain
\begin{equation}
\left [ \pdv[2]{\tau} + k^2 + a(\tau)^2\dv [2] {V}{\bar\phi} - \frac{a(\tau)''}{a(\tau)} \right ] Q(\tau, \vb k) = J(\tau, \vb k),
\label{eq:Infl_pert_eom}
\end{equation}
where 
\begin{equation} \label{J_def}
J(\tau, \vb k) \equiv \frac{a(\tau)^3}{\Lambda} \int \dd [3] \vb x\; \vb E \vdot \vb B \, e^{i \vb k \vdot \vb x}.   
\end{equation}
We recall that $a(\tau)$ is the scale factor, $\Lambda^{-1}$ is the axion-gauge field coupling, and $\vb E$ and $\vb B$ are defined in Eq. (\ref{eqn:EB}) in terms of the gauge field $\vb A$. We note that the second term on the right-hand side of Eq. (\ref{eq:fluc_eom}) has been spatially averaged and is thus dependent only on time. Thus its spatial FT gives a Dirac delta function, meaning that this term only affects modes at zero wave-number; we may therefore neglect it. 

We may now separate the mode function $Q$ into its vacuum and inverse decay components, as
\begin{equation}
Q(\tau, \vb k) = Q^\text{vac}(\tau, \vb k) + Q^\text{inv}(\tau, \vb k).    
\end{equation}
We further decompose the vacuum contribution as
\begin{equation}
Q^\text{vac}(\tau, \vb k) = b(\vb k)\varphi(\tau, k) + b^\dag(-\vb k)\varphi^*(\tau, k),     
\end{equation}
where $b(\vb k)$ and $b^\dag(\vb k)$ are the \textit{inflaton} annihilation and creation operators, respectively (as distinct from the $U(1)$ gauge quanta annihilation and creation operators $a$ and $a^\dag$). These operators obey the commutation relations 
\begin{align}
\comm{b(\vb k)}{b^\dag(\vb k')} & = (2\pi)^3 \Dird (\vb k - \vb k'), \hspace{2mm} \comm{b(\vb k)}{b(\vb k')} = 0, \\
\comm{b(\vb k)}{a_\lambda^\dag(\vb k')} & = \comm{b(\vb k)}{a_\lambda(\vb k')} = 0.\nonumber
\end{align}
$\varphi(\tau, k)$ is the homogeneous solution of Eq. (\ref{eq:Infl_pert_eom}):
\begin{equation}\label{eq:varphi}
\varphi(\tau, k) = i \frac{\sqrt{-\pi \tau}}{2} \, H_\nu^{(1)}(-k\tau), \qquad \nu \approx  \frac{3}{2} + \mathcal{O}(\epsilon, \eta),
\end{equation}
where we have assumed a Bunch-Davies vacuum and $H_\nu^{(1)}$ is a Hankel function of the first kind. $\mathcal{O}(\epsilon, \eta)$ denotes terms which are first (or higher) order in $\epsilon$ and $\eta$. The factor of $i$ is an arbitrary phase which was chosen so that $\varphi(\tau, k)$ is real in the limit as $-k\tau \to 0$. We note that $\varphi$ depends only on the magnitude of the wave vector $\vb k$. From these vacuum modes $\varphi$, we construct the retarded Green's function as:
\begin{equation} \label{eq:Greens_Function}
G(\tau, \tau', k) = i \Theta(\tau - \tau') \left [\varphi(\tau, k)\varphi^*(\tau', k) - \varphi^*(\tau, k)\varphi(\tau', k) \right ]. 
\end{equation}
Here 
$$
\Theta(\tau - \tau') \equiv
\begin{cases}
      1 & \tau' \leq \tau \\
      0 & \tau' > \tau 
\end{cases}
$$
is the Heaviside step function. This function enforces causality: a mode function at time $\tau$ is influenced only by the past behavior of the source, i.e. at times $\tau' \leq \tau$.

With this in hand, we now find the particular solution to Eq. (\ref{eq:Infl_pert_eom}) as
\begin{align}
    Q^{\text{inv}}(\tau, \vb k) = \int_{-\infty}^{\tau} \dd \tau' \,G(\tau, \tau', k)\,J(\tau', \vb k),
\end{align}
where $J(\tau, \vb k)$ was defined in Eq. (\ref{J_def}). Substituting Eqs. (\ref{eqn:EB}) and (\ref{eq:A}) into Eq. (\ref{J_def}), we find
\begin{align}
\label{eq:J}
&J(\tau, \vb k) = -\frac{1}{\Lambda a(\tau)} \int \frac{\dd [3] \vb q}{(2\pi)^{3}}\, q\, [ \bm{\hat \epsilon}_- (\widehat{\vb k- \vb q}) \vdot \bm{\hat \epsilon}_- (\vu q)] \\
&\times \bigg \{ a_-(\vb k - \vb q) A_-'(\tau, |\vb k - \vb q|) a_-(\vb q) A_-(\tau, q) +  a_-(\vb k - \vb q) A_-'(\tau, |\vb k - \vb q|) a_-^\dag(-\vb q) A_-^*(\tau, q) \nonumber \\
&+  a_-^\dag(\vb q - \vb k) A_-'^*(\tau, |\vb q - \vb k|) a_-(\vb q) A_-(\tau, q) + a_-^\dag(\vb q - \vb k) A_-'^*(\tau, |\vb q - \vb k|) a_-^\dag(-\vb q) A_-^*(\tau, q) \bigg \}.\nonumber  \end{align}
We recall that  $\bm{\hat \epsilon}_-(\vu k)$ is the negative-helicity polarization vector defined before Eq.~(\ref{eq:pol_vec}), the primes on the mode functions $A_-$ (which themselves are defined in Eq. (\ref{eq:A})) are derivatives with respect to conformal time $\tau$, and lowercase $a_-$ is the gauge field annihilation operator, with $a_-^\dag$ the creation operator. Since the production of positive helicity modes is negligible, we have not included them in the source term $J(\tau, \vb k)$.

\section{The Primordial Trispectrum}
\label{sec:corrfun}
We will now calculate the FT of the 4PCF of the primordial curvature perturbations; this is known as the primordial trispectrum. There are several methods used to study cosmological correlation functions \cite{Acquaviva_2002, Weinberg_2005, baumann2022snowmasswhitepapercosmological, Caravano_2023}. In this paper, we effectively employ the in-in formalism \cite{Schwinger_1961, Maldacena_2003, Chen_2010, Chen_2017}. In particular, we follow the method of \cite{Barnaby_2011_2}, which is equivalent to the in-in formalism for the case of a real mode function. The error associated with using the real mode approximation in the range $1.55 \leq \xi \leq 2.4$ is below 1\% as found in \cite{Niu_2023}. \cite{Barnaby_2011_2} found that inverse decay effects are negligible for $\xi \lesssim 1$ and rule out $\xi \gtrsim 2.65$ using data from the Cosmic Background Explorer. Thus the real part of the mode function dominates for most of the allowed region of the parameter space. For this reason, we consider only the real part of the gauge field mode functions throughout this work.

We choose the spatially flat gauge, for which the curvature perturbation on uniform density hyper-surfaces is $\zeta(\tau, \vb x) = -H(\tau) \delta \phi(\tau, \vb x)/ \dot \phi(\tau)$. We now write the curvature perturbation as an inverse FT:
\begin{align} 
 \zeta(\tau, \vb x) =  \int \frac{\dd^3 \vb k}{(2\pi)^{3}}\, \zeta(\tau, \vb k) \,e^{-i\vb k \vdot \vb x}.
 \end{align}
 Using Eq. (\ref{Q_MF}), we can rewrite the FT of the curvature perturbation as 
 \begin{align} \label{zeta}
 &\zeta(\tau, \vb k) = -Q(\tau, \vb k)H(\tau)/(a(\tau)\dot \phi(\tau)).
 \end{align}
The trispectrum evaluated at time $\tau$ is
\begin{align}\label{Trispectrum}
\expval{\zeta(\tau, \vb k_1) \zeta(\tau, \vb k_2) \zeta(\tau, \vb k_3) \zeta(\tau, \vb k_4)} = \frac{H(\tau)^4}{\dot \phi(\tau)^4 \hspace{0.5mm} a(\tau)^4} \bigg [ &\expval{Q^{\text{vac}}(\tau, \vb k_1) Q^{\text{vac}}(\tau, \vb k_2) Q^{\text{vac}}(\tau, \vb k_3) Q^{\text{vac}}(\tau, \vb k_4)} \nonumber \\ 
& \hspace{-20mm} + \expval{Q^{\text{vac}}(\tau, \vb k_1) Q^{\text{vac}}(\tau, \vb k_2) Q^{\text{inv}}(\tau, \vb k_3) Q^{\text{inv}}(\tau, \vb k_4)} + \text{symm}. \nonumber \\ & \hspace{-20mm} + \expval{Q^{\text{inv}}(\tau, \vb k_1) Q^{\text{inv}}(\tau, \vb k_2) Q^{\text{inv}}(\tau, \vb k_3) Q^{\text{inv}}(\tau, \vb k_4)} \bigg ], 
\end{align}
where the $<\cdots>$ represents a vacuum expectation value of the included operators and symm. denotes a symmetrization of the previous term over all possible arguments. We note that any terms including an odd number of $Q^{\text{vac}}$ operators must also have an odd number of inflaton creation and annihilation operators and therefore their vacuum expectation value must vanish since these operators commute with the gauge field creation and annihilation operators. 

The trispectrum contains both parity-even and parity-odd terms. Although parity symmetry is highly violated in the gauge field, where one of the modes grows exponentially while the other one gets largely suppressed, 
the parity-odd contribution has been found to be subdominant compared to the parity-even part of the trispectrum~\cite{Niu_2023, Fujita_2024}. The only term in Eq. (\ref{Trispectrum}) containing parity-odd modes is the last one, which contains four inverse decay mode functions.\footnote{As explained in \S\ref{Section2}, parity-violation arises from the interaction vertex $\phi \tilde{F} F$; this vertex appears only at one loop and beyond. The expectation values that contain two vacuum and two inverse decay mode functions are in fact parity-even, and we have proven this fact in Appendix \ref{AppendixE}.} To obtain the parity-odd trispectrum we thus need only to calculate this last term:
\begin{align}
&\expval{Q^{\text{inv}}(\tau, \vb k_1) Q^{\text{inv}}(\tau, \vb k_2) \nonumber Q^{\text{inv}}(\tau, \vb k_3) Q^{\text{inv}}(\tau, \vb k_4)} \\
&= \nonumber \int_{-\infty}^\tau \dd \tau_1 \int_{-\infty}^\tau \dd \tau_2 \int_{-\infty}^\tau \dd \tau_3 \int_{-\infty}^\tau \dd \tau_4 \; G(\tau, \tau_1, k_1) \,G(\tau, \tau_2, k_2) \,G(\tau, \tau_3, k_3) \, G(\tau, \tau_4, k_4) \\ &\qquad\qquad\qquad\qquad \qquad\qquad\qquad\qquad \times \expval{J(\tau_1, \vb k_2)\, J(\tau_2, \vb k_2) \,J(\tau_3, \vb k_3) \, J(\tau_4, \vb k_4)}. 
\label{eq:inv_4PCF} 
\end{align}
We will call this term the inverse decay trispectrum. Although this term contains both parity-odd and even contributions, we can always project the result onto the isotropic basis functions of \cite{Isotropic_basis} to single out the parity-odd modes.

We evaluate the trispectrum for modes well outside of the horizon; thus we take the limit $-k\tau \rightarrow 0$ (i.e. very small $k$). We may therefore employ an approximation for $\varphi(\tau, k)$ but not for $\varphi(\tau', k)$. The Green's function of Eq. (\ref{eq:Greens_Function}) then becomes
\begin{align}
G(\tau, \tau', k) &= i\Theta(\tau - \tau')\frac{a(\tau)H(\tau)}{\sqrt{2k^3}}(-k\tau)^{\frac{n_s - 1}{2}}\big[\varphi^*(\tau',k) - \varphi(\tau',k)\big] \nonumber\\
&= a(\tau) H(\tau) \sqrt{\frac{-\pi \tau'}{8k^3}} (-k\tau)^{\frac{n_s - 1}{2}}  \Theta(\tau - \tau')\left [H_{3/2}^{(1)*}(-k\tau') + H_{3/2}^{(1)}(-k\tau') \right ],
\end{align}
where we have used Eq. (\ref{eq:varphi}) as well as the approximation that
\begin{align}
\varphi(\tau, k) \approx \frac{a(\tau)H(\tau)}{\sqrt{2k^3}}(-k\tau)^{\frac{n_s - 1}{2}}     
\end{align}
for $-k\tau \rightarrow 0$. 

We now use 
\begin{align}
\Re[H_{3/2}^{(1)}(x)] = J_{3/2}(x) = \sqrt{\frac{2x}{\pi}}\,j_1(x) 
\end{align}
to rewrite the Green's function as 
\begin{align} \label{Green_function_limit}
G(\tau, \tau', k) = \Theta(\tau - \tau') \frac{a(\tau) H(\tau) \tau'}{k} (-k\tau)^{\frac{n_s - 1}{2}} j_1(-k\tau').    
\end{align}

In the next section, we will focus on evaluating the expectation value of the product of the four source functions $J(\tau, \vb k)$, which enters the integral in Eq. (\ref {eq:inv_4PCF}). 
Each $J(\tau, \vb k)$ contains creation and annihilation operators. To compute the correlation function, we must consider all possible Wick contractions of these operators.  

\subsection{Counting Contractions}
We will compute the inverse decay trispectrum at the one-loop level. We begin by
inserting Eq. (\ref{eq:J}) for $J$ into the correlation function of the source terms in Eq. (\ref{eq:inv_4PCF}),
\begin{align}
\label{eq:J_ev}
&\expval{JJJJ} \equiv \expval{J(\tau_1, \vb k_2) J(\tau_2, \vb k_2) J(\tau_3, \vb k_3) J(\tau_4, \vb k_4)}. 
\end{align}
\subsubsection{Operator Orderings}
We find that the non-vanishing pieces of the correlation function at the one-loop level contain creation and annihilation operators in the following orderings. We highlight that ordering does matter here as these are operators.  We also note that, up to overall sign, the arguments in the list below do not change as we go from I through V (up to an overall minus sign).
\begin{align} \label{eqn:op-ord}
\text{I}. \hspace{5 mm} &a_-(\vb k_1 - \vb q_1) a_-(\vb q_1) a_-(\vb k_2 - \vb q_2) a_-(\vb q_2) a_-^{\dag}(\vb q_3 - \vb k_3) a_-^{\dag}(-\vb q_3) a_-^{\dag}(\vb q_4 - \vb k_4) a_-^{\dag}(-\vb q_4) \nonumber \\ \nonumber
\text{II}. \hspace{5 mm} &a_-(\vb k_1 - \vb q_1) a_-(\vb q_1) a_-^{\dag}(\vb q_2 - \vb k_2) a_-(\vb q_2) a_-(\vb k_3 - \vb q_3) a_-^{\dag}(-\vb q_3) a_-^{\dag}(\vb q_4 - \vb k_4) a_-^{\dag}(-\vb q_4) \\ \nonumber
\text{III}. \hspace{5 mm} &a_-(\vb k_1 - \vb q_1) a_-(\vb q_1) a_-(\vb k_2 - \vb q_2) a_-^{\dag}(-\vb q_2) a_-^{\dag}(\vb q_3 - \vb k_3) a_-(\vb q_3) a_-^{\dag}(\vb q_4 - \vb k_4) a_-^{\dag}(-\vb q_4) \\ \nonumber
\text{IV}. \hspace{5 mm} &a_-(\vb k_1 - \vb q_1) a_-(\vb q_1) a_-^{\dag}(\vb q_2 - \vb k_2) a_-(\vb q_2) a_-^{\dag}(\vb q_3 - \vb k_3) a_-(\vb q_3) a_-^{\dag}(\vb q_4 - \vb k_4) a_-^{\dag}(-\vb q_4) \\ 
\text{V}. \hspace{5 mm} &a_-(\vb k_1 - \vb q_1) a_-(\vb q_1) a_-(\vb k_2 - \vb q_2) a_-^{\dag}(-\vb q_2) a_-(\vb k_3 - \vb q_3) a_-^{\dag}(-\vb q_3) a_-^{\dag}(\vb q_4 - \vb k_4) a_-^{\dag}(-\vb q_4)
\end{align}
Each term has a different ordering of creation and annihilation operators; we will term this the ``operator ordering''. 

\subsubsection{Contractions Within Each Operator Ordering}
\begin{figure}[ht]
\vspace{-2mm}
    \centering
    \includegraphics[width=0.6\textwidth]{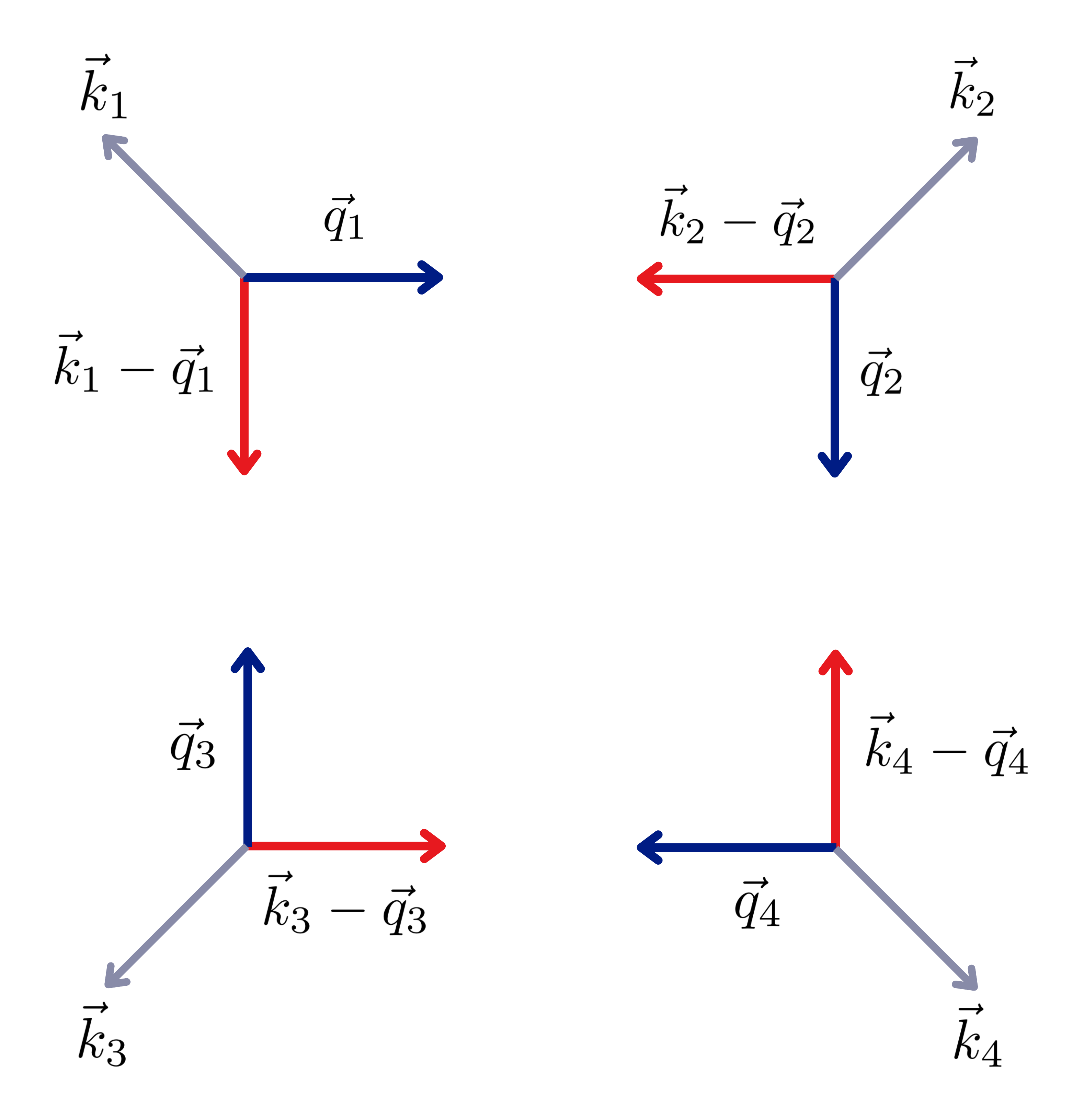}
    \vspace{-2mm}
    \caption{\footnotesize{The gray lines represent the external momenta $\vec{k}_i$. The internal lines (red and blue) of the four vertices can be connected in several ways to form a one-loop diagram. Each time we contract two operators, we can represent this pictorially by connecting internal lines from two different vertices. We will associate each blue line with a ``simple momentum'' $\Vec{q}_i$, and each red line with a ``compound momentum'' $\Vec{k}_j - \Vec{q}_j$. For a given contraction, the momentum subscript with the larger integer will always correspond to a creation operator, and the momentum momentum subscript with the smaller integer will always correspond to an annihilation operator. }} 
    \label{fig:Diagram}
\end{figure}

Within each of the five operator orderings of Eq. (\ref{eqn:op-ord}) above, we must now consider all of the possible ways we can contract the creation and annihilation operators to form a 1-loop diagram. Figure \ref{fig:Diagram} shows the four vertices corresponding to the four external momenta of the trispectrum, the internal lines of which can be connected to form a one-loop diagram. The connection of two internal lines corresponds to a contraction.

Now, to form a valid contraction of a single creation and single annihilation operator, the creation operator must be on the right-hand side of the pair. Each contraction of a creation and annihilation operator gives a Dirac delta function:
\begin{align}\label{Contraction}
\wick{
    \c1 a_-(\vb k) \c1 a_-^{\dag}(\vb k') = (2\pi)^3 \Dird(\vb k - \vb k').
}
\end{align} 
The contraction can be represented pictorially as connecting a pair of internal momentum lines in Figure \ref{fig:Diagram}. We now give one example of a contraction combination for operator ordering I.
\[
\langle
 \wick{
        \c1 a_-(\vb k_1 - \vb q_1) \c2 a_-(\vb q_1) \c3 a_-(\vb k_2 - \vb q_2) \c4 a_-(\vb q_2) 
        \c2 a_-^{\dag}(\vb q_3 - \vb k_3) \c3 a_-^{\dag}(-\vb q_3) \c4 a_-^{\dag}(\vb q_4 - \vb k_4) \c1 a_-^{\dag}(-\vb q_4) 
  }
\rangle 
\] 
To represent this contraction combination pictorially one would connect all of the internal lines in Figure \ref{fig:Diagram}. Thus we will often refer to a contraction combination, or the four Dirac delta functions that it yields, as a ``diagram''. Specifically, in this paper, the word ``diagram'' will mean that all of the internal and external momenta are specified with numerical subscripts. According to Wick's theorem, the correlation function of the $J$'s, Eq. (\ref{eq:J_ev}), receives a contribution from all such diagrams.

\begin{figure}[ht]
\vspace{-2mm}
    \centering
    \includegraphics[width=1.0\textwidth]{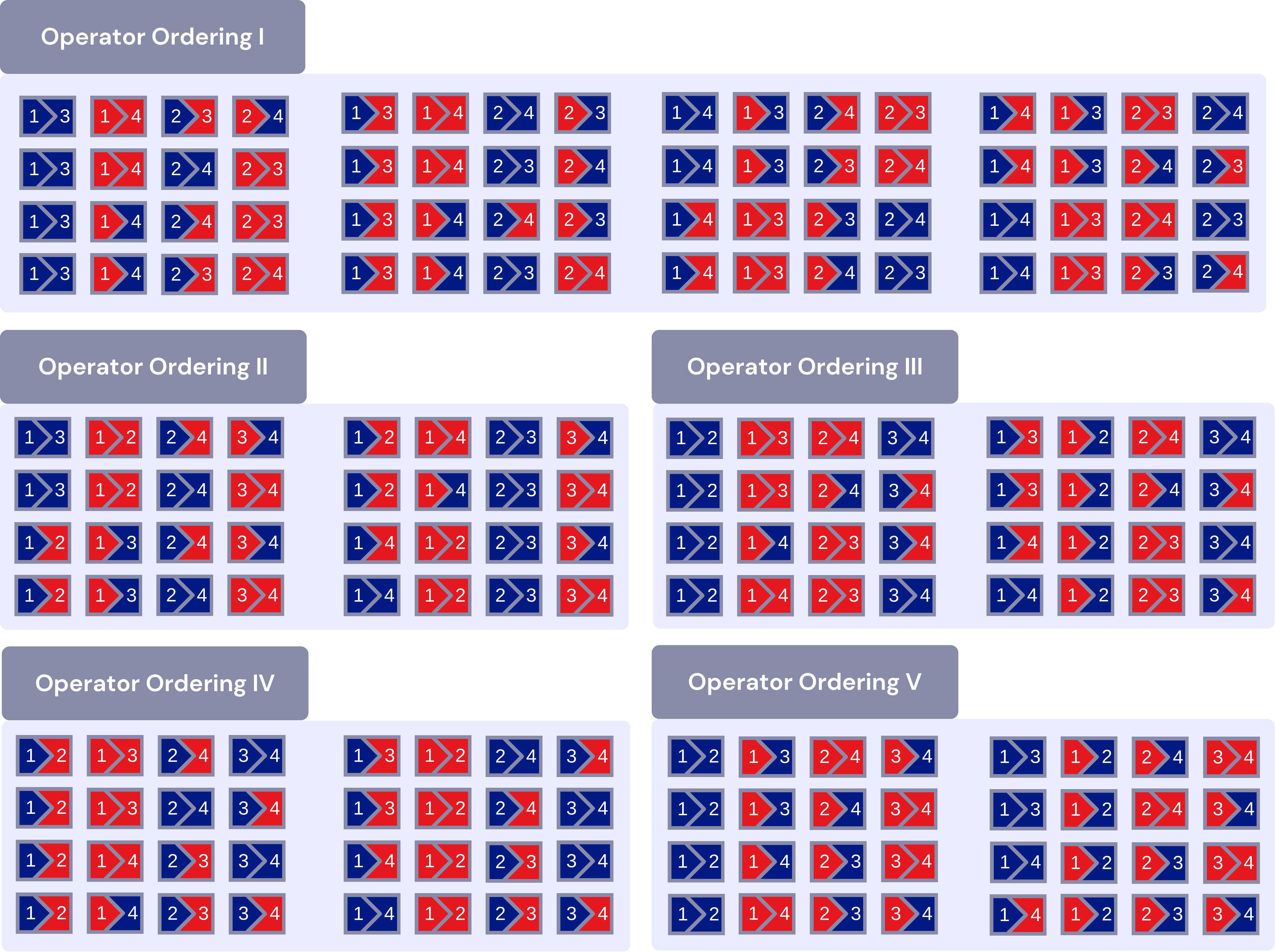}
    \vspace{-2mm}
    \caption{\footnotesize{This figure represents all one-loop diagrams produced by considering all possible contraction combinations of the creation and annihilation operators within each operator ordering. Each contraction is represented by a ``flag'' with two numbers. The color blue corresponds to an internal line with simple momentum $\vb q_i$ while the color red corresponds to an internal line with compound momentum $\vb k_i - \vb q_i$. The subscript of the momentum is determined by the numbers in the flag, e.g. \includegraphics[width = 0.03\textwidth]{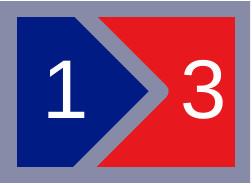} represents connecting the internal lines with momenta $\vb q_1$ and $\vb k_3 - \vb q_3$. Within each operator ordering, each row of four flags shows us how to connect the internal lines in Figure \ref{fig:Diagram}, thus giving a one-loop diagram. }} 
    \label{fig:Operator_Ordering}
\end{figure}

We will denote the sum of all diagrams that have a given operator ordering, I, II, III, IV, or V, as $\Delta_{\text I}$, $\Delta_{\text{II}}$, $\Delta_{\text{III}}$, $\Delta_{\text{IV}}$, or $\Delta_{\text{V}}$.
We now have the full correlation function given in Eq. (\ref{eq:J_ev}) as
\begin{align}
\label{J_corr}
&\expval{JJJJ} = \frac{1}{\Lambda^4 a^4(\tau)} \int \int \int \int \frac{\dd[3] \vb{q}_1}{(2\pi)^3} \, \frac{\dd[3] \vb{q}_2}{(2\pi)^3} \, \frac{\dd[3] \vb{q}_3}{(2\pi)^3} \, \frac{\dd[3] \vb{q}_4}{(2\pi)^3} \, (\Delta_{\text{I}} + \Delta_{\text{II}} + \Delta_{\text{III}} + \Delta_{\text{IV}} + \Delta_{\text{V}}) \nonumber\\
    & \left [\bm{\hat{\epsilon}}_-(\vu q_1) \vdot \bm{\hat{\epsilon}}_-(\widehat{\vb k_1 - \vb q_1}) \right] \left [\bm{\hat{\epsilon}}_-(\vu q_2) \vdot \bm{\hat{\epsilon}}_-(\widehat{\vb k_2 - \vb q_2}) \right ] \left [\bm{\hat{\epsilon}}_-(\vu q_3) \vdot \bm{\hat{\epsilon}}_-(\widehat{\vb k_3 - \vb q_3}) \right ] \left [\bm{\hat{\epsilon}}_-(\vu q_4) \vdot \bm{\hat{\epsilon}}_-(\widehat{\vb k_4 - \vb q_4}) \right]  \nonumber \\
    & \times q_1 A_-^{'}(\tau_1, |\vb k_1 - \vb q_1|) A_-(\tau_1, q_1) \times q_2 A_-^{'}(\tau_2, |\vb k_2 -  \vb q_2|) A_-(\tau_2, q_2) \nonumber\\
    &\times q_3 A_-^{'}(\tau_3, |\vb k_3 -  \vb q_3|)A_-(\tau_3, q_3) \times q_4 A_-^{'}(\tau_4, |\vb k_4 -  \vb q_4|)A_-(\tau_4, q_4), 
\end{align}
and where
\begin{align} \label{Contr_1}
&\Delta_{\text{I}} \equiv (2\pi)^{12} \Bigg \{\delta_{\rm D}^{[3]}(\vb q_2 + \vb k_4 - \vb q_4) \delta_{\rm D}^{[3]}(\vb q_1 + \vb k_3 - \vb q_3) \delta_{\rm D}^{[3]}(\vb k_1 - \vb q_1 + \vb q_4) \delta_{\rm D}^{[3]}(\vb k_2 - \vb q_2 + \vb q_3) \nonumber \\
&\hspace{24mm} + \text{all other contraction combinations for operator ordering I.} \Bigg \}
\end{align}
corresponds to the sum of all diagrams with operator ordering I. The first term in Eq. (\ref{Contr_1}) is given by the contraction combination below Eq. (\ref{Contraction}). The other contraction combinations within operator ordering I simply lead to different arrangements of the momenta in the arguments of the Dirac delta functions. The Dirac delta functions corresponding to the other diagrams in operator orderings I-V can be found in Figure \ref{fig:Operator_Ordering}. We note that since we are considering only the real mode approximation for the gauge field mode functions, we are able to ignore the complex conjugates on these mode functions and pull the mode functions out as a common factor for the $J$'s.

For operator ordering I, there are sixteen contraction combinations, each representing a one-loop diagram. For operator orderings II through V, there are eight diagrams, as we show in Figure \ref{fig:Operator_Ordering}. Adding up the number of diagrams for each of the five operator orderings in Eq. (\ref{eqn:op-ord}), we see that the correlation function of the $J$'s will be a sum of 48 different terms.  

\begin{figure}[ht]
\vspace{-2mm}
    \centering
    \includegraphics[width=1\textwidth]{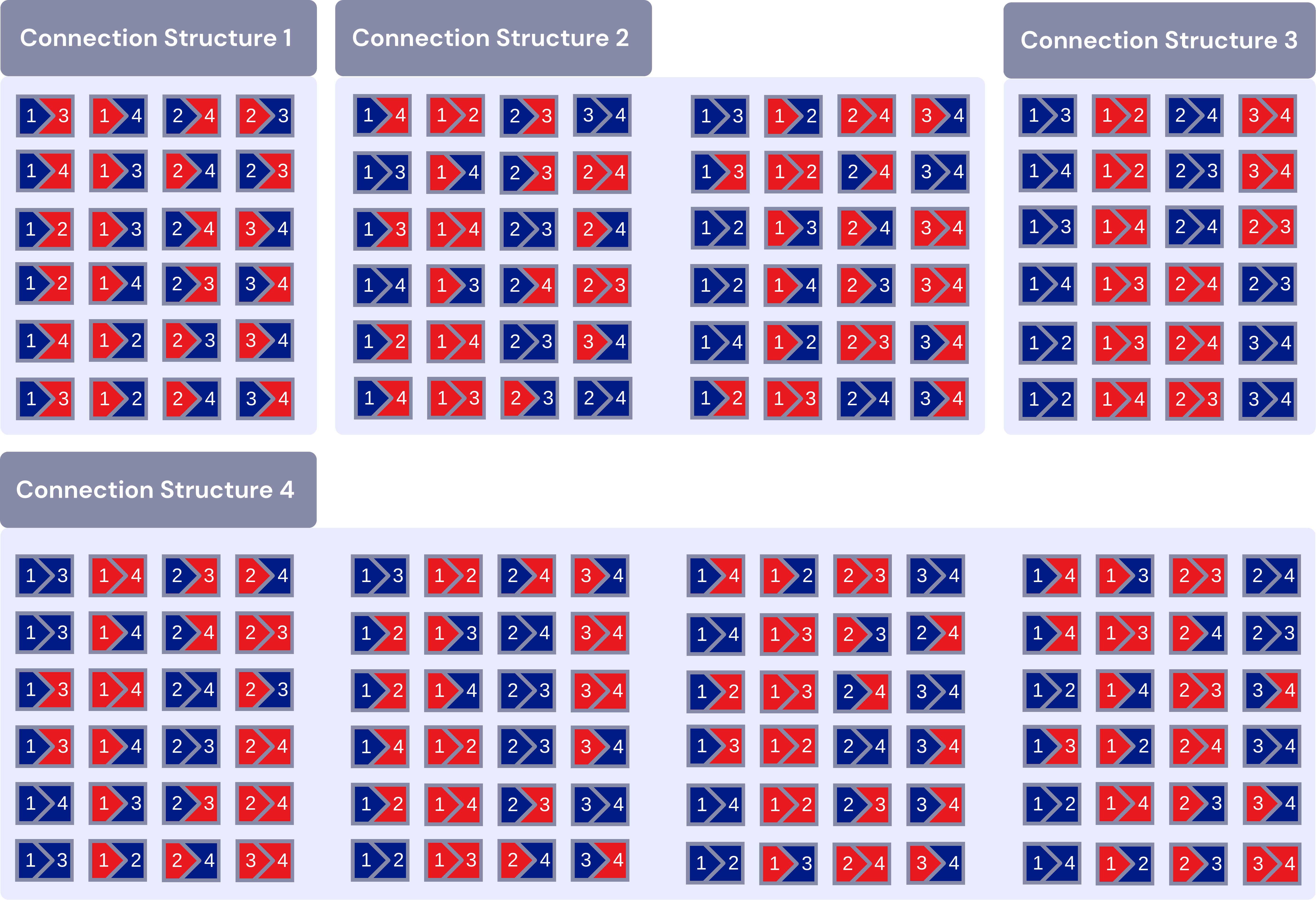}
    \vspace{-2mm}
    \caption{\footnotesize{Here we have categorized all of the terms from Figure \ref{fig:Operator_Ordering} by their connection structure. The meaning of the individual colored flags and the fact that each row of four flags corresponds to a one-loop diagram is described in the Figure \ref{fig:Operator_Ordering} caption. The four connection structures are presented in Figure \ref{fig:diagram_structure_diagrams}. The significance of these connection structures is described in \S \ref{Section5}.}}
    \label{fig:diagram_structure}
\end{figure}

\begin{figure}[ht]
\vspace{-2mm}
    \centering
    \includegraphics[width=0.7\textwidth]{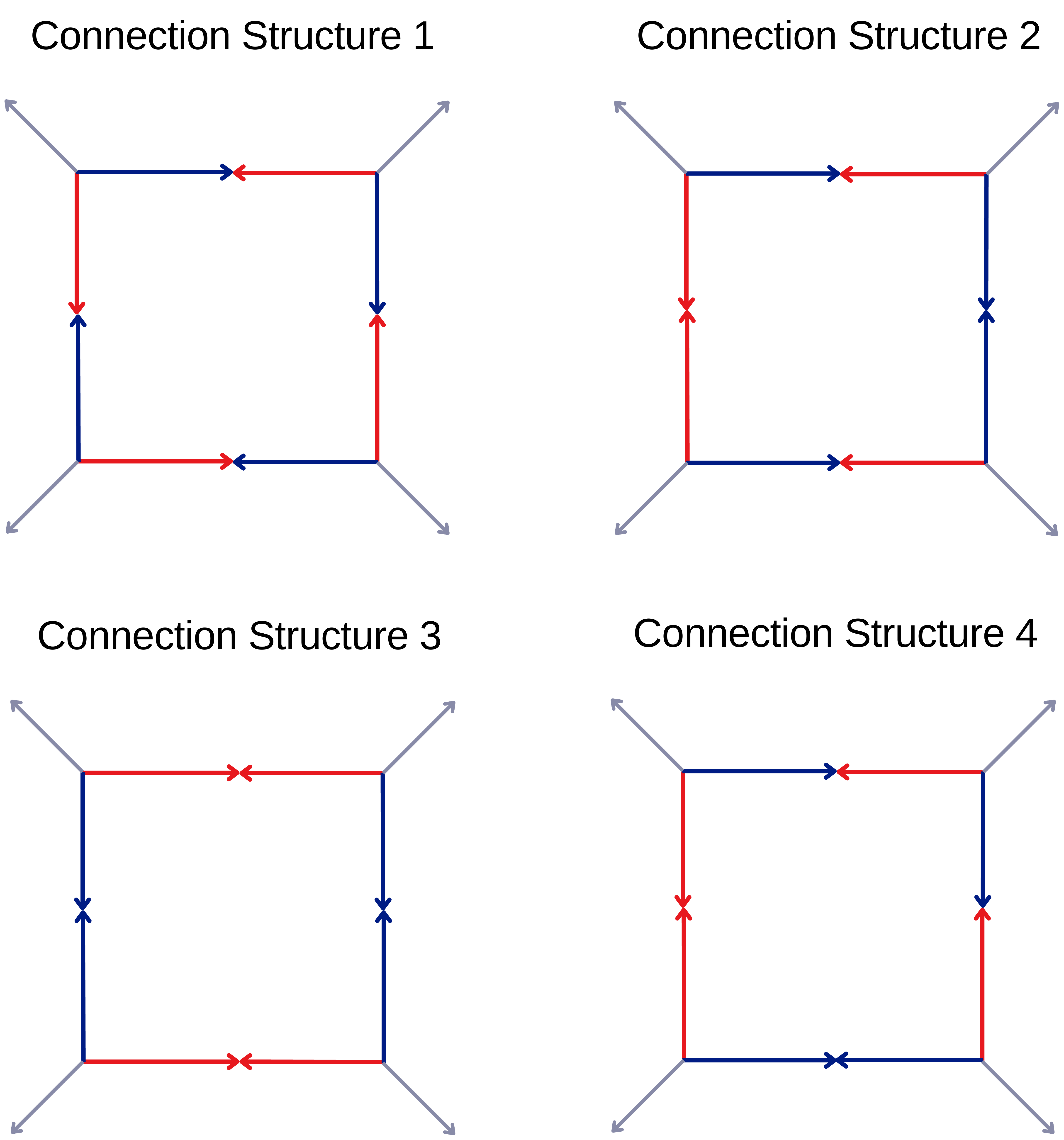}
    \vspace{-2mm}
    \caption{\footnotesize{The four connection structures are determined by the colors of the connected internal lines above (the external lines are grey). The color patterns correspond to the different ways one can contract the creation and annihilation operators with momenta $\vb k_i - \vb q_i$ and $\vb q_j$ (red-blue), $\vb q_i$ and $\vb q_j$ (blue-blue), and $\vb k_i - \vb q_i$ and $\vb k_j - \vb q_j$ (red-red). The chosen contraction combination determines the number of momentum vectors in the argument of the delta function. These four diagrams represent the four connection structures of \S\ref{Section5}.}}
    \label{fig:diagram_structure_diagrams}
\end{figure}

\begin{figure}[ht]
\vspace{-2mm}
    \centering
    \includegraphics[width=0.9\textwidth]{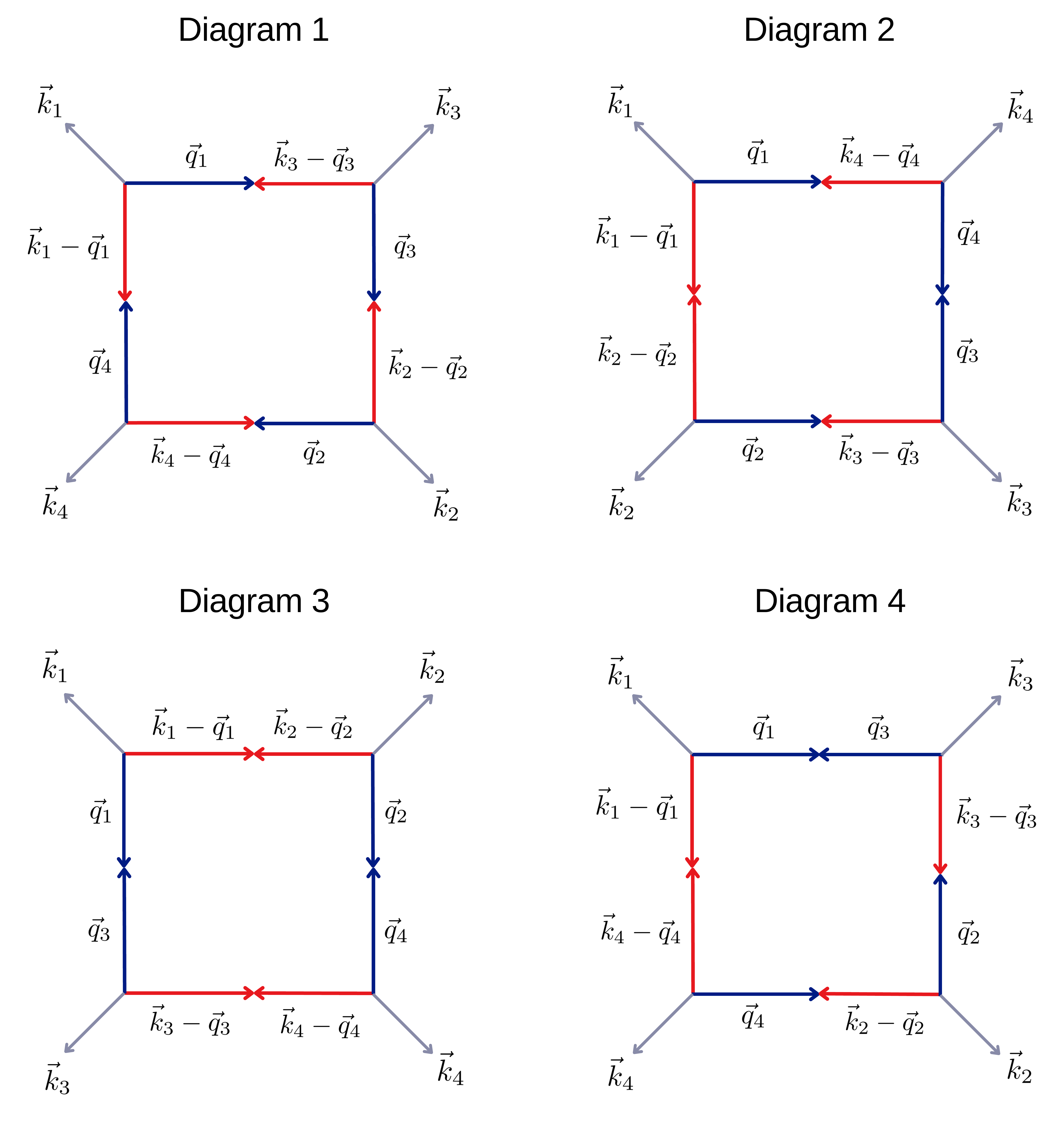}
    \vspace{-2mm}
    \caption{\footnotesize{Here we show four of the forty-eight possible diagrams, one representing each of the four connection structures. The numbering of each diagram corresponds to its connection structure. The diagrams within each connection structure can be related by interchanging the subscripts on the momenta. Due to this fact, we consider only a single diagram for each of the four connection structures when computing the trispectrum in 
    \S\ref{Section5}. We call these four diagrams ``Reference Diagrams''.}}
    \label{Diagrams1-4}
\end{figure}

\section{Classifying Diagrams into Four Connection Structures} 
\label{Section5}
The contractions outlined in \S\ref{sec:corrfun} give the whole correlation function of the $J$'s as a sum of 48 terms. Each term contains four Dirac delta functions and corresponds to one of the 48 diagrams. It is notable that each of these diagrams can be classified into one of just four types. We term these four types ``connection structures''.
In Figure \ref{fig:diagram_structure}, we have grouped the diagrams according to their connection structures. In Figure \ref{fig:diagram_structure_diagrams} we represent these connection structures pictorially.\footnote{We note that this is not the only way to compute the contractions. Alternatively, one can insert a Dirac delta function to enforce the symmetry at each vertex; this corresponds to momentum conservation at each vertex. However, in this paper, we will proceed with the first approach, since it will allow us to easily write down the contribution to the trispectrum for any of the 48 diagrams once we have calculated this result for a single diagram from each of the four connection structures.}

{\bf Diagrams with the same connection structure are related to one another by interchanging the subscripts on the external momenta $\vb k_i$.}\footnote{The subscripts of the loop momenta $\vb q_i$ can be changed as well since these are integration variables.} However, for diagrams with different connection structures it is not possible to do such an interchange of the momenta. {\bf In this sense, diagrams with different connection structures are completely independent of one another.} 

In this section, we will compute the contribution to the trispectrum for each connection structure. We may consider just a single diagram for each connection structure, since the final result for two diagrams with the same connection structure can be related by simply relabeling the subscripts of the external momenta. 

We now perform the computations for four independent diagrams, one for each connection structure. We will call these ``Reference Diagrams''; they are shown in Figure \ref{Diagrams1-4}. We have
\begin{align}    
(1) \hspace{5 mm} & \delta_{\rm D}^{[3]}(\vb q_2 + \vb k_4 - \vb q_4) \delta_{\rm D}^{[3]}(\vb q_1 + \vb k_3 - \vb q_3) \delta_{\rm D}^{[3]}(\vb q_4 + \vb k_1 - \vb q_1) \delta_{\rm D}^{[3]}(\vb q_3 + \vb k_2 - \vb q_2) \nonumber \\ \nonumber \\
(2) \hspace{5 mm} & \delta_{\rm D}^{[3]}(\vb k_1 - \vb q_1 + \vb k_2 - \vb q_2) \delta_{\rm D}^{[3]}(\vb q_1 + \vb k_4 - \vb q_4) \delta_{\rm D}^{[3]}(\vb q_2 + \vb k_3 - \vb q_3) \delta_{\rm D}^{[3]}(\vb q_3 + \vb q_4)  \nonumber \\ \nonumber \\
(3) \hspace{5 mm} & \delta_{\rm D}^{[3]}(\vb k_1 - \vb q_1 + \vb k_2 - \vb q_2) \delta_{\rm D}^{[3]}(\vb k_3 - \vb q_3 + \vb k_4 - \vb q_4) \delta_{\rm D}^{[3]}(\vb q_1 + \vb q_3) \delta_{\rm D}^{[3]}(\vb q_2 + \vb q_4)  \nonumber \\ \nonumber \\
(4) \hspace{5 mm} & \delta_{\rm D}^{[3]}(\vb k_1 - \vb q_1 + \vb k_4 - \vb q_4) \delta_{\rm D}^{[3]}(\vb q_4 + \vb k_2 - \vb q_2) \delta_{\rm D}^{[3]}(\vb q_2 + \vb k_3 - \vb q_3) \delta_{\rm D}^{[3]}(\vb q_1 + \vb q_3) . \nonumber
\end{align} 
In order to simplify the trispectrum computation, we will rewrite the Dirac delta functions as FTs of unity, and then use the plane wave expansion to separate the dependence on each argument. This Dirac delta function expansion is described in more detail in Appendix \ref{AppendixA}.

We now determine the contribution to the trispectrum from each of the four connection structures.
\subsection{First Connection Structure Example: \hyperref[Diagrams1-4]{\hyperref[Diagrams1-4]{Reference Diagram 1}}}
\label{sec:intro}
We begin by inserting the first term in Eq. (\ref{Contr_1}), which corresponds to \hyperref[Diagrams1-4]{Reference Diagram 1} in Figure \ref{Diagrams1-4}, into Eq. (\ref{J_corr}), 
\begin{align}
\label{eq:src_int1}
&\expval{J(\tau_1, \vb k_1) J(\tau_2, \vb k_2) J(\tau_3, \vb k_3) J(\tau_4, \vb k_4)}^{(1)} \equiv  \\
&\frac{1}{\Lambda^4 a^4(\tau)} \int \int \int \int \dd[3] \vb q_1 \ \dd[3] \vb q_2 \ \dd[3] \vb q_3 \ \dd[3] \vb q_4 \; q_1 \; q_2 \; q_3 \; q_4 \left [\bm{\hat{\epsilon}}_-(\vu q_1) \vdot \bm{\hat{\epsilon}}_-(\widehat{\vb k_1 - \vb q_1}) \right] \nonumber \\
&\times\left [\bm{\hat{\epsilon}}_-(\vu q_2) \vdot \bm{\hat{\epsilon}}_-(\widehat{\vb k_2 - \vb q_2}) \right ] \left [\bm{\hat{\epsilon}}_-(\vu q_3) \vdot \bm{\hat{\epsilon}}_-(\widehat{\vb k_3 - \vb q_3}) \right ] \left [\bm{\hat{\epsilon}}_-(\vu q_4) \vdot \bm{\hat{\epsilon}}_-(\widehat{\vb k_4 - \vb q_4}) \right ] 
\delta_{\rm D}^{[3]}(\vb q_2 + \vb k_4 - \vb q_4) \nonumber \\ 
&\times \delta_{\rm D}^{[3]}(\vb q_1 + \vb k_3 - \vb q_3) \delta_{\rm D}^{[3]}(\vb k_1 - \vb q_1 + \vb q_4) \delta_{\rm D}^{[3]}(\vb k_2 - \vb q_2 + \vb q_3) 
A_-^{'}(\tau_1, |\vb k_1 -  \vb q_1|) A_-(\tau_1, q_1) \nonumber \\
&\times A_-^{'}(\tau_2, |\vb k_2 -  \vb q_2|) A_-(\tau_2, q_2) A_-^{'}(\tau_3, |\vb k_3 -  \vb q_3|)A_-(\tau_3, q_3) A_-^{'}(\tau_4, |\vb k_4 -  \vb q_4|)A_-(\tau_4, q_4). \nonumber
\end{align}
Upon using the expansion (\ref{eq:3D_Dird}), these Dirac delta functions become
\begin{align}
&\delta_{\rm D}^{[3]}(\vb q_2 + \vb k_4 - \vb q_4) \delta_{\rm D}^{[3]}(\vb q_1 + \vb k_3 - \vb q_3) \delta_{\rm D}^{[3]}(\vb k_1 - \vb q_1 + \vb q_4) \delta_{\rm D}^{[3]}(\vb k_2 - \vb q_2 + \vb q_3) \label{eq:Deltas_1} \\ \nonumber 
&= 8 \sum_{\lambda_2 \ell_4 \lambda'_4} \sum_{\mu_2 m_4 \mu'_4} i^{\lambda_2 + \ell_4 + \lambda'_4} \mathcal{G}_{\lambda_2 \ell_4 \lambda'_4}^{\mu_2 m_4 \mu'_4} \mathcal{R}_{\lambda_2 \ell_4 \lambda'_4}(q_2, k_4, q_4) Y_{\lambda_2}^{\mu_2 *}(\vu q_2) Y_{\ell_4}^{m_4 *}(\vu k_4) Y_{\lambda'_4}^{\mu'_4 *}(-\vu q_4) \nonumber \\
&\times 8 \sum_{\lambda_1 \ell_3 \lambda'_3} \sum_{\mu_1 m_3 \mu'_3} i^{\lambda_1 + \ell_3 + \lambda'_3} \mathcal{G}_{\lambda_1 \ell_3 \lambda'_3}^{\mu_1 m_3 \mu'_3} \mathcal{R}_{\lambda_1 \ell_3 \lambda'_3}(q_1, k_3, q_3) Y_{\lambda_1}^{\mu_1 *}(\vu q_1) Y_{\ell_3}^{m_3 *}(\vu k_3) Y_{\lambda'_3}^{\mu'_3 *}(-\vu q_3) \nonumber \\
&\times 8 \sum_{\ell_1 \lambda'_1 \lambda_4} \sum_{m_1 \mu'_1 \mu_4} i^{\ell_1 + \lambda'_1 + \lambda_4} \mathcal{G}_{\ell_1 \lambda'_1 \lambda_4}^{m_1 \mu'_1 \mu_4} \mathcal{R}_{\ell_1 \lambda'_1 \lambda_4}(k_1, q_1, q_4) Y_{\ell_1}^{m_1 *}(\vu k_1) Y_{\lambda'_1}^{\mu'_1 *}(-\vu q_1) Y_{\lambda_4}^{\mu_4 *}(\vu q_4) \nonumber \\
&\times 8 \sum_{\ell_2 \lambda'_2 \lambda_3} \sum_{m_2 \mu'_2 \mu_3} i^{\ell_2 + \lambda'_2 + \lambda_3} \mathcal{G}_{\ell_2 \lambda'_2 \lambda_3}^{m_2 \mu'_2 \mu_3} \mathcal{R}_{\ell_2 \lambda'_2 \lambda_3}(k_2, q_2, q_3) Y_{\ell_2}^{m_2 *}(\vu k_2) Y_{\lambda'_2}^{\mu'_2 *}(-\vu q_2) Y_{\lambda_3}^{\mu_3 *}(\vu q_3). \nonumber
\end{align}
$\mathcal{G}$ is the Gaunt integral and $\mathcal{R}$ is an overlap integral of spherical Bessel functions; $\mathcal{G}$ and $\mathcal{R}$ are defined mathematically in Eqs. (\ref{eq:Gaunt}) and (\ref{eq:R_3}), respectively. We use $\ell_i, \lambda_i, \lambda'_i$ as the angular momenta corresponding to vectors $\vb k_i, \vb q_i$, and  $-\vb q_i$ respectively. In these sums, $\ell_i, \lambda_i, \lambda'_i \in \mathbb{N}_0$,\footnote{$\mathbb{N}_0$ is the set of natural numbers including zero.} and the $z$-components of the angular momenta $m_i, \mu_i, \mu'_i \in \mathbb{N}_0$. Each $z$-component is controlled by its corresponding angular momentum, i.e. $-\ell_i \leq m_i \leq \ell_i$.

Using Eq. (\ref{eq:Deltas_1}) in Eq. (\ref{eq:src_int1}), and expanding the $\mathcal{R}$'s using Eq.  (\ref{eq:R_3}), we find
\begin{align}
&\expval{J(\tau, \vb k_1) J(\tau, \vb k_2) J(\tau, \vb k_3) J(\tau, \vb k_4)}^{(1)} = \\
&\frac{4{,}096}{\Lambda^4 a^4(\tau)}\sum_{\vb \Bell, \vb m} \sum_{\bm{\lambda}, \bm{\mu}} \sum_{\bm \lambda', \bm \mu'} \Bigg [\prod_{j=1}^{4} i^{\ell_j + \lambda_j + \lambda'_j} Y_{\ell_j}^{m_j *}(\vu k_j) \Bigg ] \mathcal{G}_{\lambda_2 \ell_4 \lambda'_4}^{\mu_2 m_4 \mu'_4} \mathcal{G}_{\lambda_1 \ell_3 \lambda'_3}^{\mu_1 m_3 \mu'_3} \mathcal{G}_{\ell_1 \lambda'_1 \lambda_4}^{m_1 \mu'_1 \mu_4} \mathcal{G}_{\ell_2 \lambda'_2 \lambda_3}^{m_2 \mu'_2 \mu_3}  \nonumber \\ &\times
\int \int \int \int \dd x_1 \, \dd x_2 \, \dd x_3 \, \dd x_4 \, \, x_1^2 \, x_2^2 \, x_3^2 \, x_4^2 \, \, j_{\ell_4}(k_4 x_1) j_{\ell_3}(k_3 x_2) j_{\ell_1}(k_1 x_3) j_{\ell_2}(k_2 x_4) \nonumber \\
&\times \int \dd[3] \vb q_4 \, q_4[\bm{\hat{\epsilon}}_-(\vu q_4) \vdot \bm{\hat{\epsilon}}_-(\widehat{\vb k_4 - \vb q_4})]  Y_{\lambda'_4}^{\mu'_4 *}(-\vu{q}_4) Y_{\lambda_4}^{\mu_4 *}(\vu{q}_4) A_-^{'}(\tau_4, |\vb k_4 -  \vb q_4|)A_-(\tau_4, q_4) \nonumber \\ 
&\hspace{102mm}\times j_{\lambda'_4}(q_4 x_1) j_{\lambda_4}(q_4 x_3) \nonumber \\
&\times \int \dd[3] \vb q_3 \, q_3 [\bm{\hat{\epsilon}}_-(\vu q_3) \vdot \bm{\hat{\epsilon}}_-(\widehat{\vb k_3 - \vb q_3})] Y_{\lambda'_3}^{\mu'_3 *}(-\vu{q}_3) Y_{\lambda_3}^{\mu_3 *}(\vu{q}_3) A_-(\tau_3, |\vb k_3 -  \vb q_3|)A_-(\tau_3, q_3)\nonumber \\ 
&\hspace{102mm}\times j_{\lambda'_3}(q_3 x_2)j_{\lambda_3}(q_3 x_4) \nonumber \\
&\times \int \dd[3] \vb q_2 \, q_2 [\bm{\hat{\epsilon}}_-(\vu q_2) \vdot \bm{\hat{\epsilon}}_-(\widehat{\vb k_2 - \vb q_2})] Y_{\lambda'_2}^{\mu'_2 *}(-\vu{q}_2) Y_{\lambda_2}^{\mu_2 *}(\vu{q}_2) A_-^{'}(\tau_2, |\vb k_2 -  \vb q_2|) A_-(\tau_2, q_2) \nonumber \\ 
&\hspace{102mm}\times j_{\lambda'_2}(q_2 x_4) j_{\lambda_2}(q_2 x_1) \nonumber \\
&\times \int \dd[3] \vb q_1 \, q_1 [\bm{\hat{\epsilon}}_-(\vu q_1) \vdot \bm{\hat{\epsilon}}_-(\widehat{\vb k_1 - \vb q_1})] Y_{\lambda'_1}^{\mu'_1 *}(-\vu{q}_1) Y_{\lambda_1}^{\mu_1 *}(\vu{q}_1)  A_-^{'}(\tau_1, |\vb k_1 -  \vb q_1|) A_-(\tau_1, q_1)\nonumber \\ 
&\hspace{102mm}\times j_{\lambda'_1}(q_1 x_3) j_{\lambda_1}(q_1 x_2). \nonumber
\end{align}
Here we have used the notation $\Bell \equiv (\ell_1, \ell_2, \ell_3, \ell_4)$ and $\vb m \equiv (m_1, m_2, m_3, m_4)$. This notation also shortens the indexing of the summations, so that we have 
\begin{align}
\sum_{\vb \Bell, \vb m} \equiv \sum_{\ell_1 \ell_2 \ell_3 \ell_4} \sum_{m_1 m_2 m_3 m_4}.
\end{align}
We now linearize each product of spherical harmonics using Eqs. (\ref{eq: Y_props}) and (\ref{eq:Y_contr}). We find
\begin{align}
&\expval{J(\tau, \vb k_1) J(\tau, \vb k_2) J(\tau, \vb k_3) J(\tau, \vb k_4)}^{(1)} = \\
&\frac{4{,}096}{\Lambda^4 a^4(\tau)}\sum_{\vb \Bell, \vb m} \sum_{\bm{\lambda}, \bm{\mu}} \sum_{\bm \lambda', \bm \mu'} \Bigg [\prod_{j=1}^{4} i^{\ell_j + \lambda_j + \lambda'_j} (-1)^{\lambda'_j + \mu_j +\mu'_j} Y_{\ell_j}^{m_j *}(\vu k_j) \sum_{\Lambda_j M_j} (-1)^{M_j} \mathcal{G}_{\lambda'_j \lambda_j \Lambda_j}^{-\mu'_j -\mu_j -M_j} \Bigg ] \nonumber \\
&\times \mathcal{G}_{\lambda_2 \ell_4 \lambda'_4}^{\mu_2 m_4 \mu'_4} \mathcal{G}_{\lambda_1 \ell_3 \lambda'_3}^{\mu_1 m_3 \mu'_3} \mathcal{G}_{\ell_1 \lambda'_1 \lambda_4}^{m_1 \mu'_1 \mu_4} \mathcal{G}_{\ell_2 \lambda'_2 \lambda_3}^{m_2 \mu'_2 \mu_3}  \nonumber \\ &\times
\int \int \int \int \dd x_1 \dd x_2 \, \dd x_3 \, \dd x_4 \, \, x_1^2 x_2^2 x_3^2 x_4^2 \, \, j_{\ell_4}(k_4 x_1) j_{\ell_3}(k_3 x_2) j_{\ell_1}(k_1 x_3) j_{\ell_2}(k_2 x_4) \nonumber \\
&\times \int \dd[3] \vb q_4 \, q_4[\bm{\hat{\epsilon}}_-(\vu q_4) \vdot \bm{\hat{\epsilon}}_-(\widehat{\vb k_4 - \vb q_4})] Y_{\Lambda_4}^{M_4}(\vu q_4) A_-^{'}(\tau_4, |\vb k_4 -  \vb q_4|)A_-(\tau_4, q_4) j_{\lambda'_4}(q_4 x_1) j_{\lambda_4}(q_4 x_3) \nonumber \\
&\times \int \dd[3] \vb q_3 \, q_3 [\bm{\hat{\epsilon}}_-(\vu q_3) \vdot \bm{\hat{\epsilon}}_-(\widehat{\vb k_3 - \vb q_3})] Y_{\Lambda_3}^{M_3}(\vu q_3) A_-^{'}(\tau_3, |\vb k_3 -  \vb q_3|)A_-(\tau_3, q_3)j_{\lambda'_3}(q_3 x_2)j_{\lambda_3}(q_3 x_4) \nonumber \\
&\times \int \dd[3] \vb q_2 \, q_2 [\bm{\hat{\epsilon}}_-(\vu q_2) \vdot \bm{\hat{\epsilon}}_-(\widehat{\vb k_2 - \vb q_2})] Y_{\Lambda_2}^{M_2}(\vu q_2) A_-^{'}(\tau_2, |\vb k_2 -  \vb q_2|) A_-(\tau_2, q_2) j_{\lambda'_2}(q_2 x_4) j_{\lambda_2}(q_2 x_1) \nonumber \\
&\times \int \dd[3] \vb q_1 \, q_1 [\bm{\hat{\epsilon}}_-(\vu q_1) \vdot \bm{\hat{\epsilon}}_-(\widehat{\vb k_1 - \vb q_1})] Y_{\Lambda_1}^{M_1}(\vu q_1)  A_-^{'}(\tau_1, |\vb k_1 -  \vb q_1|) A_-(\tau_1, q_1) j_{\lambda'_1}(q_1 x_3) j_{\lambda_1}(q_1 x_2). \nonumber
\end{align}
Our expression may now be written in the simplified form
\begin{align}
&\expval{J(\tau, \vb k_1) J(\tau, \vb k_2) J(\tau, \vb k_3) J(\tau, \vb k_4)}^{(1)} = \\
&\frac{4{,}096}{\Lambda^4 a^4(\tau)}\sum_{\vb \Bell, \vb m} \sum_{\bm{\lambda}, \bm{\mu}} \sum_{\bm \lambda', \bm \mu'} \Bigg [\prod_{j=1}^{4} i^{\ell_j + \lambda_j + \lambda'_j} (-1)^{\lambda'_j + \mu_j + \mu'_j}  Y_{\ell_j}^{m_j *}(\vu k_j) \sum_{\Lambda_j M_j} (-1)^{M_j} \mathcal{G}_{\lambda'_j \lambda_j \Lambda_j}^{-\mu'_j -\mu_j -M_j} \Bigg ] \nonumber \\
&\times \mathcal{G}_{\lambda_2 \ell_4 \lambda'_4}^{\mu_2 m_4 \mu'_4} \mathcal{G}_{\lambda_1 \ell_3 \lambda'_3}^{\mu_1 m_3 \mu'_3} \mathcal{G}_{\ell_1 \lambda'_1 \lambda_4}^{m_1 \mu'_1 \mu_4} \mathcal{G}_{\ell_2 \lambda'_2 \lambda_3}^{m_2 \mu'_2 \mu_3}  \nonumber \\ &\times
\int \int \int \int \dd x_1 \, \dd x_2 \, \dd x_3 \, \dd x_4 \, \, x_1^2 \, x_2^2 \, x_3^2 \, x_4^2 \, \, j_{\ell_4}(k_4 x_1) j_{\ell_3}(k_3 x_2) j_{\ell_1}(k_1 x_3) j_{\ell_2}(k_2 x_4) \nonumber \\
&\times \text{I}_{\Lambda_1 \lambda_1 \lambda'_1}^{M_1}(\vb k_1, x_2, x_3, \tau_1) \text{I}_{\Lambda_2 \lambda_2 \lambda'_2}^{M_2}(\vb k_2, x_1, x_4, \tau_2) \text{I}_{\Lambda_3 \lambda_3 \lambda'_3}^{M_3}(\vb k_3, x_4, x_2, \tau_3) \text{I}_{\Lambda_4 \lambda_4 \lambda'_4}^{M_4}(\vb k_4, x_3, x_1, \tau_4) \nonumber
\end{align}
where
\begin{align} 
\text{I}_{\Lambda \lambda \lambda'}^{M}(\vb{k}, x, x', \tau') \equiv \int \dd[3] &\vb{q} \ q [\bm{\hat{\epsilon}}_-(\vu{q})\vdot\bm{\hat{\epsilon}}_-(\widehat{\vb k - \vb q})] Y_{\Lambda}^{M}(\vu q) A_-^{'}(\tau', |\vb{k - q}|) A_-(\tau', q) j_{\lambda}(q x) j_{\lambda'}(q x'). 
\label{eq:I}
\end{align}
Inserting this expression into Eq. (\ref{eq:inv_4PCF}), we find that the contribution to the trispectrum due to \hyperref[Diagrams1-4]{Reference Diagram 1} is 
\begin{align}
&\expval{Q^{\text{inv}}(\tau, \vb k_1) Q^{\text{inv}}(\tau, \vb k_2)  Q^{\text{inv}}(\tau, \vb k_3) Q^{\text{inv}}(\tau, \vb k_4)}^{(1)} = \label{eq:Trispectrum_1}  \\  
&\frac{4{,}096}{\Lambda^4 a^4(\tau)}\sum_{\vb \Bell, \vb m} \sum_{\bm{\lambda}, \bm{\mu}} \sum_{\bm \lambda', \bm \mu'} \Bigg [\prod_{j=1}^{4}  i^{\ell_j + \lambda_j + \lambda'_j} (-1)^{\lambda'_j + \mu_j + \mu'_j}  Y_{\ell_j}^{m_j *}(\vu k_j) \sum_{\Lambda_j M_j} (-1)^{M_j} \mathcal{G}_{\lambda'_j \lambda_j \Lambda_j}^{-\mu'_j -\mu_j -M_j} \Bigg ] \nonumber \\
&\times \mathcal{G}_{\lambda_2 \ell_4 \lambda'_4}^{\mu_2 m_4 \mu'_4} \mathcal{G}_{\lambda_1 \ell_3 \lambda'_3}^{\mu_1 m_3 \mu'_3} \mathcal{G}_{\ell_1 \lambda'_1 \lambda_4}^{m_1 \mu'_1 \mu_4} \mathcal{G}_{\ell_2 \lambda'_2 \lambda_3}^{m_2 \mu'_2 \mu_3}  \nonumber \\ &\times
\int \int \int \int \dd x_1 \, \dd x_2 \, \dd x_3 \, \dd x_4 \, \, x_1^2 \, x_2^2 \, x_3^2 \, x_4^2 \, \, j_{\ell_4}(k_4 x_1) j_{\ell_3}(k_3 x_2) j_{\ell_1}(k_1 x_3) j_{\ell_2}(k_2 x_4) \nonumber \\
&\times \int_{-\infty}^0 \dd \tau_4 \; G(\tau, \tau_4, k_4) \text{I}_{\Lambda_4 \lambda_4 \lambda'_4}^{M_4}(\vb k_4, x_3, x_1, \tau_4) \int_{-\infty}^0 \dd \tau_3 \; G(\tau, \tau_3, k_3) \text{I}_{\Lambda_3 \lambda_3 \lambda'_3}^{M_3}(\vb k_3, x_4, x_2, \tau_3) \nonumber \\ 
&\times \int_{-\infty}^0 \dd \tau_2 \; G(\tau, \tau_2, k_2) \text{I}_{\Lambda_2 \lambda_2 \lambda'_2}^{M_2}(\vb k_2, x_1, x_4, \tau_2) \int_{-\infty}^0 \dd \tau_1 \; G(\tau, \tau_1, k_1) \text{I}_{\Lambda_1 \lambda_1 \lambda'_1}^{M_1}(\vb k_1, x_2, x_3, \tau_1). \nonumber 
\end{align}
\subsection{Second Connection Structure Example: \hyperref[Diagrams1-4]{Reference Diagram 2}}
The contraction combination corresponding to \hyperref[Diagrams1-4]{Reference Diagram 2} in Figure \ref{Diagrams1-4} is

\[
\langle
 \wick{
        \c1 a_-(\vb k_1 - \vb q_1) \c2 a_-(\vb q_1) \c1 a_-^{\dag}(\vb q_2 - \vb k_2) \c3 a_-(\vb q_2)
         \c3 a_-^{\dag}(\vb q_3 - \vb k_3) \c4 a_-(\vb q_3) \c2 a_-^{\dag}(\vb q_4 - \vb k_4) \c4 a_-^{\dag}(-\vb q_4)
  }
\rangle .
\]
The corresponding source-term correlation function is
\begin{align}
&\expval{J(\tau, \vb k_1) J(\tau, \vb k_2) J(\tau, \vb k_3) J(\tau, \vb k_4)}^{(2)} = \label{eq:src_Int2}\\
&\frac{1}{\Lambda^4 a^4(\tau)}\int \int \int \int \dd[3] \vb q_1 \ \dd[3] \vb q_2 \ \dd[3] \vb q_3 \ \dd[3] \vb q_4 \; q_1 \; q_2 \; q_3 \; q_4 [\bm{\hat{\epsilon}}_-(\vu q_1) \vdot \bm{\hat{\epsilon}}_-(\widehat{\vb k_1 - \vb q_1})] [\bm{\hat{\epsilon}}_-(\vu q_2) \vdot \bm{\hat{\epsilon}}_-(\widehat{\vb k_2 - \vb q_2})] \nonumber\\  
&\times [\bm{\hat{\epsilon}}_-(\vu q_3) \vdot \bm{\hat{\epsilon}}_-(\widehat{\vb k_3 - \vb q_3})][\bm{\hat{\epsilon}}_-(\vu q_4) \vdot \bm{\hat{\epsilon}}_-(\widehat{\vb k_4 - \vb q_4})] \delta_{\rm D}^{[3]}(\vb k_1 - \vb q_1 + \vb k_2 - \vb q_2) \delta_{\rm D}^{[3]}(\vb q_1 + \vb k_4 - \vb q_4) \nonumber \\
& \times \delta_{\rm D}^{[3]}(\vb{k_3 - q_3 + q_2})\delta_{\rm D}^{[3]}(\vb q_3 + \vb q_4) A_-^{'}(\tau_1, |\vb k_1 -  \vb q_1|) A_-(\tau_1, q_1) A_-^{'}(\tau_2, |\vb k_2 -  \vb q_2|) A_-(\tau_2, q_2) \nonumber \\
&\times A_-^{'}(\tau_3, |\vb k_3 -  \vb q_3|)A_-(\tau_3, q_3) A_-^{'}(\tau_4, |\vb k_4 -  \vb q_4|)A_-^{}(\tau_4, q_4). \nonumber 
\end{align}
We proceed in precisely the same manner as in the previous case, expanding the Dirac delta functions. This time we must also employ Eqs. (\ref{eq:2D_Dird}) and (\ref{eq:4D_Dird}). We have
\begin{align}
&\delta_{\rm D}^{[3]}(\vb k_1 - \vb q_1 + \vb k_2 - \vb q_2) \delta_{\rm D}^{[3]}(\vb q_1 + \vb k_4 - \vb q_4) \delta_{\rm D}^{[3]}(\vb{k_3 - q_3 + q_2}) \delta_{\rm D}^{[3]}(\vb q_3 + \vb q_4) \label{eq:Deltas_2} \\ \nonumber
&= 32 \pi \sum_{\vb \Bell, \vb m} \sum_{\bm{\lambda}, \bm{\mu}} \sum_{\bm \lambda', \bm \mu'} \sum_{A,B} i^{\ell_1 + \ell_2 + \lambda'_1 + \lambda'_2} \mathcal{K}_{\ell_1 \lambda'_1 \ell_2 \lambda'_2 A}^{m_1 \mu'_1 m_2 \mu'_2 B} \mathcal{R}_{\ell_1 \lambda'_1 \ell_2 \lambda'_2 A}(k_1, q_1, k_2, q_2) \\ 
&\times Y_{\ell_1}^{m_1 *}(\vu k_1) Y_{\lambda'_1}^{\mu'_1 *}(-\vu q_1) Y_{\ell_2}^{m_2 *}(\vu k_2) Y_{\lambda'_2}^{\mu'_2 *}(-\vu q_2) \nonumber \\ \nonumber 
&\times 8 i^{\lambda_1 + \ell_4 + \lambda'_4} \mathcal{G}_{\lambda_1 \ell_4 \lambda'_4}^{\mu_1 m_4 \mu'_4} \mathcal{R}_{\lambda_1 \ell_4 \lambda'_4}(q_1, k_4, q_4) Y_{\lambda_1}^{\mu_1 *}(\vu q_1) Y_{\ell_4}^{m_4 *}(\vu k_4) Y_{\lambda'_4}^{\mu'_4 *}(-\vu q_4) \nonumber \\
&\times 8 i^{\ell_3 + \lambda'_3 + \lambda_2} \mathcal{G}_{\ell_3 \lambda'_3 \lambda_2}^{m_3 \mu'_3 \mu_2} \mathcal{R}_{\ell_3 \lambda'_3 \lambda_2}(k_3, q_3, q_2) Y_{\ell_3}^{m_3 *}(\vu k_3) Y_{\lambda'_3}^{\mu'_3 *}(-\vu q_3) Y_{\lambda_2}^{\mu_2 *}(\vu q_2) \nonumber \\
&\times \frac{2}{\pi} (-1)^{\lambda_3} \mathcal{R}_{\lambda_3 \lambda_4}(q_3, q_4) Y_{\lambda_3}^{\mu_3 *}(\vu q_3)Y_{\lambda_4}^{\mu_4}(\vu q_4) \kron_{\lambda_3 \lambda_4} \kron_{\mu_3 \mu_4}, \nonumber
\end{align}
where $\mathcal{K}_{\ell_1 \ell_2 \ell_3 \ell_4 L}^{m_1 m_2 m_3 m_4 M}$ is defined in Appendix \ref{AppendixA}.
After using Eq. (\ref{eq:Deltas_2}) in Eq. (\ref{eq:src_Int2}), we obtain
\begin{align}
&\expval{J(\tau, \vb k_1) J(\tau, \vb k_2) J(\tau, \vb k_3) J(\tau, \vb k_4)}^{(2)} =  \\ \nonumber
&\frac{4{,}096}{\Lambda^4 a^4(\tau)} \sum_{\vb \Bell, \vb m} \sum_{\bm{\lambda}, \bm{\mu}} \sum_{\bm \lambda', \bm \mu'} \sum_{A,B} \Bigg [\prod_{j=1}^{4} i^{\ell_j + \lambda_j + \lambda'_j} Y_{\ell_j}^{m_j *}(\vu k_j) \Bigg ] \mathcal{K}_{\ell_1 \lambda'_1 \ell_2 \lambda'_2 A}^{m_1 \mu'_1 m_2 \mu'_2 B} \mathcal{G}_{\lambda_1 \ell_4 \lambda'_4}^{\mu_1 m_4 \mu'_4} \mathcal{G}_{\ell_3 \lambda'_3 \lambda_2}^{m_3 \mu'_3 \mu_2} \kron_{\lambda_3 \lambda_4} \kron_{\mu_3 \mu_4} \\ \nonumber 
&\times \int \int \int \int \dd x_1 \, \dd x_2 \, \dd x_3 \, \dd x_4 \, \, x_1^2 \, x_2^2 \, x_3^2 \, x_4^2 \, \, j_{\ell_1}(k_1 x_1) j_{\ell_2}(k_2 x_1) j_{\ell_4}(k_4 x_2) j_{\ell_3}(k_3 x_3)\\
&\times \int \dd[3] \vb q_4 \, q_4[\bm{\hat{\epsilon}}_-(\vu q_4) \vdot \bm{\hat{\epsilon}}_-(\widehat{\vb k_4 - \vb q_4})]  Y_{\lambda'_4}^{\mu'_4 *}(-\vu{q}_4) Y_{\lambda_4}^{\mu_4}(\vu{q}_4) A_-^{'}(\tau_4, |\vb k_4 -  \vb q_4|)A_-(\tau_4, q_4)\nonumber \\ 
&\hspace{102mm}\times j_{\lambda'_4}(q_4 x_2) j_{\lambda_4}(q_4 x_4) \nonumber \\
&\times \int \dd[3] \vb q_3 \, q_3 [\bm{\hat{\epsilon}}_-(\vu q_3) \vdot \bm{\hat{\epsilon}}_-(\widehat{\vb k_3 - \vb q_3})] Y_{\lambda'_3}^{\mu'_3 *}(-\vu{q}_3) Y_{\lambda_3}^{\mu_3 *}(\vu{q}_3) A_-^{'}(\tau_3, |\vb k_3 -  \vb q_3|)A_-(\tau_3, q_3)\nonumber \\ 
&\hspace{102mm}\times j_{\lambda'_3}(q_3 x_3)j_{\lambda_3}(q_3 x_4) \nonumber \\
&\times \int \dd[3] \vb q_2 \, q_2 [\bm{\hat{\epsilon}}_-(\vu q_2) \vdot \bm{\hat{\epsilon}}_-(\widehat{\vb k_2 - \vb q_2})] Y_{\lambda'_2}^{\mu'_2 *}(-\vu{q}_2) Y_{\lambda_2}^{\mu_2 *}(\vu{q}_2) A_-^{'}(\tau_2, |\vb k_2 -  \vb q_2|) A_-(\tau_2, q_2)\nonumber \\ 
&\hspace{102mm}\times j_{\lambda'_2}(q_2 x_1) j_{\lambda_2}(q_2 x_3) \nonumber \\
&\times \int \dd[3] \vb q_1 \, q_1 [\bm{\hat{\epsilon}}_-(\vu q_1) \vdot \bm{\hat{\epsilon}}_-(\widehat{\vb k_1 - \vb q_1})] Y_{\lambda'_1}^{\mu'_1 *}(-\vu{q}_1) Y_{\lambda_1}^{\mu_1 *}(\vu{q}_1)  A_-^{'}(\tau_1, |\vb k_1 -  \vb q_1|) A_-(\tau_1, q_1)\nonumber \\ 
&\hspace{102mm}\times j_{\lambda'_1}(q_1 x_1) j_{\lambda_1}(q_1 x_2). \nonumber
\end{align}
We again linearize the spherical harmonics, finding
\begin{align}
&\expval{J(\tau, \vb k_1) J(\tau, \vb k_2) J(\tau, \vb k_3) J(\tau, \vb k_4)}^{(2)} = \\ \nonumber
&\frac{4{,}096}{\Lambda^4 a^4(\tau)} \sum_{\vb \Bell, \vb m} \sum_{\bm{\lambda}, \bm{\mu}} \sum_{\bm \lambda', \bm \mu'} \sum_{A,B} \Bigg [\prod_{j=1}^{4} i^{\ell_j + \lambda_j + \lambda'_j} Y_{\ell_j}^{m_j *}(\vu k_j) \sum_{\Lambda_j} \sum_{M_j} (-1)^{M_j} \Bigg ] \mathcal{K}_{\ell_1 \lambda'_1 \ell_2 \lambda'_2 A}^{m_1 \mu'_1 m_2 \mu'_2 B} \mathcal{G}_{\lambda_1 \ell_4 \lambda'_4}^{\mu_1 m_4 \mu'_4} \mathcal{G}_{\ell_3 \lambda'_3 \lambda_2}^{m_3 \mu'_3 \mu_2} \\
&\times \kron_{\lambda_3 \lambda_4} \kron_{\mu_3 \mu_4} (-1)^{\lambda'_4 + \mu'_4}  (-1)^{\lambda'_3 + \mu_3 + \mu'_3} (-1)^{\lambda'_2 + \mu_2 + \mu'_2} (-1)^{\lambda'_1 + \mu_1 + \mu'_1}  \nonumber\\ \nonumber
& \times \mathcal{G}_{\lambda'_4 \lambda_4 \Lambda_4}^{-\mu'_4 \mu_4 -M_4}\mathcal{G}_{\lambda'_3 \lambda_3 \Lambda_3}^{-\mu'_3 -\mu_3 -M_3} \mathcal{G}_{\lambda'_2 \lambda_2 \Lambda_2}^{-\mu'_2 -\mu_2 -M_2} \mathcal{G}_{\lambda'_1 \lambda_1 \Lambda_1}^{-\mu'_1 -\mu_1 -M_1}\\
&\times \int \int \int \int \dd x_1 \, \dd x_2 \, \dd x_3 \, \dd x_4 \, \, x_1^2 \, x_2^2 \, x_3^2 \, x_4^2 \, \, j_{\ell_1}(k_1 x_1) j_{\ell_2}(k_2 x_1) j_{\ell_4}(k_4 x_2) j_{\ell_3}(k_3 x_3) \nonumber\\
&\times \int \dd[3] \vb q_4 \, q_4[\bm{\hat{\epsilon}}_-(\vu q_4) \vdot \bm{\hat{\epsilon}}_-(\widehat{\vb k_4 - \vb q_4})] Y_{\Lambda_4}^{M_4}(\vu{q}_4) A_-^{'}(\tau_4, |\vb k_4 -  \vb q_4|)A_-(\tau_4, q_4) j_{\lambda'_4}(q_4 x_2) j_{\lambda_4}(q_4 x_4) \nonumber \\
&\times \int \dd[3] \vb q_3 \, q_3 [\bm{\hat{\epsilon}}_-(\vu q_3) \vdot \bm{\hat{\epsilon}}_-(\widehat{\vb k_3 - \vb q_3})] Y_{\Lambda_3}^{M_3}(\vu{q}_3)  A_-^{'}(\tau_3, |\vb k_3 -  \vb q_3|)A_-(\tau_3, q_3)j_{\lambda'_3}(q_3 x_3)j_{\lambda_3}(q_3 x_4) \nonumber \\
&\times \int \dd[3] \vb q_2 \, q_2 [\bm{\hat{\epsilon}}_-(\vu q_2) \vdot \bm{\hat{\epsilon}}_-(\widehat{\vb k_2 - \vb q_2})] Y_{\Lambda_2}^{M_2}(\vu{q}_2)  A_-^{'}(\tau_2, |\vb k_2 -  \vb q_2|) A_-(\tau_2, q_2) j_{\lambda'_2}(q_2 x_1) j_{\lambda_2}(q_2 x_3) \nonumber \\
&\times \int \dd[3] \vb q_1 \, q_1 [\bm{\hat{\epsilon}}_-(\vu q_1) \vdot \bm{\hat{\epsilon}}_-(\widehat{\vb k_1 - \vb q_1})] Y_{\Lambda_1}^{M_1}(\vu{q}_1)  A_-^{'}(\tau_1, |\vb k_1 -  \vb q_1|) A_-(\tau_1, q_1) j_{\lambda'_1}(q_1 x_1) j_{\lambda_1}(q_1 x_2). \nonumber   
\end{align}
Thus the contribution to the trispectrum due to \hyperref[Diagrams1-4]{Reference Diagram 2} is 
\begin{align}
&\expval{Q^{\text{inv}}(\tau, \vb k_1) Q^{\text{inv}}(\tau, \vb k_2)  Q^{\text{inv}}(\tau, \vb k_3) Q^{\text{inv}}(\tau, \vb k_4)}^{(2)} = \\
&\frac{4{,}096}{\Lambda^4 a^4(\tau)} \sum_{\vb \Bell, \vb m} \sum_{\bm{\lambda}, \bm{\mu}} \sum_{\bm \lambda', \bm \mu'} \sum_{A,B} \Bigg [\prod_{j=1}^{4} i^{\ell_j + \lambda_j + \lambda'_j} (-1)^{\lambda'_j + \mu_j + \mu'_j} Y_{\ell_j}^{m_j *}(\vu k_j) \sum_{\Lambda_j} \sum_{M_j} (-1)^{M_j} \Bigg ](-1)^{\mu_4} \nonumber \\
&\times \mathcal{K}_{\ell_1 \lambda'_1 \ell_2 \lambda'_2 A}^{m_1 \mu'_1 m_2 \mu'_2 B} \mathcal{G}_{\lambda_1 \ell_4 \lambda'_4}^{\mu_1 m_4 \mu'_4} \mathcal{G}_{\ell_3 \lambda'_3 \lambda_2}^{m_3 \mu'_3 \mu_2} \kron_{\lambda_3 \lambda_4} \kron_{\mu_3 \mu_4} \nonumber \mathcal{G}_{\lambda'_4 \lambda_4 \Lambda_4}^{-\mu'_4 \mu_4 -M_4}\mathcal{G}_{\lambda'_3 \lambda_3 \Lambda_3}^{-\mu'_3 -\mu_3 -M_3} \mathcal{G}_{\lambda'_2 \lambda_2 \Lambda_2}^{-\mu'_2 -\mu_2 -M_2} \mathcal{G}_{\lambda'_1 \lambda_1 \Lambda_1}^{-\mu'_1 -\mu_1 -M_1}\\
&\times \int \int \int \int \dd x_1 \, \dd x_2 \, \dd x_3 \, \dd x_4 \, \, x_1^2 \, x_2^2 \, x_3^2 \, x_4^2 \, \, j_{\ell_1}(k_1 x_1) j_{\ell_2}(k_2 x_1) j_{\ell_4}(k_4 x_2) j_{\ell_3}(k_3 x_3) \nonumber\\
&\times \int_{-\infty}^0 \dd \tau_4 \; G(\tau, \tau_4, k_4) \text{I}_{\Lambda_4 \lambda_4 \lambda'_4}^{M_4}(\vb k_4, x_4, x_2, \tau_4) \int_{-\infty}^0 \dd \tau_3 \; G(\tau, \tau_3, k_3) \text{I}_{\Lambda_3 \lambda_3 \lambda'_3}^{M_3}(\vb k_3, x_4, x_3, \tau_3) \nonumber \\ 
&\times \int_{-\infty}^0 \dd \tau_2 \; G(\tau, \tau_2, k_2) \text{I}_{\Lambda_2 \lambda_2 \lambda'_2}^{M_2}(\vb k_2, x_3, x_1, \tau_2) \int_{-\infty}^0 \dd \tau_1 \; G(\tau, \tau_1, k_1) \text{I}_{\Lambda_1 \lambda_1 \lambda'_1}^{M_1}(\vb k_1, x_2, x_1, \tau_1). \nonumber 
\end{align}

\subsection{Third Connection Structure Example: \hyperref[Diagrams1-4]{Reference Diagram 3}}
\hyperref[Diagrams1-4]{Reference Diagram 3} in Figure \ref{Diagrams1-4} arises from the contraction combination
\[
\langle
 \wick{
        \c1 a_-(\vb k_1 - \vb q_1) \c3 a_-(\vb q_1) \c1 a_-^{\dag}(\vb q_2 - \vb k_2) \c4 a_-(\vb q_2) 
        \c2 a_-(\vb k_3 - \vb q_3) \c3 a_-^{\dag}(-\vb q_3) \c2 a_-^{\dag}(\vb q_4 - \vb k_4) \c4 a_-^{\dag}(-\vb q_4)
  }
\rangle .
\]
The corresponding source-term correlation function is
\begin{align}
&\expval{J(\tau, \vb k_1) J(\tau, \vb k_2) J(\tau, \vb k_3) J(\tau, \vb k_4)}^{(3)} = \label{eq:src_Int3}\\
&\frac{1}{\Lambda^4 a^4(\tau)}\int \int \int \int \dd[3] \vb q_1 \ \dd[3] \vb q_2 \ \dd[3] \vb q_3 \ \dd[3] \vb q_4 \; q_1 \; q_2 \; q_3 \; q_4 [\bm{\hat{\epsilon}}_-(\vu q_1) \vdot \bm{\hat{\epsilon}}_-(\widehat{\vb k_1 - \vb q_1})] [\bm{\hat{\epsilon}}_-(\vu q_2) \vdot \bm{\hat{\epsilon}}_-(\widehat{\vb k_2 - \vb q_2})] \nonumber \\
&\times[\bm{\hat{\epsilon}}_-(\vu q_3) \vdot \bm{\hat{\epsilon}}_-(\widehat{\vb k_3 - \vb q_3})]  [\bm{\hat{\epsilon}}_-(\vu q_4) \vdot \bm{\hat{\epsilon}}_-(\widehat{\vb k_4 - \vb q_4})] 
\delta_{\rm D}^{[3]}(\vb k_1 - \vb q_1 + \vb k_2 - \vb q_2) \delta_{\rm D}^{[3]}(\vb k_3 - \vb q_3 + \vb k_4 - \vb q_4) \nonumber \\ 
&\times \delta_{\rm D}^{[3]}(\vb q_1 + \vb q_3) \delta_{\rm D}^{[3]}(\vb q_2 + \vb q_4)  A_-^{'}(\tau_1, |\vb k_1 -  \vb q_1|) A_-(\tau_1, q_1) A_-^{'}(\tau_2, |\vb k_2 -  \vb q_2|) A_-(\tau_2, q_2) \nonumber \\
&\times A_-^{'}(\tau_3, |\vb k_3 -  \vb q_3|)A_-(\tau_3, q_3) A_-^{'}(\tau_4, |\vb k_4 -  \vb q_4|)A_-(\tau_4, q_4). \nonumber 
\end{align}
We again use the Dirac delta function expansions given in Appendix \ref{AppendixA} and find
\begin{align}
&\expval{J(\tau, \vb k_1) J(\tau, \vb k_2) J(\tau, \vb k_3) J(\tau, \vb k_4)}^{(3)} = \\
&\frac{4{,}096}{\Lambda^4 a^4(\tau)} \sum_{\vb \Bell, \vb m} \sum_{\bm{\lambda}, \bm{\mu}} \sum_{\bm \lambda', \bm \mu'} \sum_{A,B} \sum_{A',B'} \Bigg [\prod_{j=1}^{4} i^{\ell_j + \lambda_j + \lambda'_j} Y_{\ell_j}^{m_j *}(\vu k_j) \Bigg ] \nonumber \\
& \times \mathcal{K}_{\ell_1 \lambda'_1 \ell_2 \lambda'_2 A}^{m_1 \mu'_1 m_2 \mu'_2 B} \mathcal{K}_{\ell_3 \lambda'_3 \ell_4 \lambda'_4 A'}^{m_3 \mu'_3 m_4 \mu'_4 B'} \kron_{\lambda_1 \lambda_3} \kron_{\mu_1 \mu_3} \kron_{\lambda_2 \lambda_4} \kron_{\mu_2 \mu_4} \nonumber \\ \nonumber 
&\times \int \int \int \int \dd[3] \vb x_1  \dd[3] \vb x_2  \dd[3] \vb x_3 \dd[3] \vb x_4 \, \, x_1^2 x_2^2 x_3^2 x_4^2 \, \, j_{\ell_1}(k_1 x_1) j_{\ell_2}(k_2 x_1) j_{\ell_3}(k_3 x_2) j_{\ell_4}(k_4 x_2)\\
&\times \int \dd[3] \vb q_4 \, q_4[\bm{\hat{\epsilon}}_-(\vu q_4) \vdot \bm{\hat{\epsilon}}_-(\widehat{\vb k_4 - \vb q_4})]  Y_{\lambda'_4}^{\mu'_4 *}(-\vu{q}_4) Y_{\lambda_4}^{\mu_4}(\vu{q}_4) A_-^{'}(\tau_4, |\vb k_4 -  \vb q_4|)A_-(\tau_4, q_4)\nonumber \\ 
&\hspace{102mm}\times j_{\lambda'_4}(q_4 x_2) j_{\lambda_4}(q_4 x_4) \nonumber \\
&\times \int \dd[3] \vb q_3 \, q_3 [\bm{\hat{\epsilon}}_-(\vu q_3) \vdot \bm{\hat{\epsilon}}_-(\widehat{\vb k_3 - \vb q_3})] Y_{\lambda'_3}^{\mu'_3 *}(-\vu{q}_3) Y_{\lambda_3}^{\mu_3 }(\vu{q}_3) A_-^{'}(\tau_3, |\vb k_3 -  \vb q_3|)A_-(\tau_3, q_3)\nonumber \\ 
&\hspace{102mm}\times j_{\lambda'_3}(q_3 x_2)j_{\lambda_3}(q_3 x_3) \nonumber \\
&\times \int \dd[3] \vb q_2 \, q_2 [\bm{\hat{\epsilon}}_-(\vu q_2) \vdot \bm{\hat{\epsilon}}_-(\widehat{\vb k_2 - \vb q_2})] Y_{\lambda'_2}^{\mu'_2 *}(-\vu{q}_2) Y_{\lambda_2}^{\mu_2 *}(\vu{q}_2) A_-^{'}(\tau_2, |\vb k_2 -  \vb q_2|) A_-(\tau_2, q_2)\nonumber \\ 
&\hspace{102mm}\times j_{\lambda'_2}(q_2 x_1) j_{\lambda_2}(q_2 x_4) \nonumber \\
&\times \int \dd[3] \vb q_1 \, q_1 [\bm{\hat{\epsilon}}_-(\vu q_1) \vdot \bm{\hat{\epsilon}}_-(\widehat{\vb k_1 - \vb q_1})] Y_{\lambda'_1}^{\mu'_1 *}(-\vu{q}_1) Y_{\lambda_1}^{\mu_1 *}(\vu{q}_1)  A_-^{'}(\tau_1, |\vb k_1 -  \vb q_1|) A_-(\tau_1, q_1)\nonumber \\ 
&\hspace{102mm}\times j_{\lambda'_1}(q_1 x_1) j_{\lambda_1}(q_1 x_3). \nonumber
\end{align}
Finally, the contribution to the trispectrum due to \hyperref[Diagrams1-4]{Reference Diagram 3} is
\begin{align}
&\expval{Q^{\text{inv}}(\tau, \vb k_1) Q^{\text{inv}}(\tau, \vb k_2)  Q^{\text{inv}}(\tau, \vb k_3) Q^{\text{inv}}(\tau, \vb k_4)}^{(3)} \equiv \label{Trispectrum_Type3} \\
&\frac{4{,}096}{\Lambda^4 a^4(\tau)} \sum_{\vb \Bell, \vb m} \sum_{\bm{\lambda}, \bm{\mu}} \sum_{\bm \lambda', \bm \mu'} \sum_{A,B} \sum_{A',B'} \Bigg [\prod_{j=1}^{4} i^{\ell_j + \lambda_j + \lambda'_j} (-1)^{\lambda'_j + \mu_j + \mu'_j} Y_{\ell_j}^{m_j *}(\vu k_j) \sum_{\Lambda_j} \sum_{M_j} (-1)^{M_j} \Bigg ](-1)^{\mu_3 + \mu_4} \nonumber \\
&\times \mathcal{K}_{\ell_1 \lambda'_1 \ell_2 \lambda'_2 A}^{m_1 \mu'_1 m_2 \mu'_2 B} \mathcal{K}_{\ell_3 \lambda'_3 \ell_4 \lambda'_4 A'}^{m_3 \mu'_3 m_4 \mu'_4 B'} \kron_{\lambda_1 \lambda_3} \kron_{\mu_1 \mu_3} \kron_{\lambda_2 \lambda_4} \kron_{\mu_2 \mu_4} \nonumber\\
&\times \nonumber \mathcal{G}_{\lambda'_4 \lambda_4 \Lambda_4}^{-\mu'_4 \mu_4 -M_4}\mathcal{G}_{\lambda'_3 \lambda_3 \Lambda_3}^{-\mu'_3 \mu_3 -M_3} \mathcal{G}_{\lambda'_2 \lambda_2 \Lambda_2}^{-\mu'_2 -\mu_2 -M_2} \mathcal{G}_{\lambda'_1 \lambda_1 \Lambda_1}^{-\mu'_1 -\mu_1 -M_1}\\
&\times \int \int \int \int \dd x_1 \, \dd x_2 \, \dd x_3 \, \dd x_4 \, \, x_1^2 \, x_2^2 \, x_3^2 \, x_4^2 \, \, j_{\ell_1}(k_1 x_1) j_{\ell_2}(k_2 x_1) j_{\ell_4}(k_4 x_2) j_{\ell_3}(k_3 x_2) \nonumber\\
&\times \int_{-\infty}^0 \dd \tau_4 \; G(\tau, \tau_4, k_4) \text{I}_{\Lambda_4 \lambda_4 \lambda'_4}^{M_4}(\vb k_4, x_4, x_2, \tau_4) \int_{-\infty}^0 \dd \tau_3 \; G(\tau, \tau_3, k_3) \text{I}_{\Lambda_3 \lambda_3 \lambda'_3}^{M_3}(\vb k_3, x_3, x_2, \tau_3) \nonumber \\ 
&\times \int_{-\infty}^0 \dd \tau_2 \; G(\tau, \tau_2, k_2) \text{I}_{\Lambda_2 \lambda_2 \lambda'_2}^{M_2}(\vb k_2, x_4, x_1, \tau_2) \int_{-\infty}^0 \dd \tau_1 \; G(\tau, \tau_1, k_1) \text{I}_{\Lambda_1 \lambda_1 \lambda'_1}^{M_1}(\vb k_1, x_3, x_1, \tau_1). \nonumber 
\end{align}

\subsection{Fourth Connection Structure Example: \hyperref[Diagrams1-4]{Reference Diagram 4}}
\hyperref[Diagrams1-4]{Reference Diagram 4} in Figure \ref{Diagrams1-4} arises from the contraction combination
\[
\langle
 \wick{
        \c1 a_-(\vb k_1 - \vb q_1) \c4 a_-(\vb q_1) \c2 a_-(\vb k_2 - \vb q_2) \c3 a_-(\vb q_2) 
        \c3 a_-^{\dag}(\vb q_3 - \vb k_3) \c4 a_-^{\dag}(-\vb q_3) \c1 a_-^{\dag}(\vb q_4 - \vb k_4) \c2 a_-^{\dag}(-\vb q_4) 
  }
\rangle. 
\] 
The corresponding source-term correlation function is 
\begin{align}
&\expval{J(\tau, \vb k_1) J(\tau, \vb k_2) J(\tau, \vb k_3) J(\tau, \vb k_4)}^{(4)} = \label{eq:src_Int4}\\
&\int \int \int \int \dd[3] \vb q_1 \ \dd[3] \vb q_2 \ \dd[3] \vb q_3 \ \dd[3] \vb q_4 [\bm{\hat{\epsilon}}_-(\vu q_1) \vdot \bm{\hat{\epsilon}}_-(\widehat{\vb k_1 - \vb q_1})] [\bm{\hat{\epsilon}}_-(\vu q_2) \vdot \bm{\hat{\epsilon}}_-(\widehat{\vb k_2 - \vb q_2})] [\bm{\hat{\epsilon}}_-(\vu q_3) \vdot \bm{\hat{\epsilon}}_-(\widehat{\vb k_3 - \vb q_3})] \nonumber \\ &\times [\bm{\hat{\epsilon}}_-(\vu q_4) \vdot \bm{\hat{\epsilon}}_-(\widehat{\vb k_4 - \vb q_4})] 
\delta_{\rm D}^{[3]}(\vb k_1 - \vb q_1 + \vb k_4 - \vb q_4) \delta_{\rm D}^{[3]}(\vb q_4 + \vb k_2 - \vb q_2) \delta_{\rm D}^{[3]}(\vb q_2 + \vb k_3 - \vb q_3) \delta_{\rm D}^{[3]}(\vb q_1 + \vb q_3) \nonumber \\ 
&\times A_-^{'}(\tau_1, |\vb k_1 -  \vb q_1|) A_-(\tau_1, q_1) A_-^{'}(\tau_2, |\vb k_2 -  \vb q_2|) A_-(\tau_2, q_2) \nonumber \\
&\times A_-^{'}(\tau_3, |\vb k_3 -  \vb q_3|)A_-(\tau_3, q_3) A_-^{'}(\tau_4, |\vb k_4 -  \vb q_4|)A_-(\tau_4, q_4). \nonumber 
\end{align}
We again expand the Dirac delta functions and obtain
\begin{align}
&\delta_{\rm D}^{[3]}(\vb k_1 - \vb q_1 + \vb k_4 - \vb q_4) \delta_{\rm D}^{[3]}(\vb q_4 + \vb k_2 - \vb q_2) \delta_{\rm D}^{[3]}(\vb q_2 + \vb k_3 - \vb q_3) \delta_{\rm D}^{[3]}(\vb q_1 + \vb q_3) \\ \nonumber
&= 32\pi \sum_{\vb \Bell, \vb m} \sum_{\bm{\lambda}, \bm{\mu}} \sum_{\bm \lambda', \bm \mu'} \sum_{A,B} i^{\ell_1 + \ell_4 + \lambda'_1 + \lambda'_4} \mathcal{K}_{\ell_1 \lambda'_1 \ell_4 \lambda'_4 A}^{m_1 \mu'_1 m_4 \mu'_4 B} \mathcal{R}_{\ell_1 \lambda'_1 \ell_4 \lambda'_4 A}(k_1, q_1, k_4, q_4) \\ 
&\times Y_{\ell_1}^{m_1 *}(\vu k_1) Y_{\lambda'_1}^{\mu'_1 *}(-\vu q_1) Y_{\ell_4}^{m_4 *}(\vu k_4) Y_{\lambda'_4}^{\mu'_4 *}(-\vu q_4) \nonumber \\ \nonumber 
&\times 8 i^{\lambda_4 + \ell_2 + \lambda'_2} \mathcal{G}_{\lambda_4 \ell_2 \lambda'_2}^{\mu_4 m_2 \mu'_2} \mathcal{R}_{\lambda_4 \ell_2 \lambda'_2}(q_4, k_2, q_2) Y_{\lambda_4}^{\mu_4 *}(\vu q_4) Y_{\ell_2}^{m_2 *}(\vu k_2) Y_{\lambda'_2}^{\mu'_2 *}(-\vu q_2) \nonumber \\
&\times 8 i^{\ell_3 + \lambda'_3 + \lambda_2} \mathcal{G}_{\ell_3 \lambda'_3 \lambda_2}^{m_3 \mu'_3 \mu_2} \mathcal{R}_{\ell_3 \lambda'_3 \lambda_2}(k_3, q_3, q_2) Y_{\ell_3}^{m_3 *}(\vu k_3) Y_{\lambda'_3}^{\mu'_3 *}(-\vu q_3) Y_{\lambda_2}^{\mu_2 *}(\vu q_2) \nonumber \\
&\times \frac{2}{\pi} (-1)^{\lambda_3} \mathcal{R}_{\lambda_1 \lambda_3}(q_1, q_3) Y_{\lambda_1}^{\mu_1 *}(\vu q_1)Y_{\lambda_3}^{\mu_3}(\vu q_3) \kron_{\lambda_1 \lambda_3} \kron_{\mu_1 \mu_3}. \nonumber
\end{align}
Now we have
\begin{align}
&\expval{J(\tau, \vb k_1) J(\tau, \vb k_2) J(\tau, \vb k_3) J(\tau, \vb k_4)}^{(4)} =  \\ \nonumber
&\frac{4{,}096}{\Lambda^4 a^4(\tau)} \sum_{\vb \Bell, \vb m} \sum_{\bm{\lambda}, \bm{\mu}} \sum_{\bm \lambda', \bm \mu'} \sum_{A,B} \Bigg [\prod_{j=1}^{4} i^{\ell_j + \lambda_j + \lambda'_j} Y_{\ell_j}^{m_j *}(\vu k_j) \Bigg ] \mathcal{K}_{\ell_1 \lambda'_1 \ell_4 \lambda'_4 A}^{m_1 \mu'_1 m_4 \mu'_4 B} \mathcal{G}_{\lambda_4 \ell_2 \lambda'_2}^{\mu_4 m_2 \mu'_2} \mathcal{G}_{\ell_3 \lambda'_3 \lambda_2}^{m_3 \mu'_3 \mu_2}    \kron_{\lambda_1 \lambda_3} \kron_{\mu_1 \mu_3} \\ \nonumber 
&\times \int \int \int \int \dd x_1 \, \dd x_2 \, \dd x_3 \, \dd x_4 \, \, x_1^2 \, x_2^2 \, x_3^2 \, x_4^2 \, \, j_{\ell_1}(k_1 x_1) j_{\ell_4}(k_4 x_1) j_{\ell_2}(k_2 x_2) j_{\ell_3}(k_3 x_3)\\
&\times \int \dd[3] \vb q_4 \, q_4[\bm{\hat{\epsilon}}_-(\vu q_4) \vdot \bm{\hat{\epsilon}}_-(\widehat{\vb k_4 - \vb q_4})]  Y_{\lambda'_4}^{\mu'_4 *}(-\vu{q}_4) Y_{\lambda_4}^{\mu_4 *}(\vu{q}_4) A_-^{'}(\tau_4, |\vb k_4 -  \vb q_4|)A_-(\tau_4, q_4)\nonumber \\ 
&\hspace{102mm}\times j_{\lambda'_4}(q_4 x_1) j_{\lambda_4}(q_4 x_2) \nonumber \\
&\times \int \dd[3] \vb q_3 \, q_3 [\bm{\hat{\epsilon}}_-(\vu q_3) \vdot \bm{\hat{\epsilon}}_-(\widehat{\vb k_3 - \vb q_3})] Y_{\lambda'_3}^{\mu'_3 *}(-\vu{q}_3) Y_{\lambda_3}^{\mu_3 }(\vu{q}_3) A_-^{'}(\tau_3, |\vb k_3 -  \vb q_3|)A_-(\tau_3, q_3)\nonumber \\ 
&\hspace{102mm}\times j_{\lambda'_3}(q_3 x_3)j_{\lambda_3}(q_3 x_4) \nonumber \\
&\times \int \dd[3] \vb q_2 \, q_2 [\bm{\hat{\epsilon}}_-(\vu q_2) \vdot \bm{\hat{\epsilon}}_-(\widehat{\vb k_2 - \vb q_2})] Y_{\lambda'_2}^{\mu'_2 *}(-\vu{q}_2) Y_{\lambda_2}^{\mu_2 *}(\vu{q}_2) A_-^{'}(\tau_2, |\vb k_2 -  \vb q_2|) A_-(\tau_2, q_2)\nonumber \\ 
&\hspace{102mm}\times j_{\lambda'_2}(q_2 x_2) j_{\lambda_2}(q_2 x_3) \nonumber \\
&\times \int \dd[3] \vb q_1 \, q_1 [\bm{\hat{\epsilon}}_-(\vu q_1) \vdot \bm{\hat{\epsilon}}_-(\widehat{\vb k_1 - \vb q_1})] Y_{\lambda'_1}^{\mu'_1 *}(-\vu{q}_1) Y_{\lambda_1}^{\mu_1 *}(\vu{q}_1)  A_-^{'}(\tau_1, |\vb k_1 -  \vb q_1|) A_-(\tau_1, q_1)\nonumber \\ 
&\hspace{102mm}\times j_{\lambda'_1}(q_1 x_1) j_{\lambda_1}(q_1 x_4). \nonumber
\end{align}
We again linearize the spherical harmonics, finding
\begin{align}
&\expval{J(\tau, \vb k_1) J(\tau, \vb k_2) J(\tau, \vb k_3) J(\tau, \vb k_4)}^{(4)} = \\ \nonumber
&\frac{4{,}096}{\Lambda^4 a^4(\tau)} \sum_{\vb \Bell, \vb m} \sum_{\bm{\lambda}, \bm{\mu}} \sum_{\bm \lambda', \bm \mu'} \sum_{A,B} \Bigg [\prod_{j=1}^{4} i^{\ell_j + \lambda_j + \lambda'_j} Y_{\ell_j}^{m_j *}(\vu k_j) \sum_{\Lambda_j} \sum_{M_j} (-1)^{M_j} \Bigg ] \mathcal{K}_{\ell_1 \lambda'_1 \ell_4 \lambda'_4 A}^{m_1 \mu'_1 m_4 \mu'_4 B} \mathcal{G}_{\lambda_4 \ell_2 \lambda'_2}^{\mu_4 m_2 \mu'_2} \mathcal{G}_{\ell_3 \lambda'_3 \lambda_2}^{m_3 \mu'_3 \mu_2} \\
&\times \kron_{\lambda_1 \lambda_3} \kron_{\mu_1 \mu_3} (-1)^{\lambda'_4 + \mu_4 + \mu'_4}  (-1)^{\lambda'_3 + \mu'_3} (-1)^{\lambda'_2 + \mu_2 + \mu'_2} (-1)^{\lambda'_1 + \mu_1 + \mu'_1}  \nonumber\\ \nonumber
& \times \mathcal{G}_{\lambda'_4 \lambda_4 \Lambda_4}^{-\mu'_4 -\mu_4 -M_4}\mathcal{G}_{\lambda'_3 \lambda_3 \Lambda_3}^{-\mu'_3  \mu_3 -M_3} \mathcal{G}_{\lambda'_2 \lambda_2 \Lambda_2}^{-\mu'_2 -\mu_2 -M_2} \mathcal{G}_{\lambda'_1 \lambda_1 \Lambda_1}^{-\mu'_1 -\mu_1 -M_1}\nonumber\\
&\times \int \int \int \int \dd x_1 \, \dd x_2 \, \dd x_3 \, \dd x_4 \, \, x_1^2 \, x_2^2 \, x_3^2 \, x_4^2 \, \, j_{\ell_1}(k_1 x_1) j_{\ell_4}(k_4 x_1) j_{\ell_2}(k_2 x_2) j_{\ell_3}(k_3 x_3)\nonumber\\
&\times \int \dd[3] \vb q_4 \, q_4[\bm{\hat{\epsilon}}_-(\vu q_4) \vdot \bm{\hat{\epsilon}}_-(\widehat{\vb k_4 - \vb q_4})]  Y_{\Lambda_4}^{M_4}(\vu{q}_4) A_-^{'}(\tau_4, |\vb k_4 -  \vb q_4|)A_-(\tau_4, q_4) j_{\lambda'_4}(q_4 x_1) j_{\lambda_4}(q_4 x_2) \nonumber \\
&\times \int \dd[3] \vb q_3 \, q_3 [\bm{\hat{\epsilon}}_-(\vu q_3) \vdot \bm{\hat{\epsilon}}_-(\widehat{\vb k_3 - \vb q_3})] Y_{\Lambda_3}^{M_3}(\vu{q}_3) A_-^{'}(\tau_3, |\vb k_3 -  \vb q_3|)A_-(\tau_3, q_3)j_{\lambda'_3}(q_3 x_3)j_{\lambda_3}(q_3 x_4) \nonumber \\
&\times \int \dd[3] \vb q_2 \, q_2 [\bm{\hat{\epsilon}}_-(\vu q_2) \vdot \bm{\hat{\epsilon}}_-(\widehat{\vb k_2 - \vb q_2})] Y_{\Lambda_2}^{M_2}(\vu{q}_2) A_-^{'}(\tau_2, |\vb k_2 -  \vb q_2|) A_-(\tau_2, q_2) j_{\lambda'_2}(q_2 x_2) j_{\lambda_2}(q_2 x_3) \nonumber \\
&\times \int \dd[3] \vb q_1 \, q_1 [\bm{\hat{\epsilon}}_-(\vu q_1) \vdot \bm{\hat{\epsilon}}_-(\widehat{\vb k_1 - \vb q_1})] Y_{\Lambda_1}^{M_1}(\vu{q}_1) A_-^{'}(\tau_1, |\vb k_1 -  \vb q_1|) A_-(\tau_1, q_1) j_{\lambda'_1}(q_1 x_1) j_{\lambda_1}(q_1 x_4). \nonumber
\end{align}
Thus the contribution to the trispectrum due to Diagram 4 is 
\begin{align}
&\expval{Q^{\text{inv}}(\tau, \vb k_1) Q^{\text{inv}}(\tau, \vb k_2)  Q^{\text{inv}}(\tau, \vb k_3) Q^{\text{inv}}(\tau, \vb k_4)}^{(4)} = \\
&\frac{4{,}096}{\Lambda^4 a^4(\tau)} \sum_{\vb \Bell, \vb m} \sum_{\bm{\lambda}, \bm{\mu}} \sum_{\bm \lambda', \bm \mu'} \sum_{A,B} \Bigg [\prod_{j=1}^{4} i^{\ell_j + \lambda_j + \lambda'_j} (-1)^{\lambda'_j + \mu_j + \mu'_j} Y_{\ell_j}^{m_j *}(\vu k_j) \sum_{\Lambda_j} \sum_{M_j} (-1)^{M_j} \Bigg ](-1)^{\mu_3} \nonumber \\
&\times \mathcal{K}_{\ell_1 \lambda'_1 \ell_4 \lambda'_4 A}^{m_1 \mu'_1 m_4 \mu'_4 B} \mathcal{G}_{\lambda_4 \ell_2 \lambda'_2}^{\mu_4 m_2 \mu'_2} \mathcal{G}_{\ell_3 \lambda'_3 \lambda_2}^{m_3 \mu'_3 \mu_2} \kron_{\lambda_1 \lambda_3} \kron_{\mu_1 \mu_3} \nonumber \mathcal{G}_{\lambda'_4 \lambda_4 \Lambda_4}^{-\mu'_4 -\mu_4 -M_4}\mathcal{G}_{\lambda'_3 \lambda_3 \Lambda_3}^{-\mu'_3  \mu_3 -M_3} \mathcal{G}_{\lambda'_2 \lambda_2 \Lambda_2}^{-\mu'_2 -\mu_2 -M_2} \mathcal{G}_{\lambda'_1 \lambda_1 \Lambda_1}^{-\mu'_1 -\mu_1 -M_1}\\
&\times \int \int \int \int \dd x_1 \, \dd x_2 \, \dd x_3 \, \dd x_4 \, \, x_1^2 \, x_2^2 \, x_3^2 \, x_4^2 \, \, j_{\ell_1}(k_1 x_1) j_{\ell_4}(k_4 x_1) j_{\ell_2}(k_2 x_2) j_{\ell_3}(k_3 x_3) \nonumber\\
&\times \int_{-\infty}^0 \dd \tau_4 \; G(\tau, \tau_4, k_4) \text{I}_{\Lambda_4 \lambda_4 \lambda'_4}^{M_4}(\vb k_4, x_4, x_2, \tau_4) \int_{-\infty}^0 \dd \tau_3 \; G(\tau, \tau_3, k_3) \text{I}_{\Lambda_3 \lambda_3 \lambda'_3}^{M_3}(\vb k_3, x_4, x_3, \tau_3) \nonumber \\ 
&\times \int_{-\infty}^0 \dd \tau_2 \; G(\tau, \tau_2, k_2) \text{I}_{\Lambda_2 \lambda_2 \lambda'_2}^{M_2}(\vb k_2, x_3, x_1, \tau_2) \int_{-\infty}^0 \dd \tau_1 \; G(\tau, \tau_1, k_1) \text{I}_{\Lambda_1 \lambda_1 \lambda'_1}^{M_1}(\vb k_1, x_2, x_1, \tau_1). \nonumber 
\end{align}
Analytic calculation of the $\text{I}_{\Lambda \lambda \lambda'}^{M}(\vb k, x, x', \tau)$ integrals would enable further reduction of the results above. We now turn to this problem.

\section{Computing the Integrals for Each Diagram}
\label{Section6}
We wish to reduce analytically the integral
\begin{align}
\text{I}_{\Lambda \lambda \lambda'}^{M}(\vb{k}, x, x', \tau) = \label{eq:I_int} \int \dd[3] \vb{q} \ q [\bm{\hat{\epsilon}}_-(\vu{q})\vdot\bm{\hat{\epsilon}}_-(\widehat{\vb k - \vb q})]Y_{\Lambda}^{M}(\vu q) &A_-^{'}(\tau, |\vb{k - q}|)A_-(\tau, q)  j_{\lambda}(q x) j_{\lambda'}(q x'). 
\end{align}
We first show that this integral is actually a convolution. We then use the Convolution Theorem, given in Eq. (\ref{eq:conv_thm}), to turn the integral (\ref{eq:I_int}) into a sequence of FTs.

We first need to expand out the dot product of the two polarization vectors $\bm{\hat{\epsilon}}_-(\vu{q})$ and $\bm{\hat{\epsilon}}_-(\widehat{\vb k- \vb q})$ in spherical coordinates. For an arbitrary vector $\vb q  =  q(\sin\theta \cos\phi, \sin\theta \sin\phi, \cos\theta)$, the polarization vector $\bm{\hat{\epsilon}}_-(\vu q) = (\cos\theta \cos\phi + i\sin\phi, \cos\theta \sin\phi - i\cos\phi, -\sin\theta)/\sqrt{2}$. Thus we have 
\begin{multline} \label{dot_prod}
\bm{\hat{\epsilon}}_-(\vu{q})\vdot\bm{\hat{\epsilon}}_-(\widehat{\vb k - \vb q}) = \frac{1}{2} \bigg [\cos\alpha \cos\theta \cos\phi \cos\beta + \cos\alpha \cos\theta \sin\phi \sin\beta - \cos\phi \cos\beta - \sin\phi \sin\beta \\
+ \sin\alpha \sin \theta - i\bigg(\vu q \cross \frac{\vb{k-q}}{|\vb{k-q}|} \vdot \vu z \bigg ) \frac{\cos\alpha - \cos\theta}{\sin\alpha \sin\theta} \bigg],    
\end{multline}
where $\alpha \in [0, \pi]$ and $\beta \in [0, 2\pi)$ refer, respectively, to the polar and the azimuthal angles of the vector $\widehat{\vb k - \vb q}$.
\subsection{Recognizing the Convolutions}
When we insert Eq. (\ref{dot_prod}) into Eq. (\ref{eq:I_int}), we may write the result as a sum of seven terms,
\begin{align}\label{I_sum}
&\text{I}_{\Lambda \lambda \lambda'}^{M}(\vb{k}, x, x', \tau) = \sum_{j=1}^{7} \text{I}_{\Lambda \lambda \lambda'}^{M (j)}(\vb{k}, x, x', \tau),
\end{align}
where $\text{I}_{\Lambda \lambda \lambda'}^{M (j)}(\vb{k}, x, x', \tau)$ is the integral over the $j^{th}$ term in the polarization vector dot product. We note that we have split the imaginary part of Eq. (\ref{dot_prod}) into two terms, one corresponding to $\cos\alpha$ and one to $\cos\theta$. 

We now consider as examples the first and last terms in the sum (\ref{I_sum}). The other terms can be computed in a similar fashion. We have for the first and last terms in (\ref{I_sum}):
\begin{flalign}
\label{I_1}
\qquad &\text{I}_{\Lambda \lambda \lambda'}^{M (1)}(\vb{k}, x, x', \tau) \equiv \frac{1}{2} \int \dd[3] \vb{q} \ \cos\alpha \cos\theta \cos\phi \cos\beta \, q A_-^{'}(\tau, |\vb{k - q}|)A_-(\tau, q) & \\ 
&\hspace{82mm} \times Y_{\Lambda}^M(\vu q)j_{\lambda}(q x) j_{\lambda'}(q x'), \nonumber &
\end{flalign}
\begin{flalign} 
\label{I_7}
&\hspace{7.2mm} \text{I}_{\Lambda \lambda \lambda'}^{M (7)} (\vb{k}, x, x', \tau) \equiv \frac{i}{2} \int \dd[3] \vb{q} \; \bigg(\vu q \cross \frac{\vb{k-q}}{|\vb{k-q}|} \vdot \vu z \bigg ) \frac{q\cos\theta}{\sin\alpha \sin\theta} A_-^{'}(\tau, |\vb{k - q}|)A_-(\tau, q) & \\
&\hspace{102mm}\times Y_{\Lambda}^M(\vu q) j_{\lambda}(q x) j_{\lambda'}(q x').  \nonumber &
\end{flalign}
We note that 
\begin{align}
&\cos\alpha = \vu z \vdot \frac{\vb k - \vb q}{|\vb k - \vb q|}, \hspace{10 mm} \sin\alpha = \sqrt{1-\cos^2\alpha}. \\
&\cos\beta = \frac{\vb k - \vb q}{|\vb k - \vb q|} \vdot \frac{\vu x}{\sin\alpha}. \nonumber
\end{align}
We have written the trigonometric functions in the forms above to make it transparent that the integrals (\ref{I_1}) and (\ref{I_7}) are convolutions.

We now write the integral (\ref{I_1}) as
\begin{align} \label{I1_conv}
&\text{I}_{\Lambda \lambda \lambda'}^{M (1)}(\vb{k}, x, x', \tau) =  \frac{1}{2}\left [ \left (\cos\theta\cos\phi Y_{\Lambda}^{M} (\vu q) q A_-(\tau, q) j_{\lambda}(q x) j_{\lambda'}(q x') \right ) \star \left (\cos\theta \cos\phi A_-^{'}(\tau, q) \right ) \right ](\vb k),     
\end{align}
where the integration is over the loop momentum $\vb q$, and the result is then a function of the external momentum $\vb k$.

The fact that the  integral (\ref{I_7}) is a convolution is more apparent if we write the scalar triple product (the first term in the integrand of \ref{I_7}) in a different form.\footnote{https://mathworld.wolfram.com/ScalarTripleProduct.html} We may rewrite the scalar triple product in terms of the three-argument isotropic basis functions $\mathcal{P}_{\ell_1 \ell_2 \ell_3}$ of \citep{Isotropic_basis}, in particular the lowest-lying parity odd one:
\begin{equation}
\mathcal{P}_{1 1 1}(\vu z, \vu k, \vu q) \equiv -\frac{3i}{\sqrt{2}(4\pi)^{3/2}}\,\vu z \vdot (\vu k \cross \vu q). 
\end{equation}
$\mathcal{P}_{1 1 1}$ may be further rewritten as a weighted sum of products of three spherical harmonics, giving
\begin{equation}
\mathcal{P}_{1 1 1}(\vu z, \vu k, \vu q) = \sum_{\nu_1 \nu_2 \nu_3} C_{\nu_1 \nu_2 \nu_3}^{1 1 1}\, Y_1^{\nu_1}(\vu z) Y_1^{\nu_2}(\vu k) Y_1^{\nu_3}(\vu q), 
\end{equation}
where $C_{\nu_1 \nu_2 \nu_3}^{1 1 1} \equiv -\mqty(1 & 1 & 1 \\ \nu_1 & \nu_2 & \nu_3)$ is a Wigner 3-$j$ symbol \cite{Edmonds} with a phase. The $z$-components of the angular momenta, $\nu_i$, are bounded by the total angular momenta, and must sum to zero.  
Using these results we may rewrite the integral (\ref{I_7}) as
\begin{align}
&\text{I}_{\Lambda \lambda \lambda'}^{M (7)} (\vb{k}, x, x', \tau) = -\frac{\sqrt{2}(4\pi)^{3/2}}{6} \sum_{\nu_1 \nu_2 \nu_3} C_{\nu_1 \nu_2 \nu_3}^{1 1 1}\, Y_1^{\nu_1}(\vu z)  \nonumber \\ 
 &\times \int \dd[3] \vb{q} \; \frac{q\cos\theta}{\sin\alpha \sin\theta} \; Y_{\Lambda}^M(\vu q) Y_1^{\nu_2}(\vu q) Y_1^{\nu_3}\bigg(\frac{\vb k - \vb q}{|\vb k - \vb q|}\bigg) 
 A_-^{'}(\tau, |\vb{k - q}|)A_-(\tau, q) j_{\lambda}(q x) j_{\lambda'}(q x') \nonumber \\
&= -\frac{4\pi}{\sqrt{6}} \sum_{\nu_2 \nu_3}  C_{0 \nu_2 \nu_3}^{1 1 1} \; \sum_{L', M'}^\infty (-1)^{M'} \mathcal{G}_{\Lambda 1 L'}^{M \nu_2 -M'} \nonumber \\
 &\times \int \dd[3] \vb{q} \; \frac{q\cos\theta}{\sin\alpha \sin\theta} \; Y_{L'}^{M'}(\vu q) Y_1^{\nu_3}\bigg(\frac{\vb k - \vb q}{|\vb k - \vb q|}\bigg) 
 A_-^{'}(\tau, |\vb{k - q}|)A_-(\tau, q) j_{\lambda}(q x) j_{\lambda'}(q x'), 
\end{align}
where we have used the fact that $Y_{\ell}^{m}(\vu z) = \kron_{m 0} \sqrt{(2\ell + 1)/4\pi} $ and we again used Eq. (\ref{eq:Y_contr}) to linearize the spherical harmonics. The integral (\ref{I_7}) is also a convolution:
\begin{align} \label{I7_conv}
&\text{I}_{\Lambda \lambda \lambda'}^{M (7)} (\vb{k}, x, x', \tau) = -\frac{4\pi}{\sqrt{6}} \sum_{\nu_2 \nu_3}  C_{0 \nu_2 \nu_3}^{1 1 1} \sum_{L', M'}^\infty (-1)^{M'} \mathcal{G}_{\Lambda 1 L'}^{M \nu_2 -M'} \\
&\times \left [ \left (\frac{\cos\theta}{\sin\theta} Y_{L'}^{M'} (\vu q) q A_-(\tau, q) j_{\lambda}(q x) j_{\lambda'}(q x') \right ) \star \left ( \frac{1}{\sin\theta} Y_1^{\nu_3}(\vu q)A_-^{'}(\tau, q) \right ) \right ](\vb k). \nonumber  
\end{align}
We may now rewrite Eqs. (\ref{I1_conv}) and (\ref{I7_conv}) as FTs using the Convolution Theorem, given in Appendix \ref{eq:conv_thm}. We find 
\begin{align}
&\text{I}_{\Lambda \lambda \lambda'}^{M (1)}(\vb{k}, x, x', \tau) = \frac{1}{2} \mathcal{F} \Bigg \{ \mathcal{F}^{-1} \bigg \{\cos\theta \cos\phi \, Y_{\Lambda}^{M} (\vu q) q A_-(\tau, q) j_{\lambda}(q x) j_{\lambda'}(q x') \bigg \}(\vb r) \nonumber \\
&\hspace{37mm} \times \mathcal{F}^{-1} \left \{ \cos\theta \cos\phi \, A_-^{'}(\tau, q) \right \}(\vb r) \Bigg \}(\vb k), \label{eq:H1} \\  
&\text{I}_{\Lambda \lambda \lambda'}^{M (7)} (\vb{k}, x, x', \tau) = -\frac{4\pi}{\sqrt{6}} \sum_{\nu_3}  C_{0 -\nu_3 \nu_3}^{1 1 1} \sum_{L', M'}^\infty (-1)^{M'} \mathcal{G}_{\Lambda 1 L'}^{M -\nu_3 -M'} \nonumber\\ 
&\hspace{28mm} \times \mathcal{F} \Bigg \{\mathcal{F}^{-1} \bigg \{\frac{\cos\theta}{\sin\theta} Y_{L'}^{M'} (\vu q) q A_-(\tau, q) j_{\lambda}(q x) j_{\lambda'}(q x') \bigg \}(\vb r) \label{eq:H7} \nonumber\\ 
&\hspace{33mm} \times \; \mathcal{F}^{-1} \left \{ \frac{1}{\sin\theta} Y_1^{\nu_3}(\vu q)A_-^{'}(\tau, q) \right \}(\vb r) \Bigg \}(\vb k), 
\end{align}
where we have used the fact that the z-components of the angular momenta in the Wigner $3$-$j$ symbol must sum to zero. Here $\mathcal{F}$ represents an FT and $\mathcal{F}^{-1}$ an inverse FT. Each of the other five terms in Eq. (\ref{I_sum}) can also be written as FTs; they are listed in Appendix \ref{AppendixB}. 
Thus the integral (\ref{eq:I_int}) can be written as a sum of seven terms, where each term $\text{I}_{\Lambda \lambda \lambda'}^{M (j)}(\vb{k}, x, x', \tau)$ is an FT. We may now write
\begin{align}
&\text{I}_{\Lambda \lambda \lambda'}^{M (j)}(\vb{k}, x, x', \tau) = \frac{1}{2} (-1)^{\kron_{j 3} + \kron_{j 4}} \text{H}_{\Lambda \lambda \lambda'}^{M (j)} (\vb{k}, x, x', \tau) \hspace{10 mm} \text{for} \hspace{2mm} j \in \{1,2,3,4,5\}, \\    
&\text{I}_{\Lambda \lambda \lambda'}^{M (j)} (\vb{k}, x, x', \tau) = \frac{4\pi}{\sqrt{6}} (-1)^{j} \sum_{\nu_3 }  C_{0 -\nu_3 \nu_3}^{1 1 1} \sum_{L', M'}^\infty (-1)^{M'} \mathcal{G}_{\Lambda 1 L'}^{M -\nu_3 -M'} \nonumber\\ 
&\hspace{30mm} \times \tilde{\text{H}}_{L' \lambda \lambda'}^{M \nu_3 (j)} (\vb{k}, x, x', \tau) \hspace{10 mm} \text{for} \hspace{2mm} j \in \{6,7\}
\end{align}
where we note that
\begin{equation}
(-1)^{\kron_{j 3} + \kron_{j 4}} =
\left\{
    \begin{array}{lr}
        -1, & \text{if } j = 3,4\\
         1, & \text{otherwise}
    \end{array}
\right\}
\end{equation}
must be included to account for the minus sign on terms $3$ and $4$ in the polarization vector dot product (\ref{dot_prod}), we have defined
\begin{flalign}
&\text{H}_{\Lambda \lambda \lambda'}^{M (j)} (\vb{k}, x, x', \tau) \equiv \label{eq:Gen_FT_1} &\\
&\mathcal{F} \left \{ \mathcal{F}^{-1} \left \{\vb f_1^{(j)}(\vu q) Y_{\Lambda}^{M} (\vu q) q A_-(\tau, q) j_{\lambda}(q x) j_{\lambda'}(q x') \right \}(\vb r) \times \mathcal{F}^{-1} \left \{ \vb f_2^{(j)}(\vu q) A_-^{'}(\tau, q) \right \}(\vb r) \right \}(\vb k), &\nonumber
\end{flalign}
and
\begin{align}
&\tilde{\text{H}}_{L' \lambda \lambda'}^{M \nu_3 (j)} (\vb{k}, x, x', \tau) \equiv \label{eq:Gen_FT_2} \\
&\mathcal{F} \left \{ \mathcal{F}^{-1} \left \{\vb f_1^{(j)}(\vu q) Y_{L'}^{M'} (\vu q) q A_-(\tau, q) j_{\lambda}(q x) j_{\lambda'}(q x') \right \}(\vb r) \times \mathcal{F}^{-1} \left \{ \vb f_2^{(j)}(\vu q) Y_{1}^{\nu_3}(\vu q) A_-^{'}(\tau, q) \right \}(\vb r) \right \}(\vb k). \nonumber     
\end{align}
We have introduced the seven-component vectors
\begin{align}
&\vb f_1(\vu q) \equiv (\cos\theta \cos\phi, \cos\theta\sin\phi, \cos\phi, \sin\phi, \sin\theta, \csc\theta,\cot\theta), \nonumber\\
&\vb f_2(\vu q) \equiv (\cos\theta \cos\phi, \cos\theta\sin\phi, \cos\phi, \sin\phi, \sin\theta, \cot\theta, \csc\theta) \label{f}.
\end{align}
Finally, we may rewrite Eq. (\ref{I_sum}) as 
\begin{align}
&\text{I}_{\Lambda \lambda \lambda'}^{M}(\vb{k}, x, x', \tau) =  \frac{1}{2} \sum_{j=1}^5 (-1)^{\kron_{j 3} + \kron_{j 4}}\text{H}_{\Lambda \lambda \lambda'}^{M (j)} (\vb{k}, x, x', \tau) \label{I_int_simpl}\\
&+ \frac{4\pi}{\sqrt{6}} \sum_{\nu_3}  C_{0 -\nu_3 \nu_3}^{1 1 1} \sum_{L', M'}^\infty (-1)^{M'} \mathcal{G}_{\Lambda 1 L'}^{M -\nu_3 -M'} \sum_{j=6}^7 (-1)^{j} \tilde{\text{H}}_{L' \lambda \lambda'}^{M \nu_3 (j)} (\vb{k}, x, x', \tau). \nonumber
\end{align}

\subsection{Analytic Reduction of Angular Integrals via Convolution Theorem} 
\label{FTs}
In this section, we separate the angular and radial parts of the FTs we encountered in \S\ref{Section6}. 
Each inverse FT of the previous section contains one of the following three forms: 
\begin{align}\label{FT_type1}
&\mathcal{F}^{-1} \left \{\vb f_1^{(j)}(\vu q) Y_{\Lambda}^{M} (\vu q) q A_-(\tau, q) j_{\lambda}(q x) j_{\lambda'}(q x') \right \}(\vb r), \\ \nonumber \\ \label{FT_type2}
&\mathcal{F}^{-1} \left \{ \vb f_2^{(j)}(\vu q) Y_{1}^{\nu_3}(\vu q) A_-^{'}(\tau, q) \right \}(\vb r), \\ \nonumber \\ \label{FT_type3}
&\mathcal{F}^{-1} \left \{ \vb f_2^{(j)}(\vu q) A_-^{'}(\tau, q) \right \}(\vb r). 
\end{align}

We now rewrite Eq. (\ref{FT_type1}) by using the plane wave expansion (\ref{eq:plane_exp}) and the linearization rule (\ref{eq:Y_contr}), to find    
\begin{align} \label{FT_type1_simpl}
&\mathcal{F}^{-1} \left \{\vb f_1^{(j)}(\vu q) Y_{\Lambda}^{M} (\vu q) q A_-(\tau, q) j_{\lambda}(q x) j_{\lambda'}(q x') \right \}(\vb r) \\
&=\frac{1}{(2\pi)^3}\int \dd [3] \vb q \ e^{-i\vb q \vdot \vb r} \vb f_1^{(j)}(\vu q) Y_{\Lambda}^{M} (\vu q) q A_-(\tau, q) j_{\lambda}(q x) j_{\lambda'}(q x')\nonumber \\
&= \frac{4\pi}{(2\pi)^3} \sum_{a = 0}^\infty \sum_{b = -a}^a (-1)^a i^a Y_a^{b *}(\vu r) \sum_{\alpha = 0}^\infty \sum_{\beta = -\alpha}^\alpha (-1)^\beta \mathcal{G}_{a \Lambda \alpha}^{b M -\beta} \nonumber \\ &\times \int \dd \Omega_q \ Y_\alpha^\beta(\vu q) \vb f_1^{(j)}(\vu q) 
 \int_0^{\infty} \dd q \ q^3 A_-(\tau, q) j_{\lambda}(q x) j_{\lambda'}(q x') j_a(qr) \nonumber \\ 
&\equiv \frac{1}{2\pi^2} \sum_{a = 0}^\infty \sum_{b = -a}^a (-1)^a i^a Y_a^{b *}(\vu r) \mathcal{C}_{(1) a \Lambda}^{(j) b M} \int_0^{\infty} \dd q \; q^3 A_-(\tau, q) j_{\lambda}(q x) j_{\lambda'}(q x') j_a(qr), \nonumber
\end{align}
where $\dd \Omega_q \equiv \sin \theta \dd \theta \dd \phi$  is the differential solid angle explored by the vector $\vu q$, and
\begin{align}
&\mathcal{C}_{(i) a L}^{(j) b M} \equiv \sum_{\alpha = 0}^\infty \sum_{\beta = -\alpha}^\alpha (-1)^\beta \mathcal{G}_{a L \alpha}^{b M -\beta} C_{(i) \alpha}^{(j) \beta} 
\end{align}
with
\begin{align}
&C_{(i) \alpha}^{(j) \beta} \equiv \int \dd \Omega_q \ Y_\alpha^\beta(\vu q) \vb f_i^{(j)}(\vu q) \label{eq:ang_coeffs}.
\end{align}
For each fixed valued of $\alpha, \beta, i$, and $j$, the angular integral in Eq. (\ref{eq:ang_coeffs}) yields a constant. We show explicitly how to compute these angular integrals, and list the final values, in Appendix \ref{AppendixC}. 
 Similarly, Eqs. (\ref{FT_type2}) and (\ref{FT_type3}) can be rewritten as 
\begin{align} \label{FT_type2_simpl}
&\mathcal{F}^{-1} \left \{ \vb f_2^{(j)}(\vu q) Y_{1}^{\nu_3}(\vu q) A_-^{'}(\tau, q) \right \}(\vb r) = \frac{1}{2\pi^2} \sum_{a = 0}^\infty \sum_{b = -a}^a (-1)^a i^a Y_a^{b *}(\vu r) \mathcal{C}_{(2) a 1}^{(j) b \nu_3} \int_0^{\infty} \dd q \; q^2  A_-^{'}(\tau, q) j_a(qr),
\end{align}
\begin{align}
&\mathcal{F}^{-1} \left \{ \vb f_2^{(j)}(\vu q) A_-^{'}(\tau, q) \right \}(\vb r) = \frac{1}{2\pi^2} \sum_{a = 0}^\infty \sum_{b = -a}^a (-1)^a i^a Y_a^{b *}(\vu r) C_{(2) a}^{(j) b} \int_0^{\infty} \dd q \ q^2 A_-^{'}(\tau, q) j_a(qr).
\end{align} 
We now use Eqs. (\ref{FT_type1_simpl}) and (\ref{FT_type2_simpl}) to rewrite Eq. (\ref{eq:Gen_FT_2}) as 
\begin{align} \label{FT_type3_simpl}
&\tilde{\text{H}}_{L' \lambda \lambda'}^{M \nu_3 (j)} (\vb{k}, x, x', \tau) = \frac{1}{4\pi^4} \mathcal{F} \Bigg \{ \sum_{a, b} \sum_{a', b'} (-1)^{a+a'} i^{a+a'} Y_a^{b *}(\vu r) Y_{a'}^{b' *}(\vu r) \mathcal{C}_{(1)a L'}^{(j) b M'} \mathcal{C}_{(2) a' 1}^{(j) b' \nu_3} \nonumber \\ &\times \int_0^{\infty} \dd q \ q^3  A_-(\tau, q) j_{\lambda'}(q x) j_{\lambda'}(q x') j_a(qr)  \int_0^{\infty} \dd q' \ q'^2  A_-^{'}(\tau, q') j_{a'}(q'r)  \Bigg \} 
\end{align}
\begin{align}
&= \frac{1}{4\pi^4} \sum_{a, b} \sum_{a', b'} (-1)^{a+a'} i^{a+a'} \mathcal{C}_{(1)a L'}^{(j) b M'} \mathcal{C}_{(2) a' 1}^{(j) b' \nu_3} \int \dd [3] \vb r \, \bigg \{e^{i \vb k \vdot \vb r} Y_a^{b *}(\vu r) Y_{a'}^{b' *}(\vu r) \nonumber \\
&\times \int_0^{\infty} \dd q \ q^3  A_-(\tau, q) j_{\lambda}(q x) j_{\lambda'}(q x') j_a(qr) \int_0^{\infty} \dd q' \ q'^2  A_-^{'}(\tau, q') j_{a'}(q'r) \bigg \}. \nonumber
\end{align}
We again use the plane wave expansion (\ref{eq:plane_exp}) and obtain

\begin{align}\label{eq:rad_ints}
&\tilde{\text{H}}_{L' \lambda \lambda'}^{M \nu_3 (j)} (\vb{k}, x, x', \tau) = \frac{1}{\pi^3} \sum_{a, b} \sum_{a', b'} \sum_{c,d} (-1)^{a+a'} i^{a+a'+c} \mathcal{C}_{(1)a L'}^{(j) b M'} \mathcal{C}_{(2) a' 1}^{(j) b' \nu_3} Y_c^{d}(\vu k) \\
&\times \int \dd \Omega_r \, Y_a^{b *}(\vu r) Y_{a'}^{b' *}(\vu r) Y_c^{d *}(\vu r) \nonumber \int_0^{\infty} \dd q \ q^3  A_-(\tau, q) j_{\lambda}(q x) j_{\lambda'}(q x') \\ 
&\times \int_0^{\infty} \dd q' \ q'^2  A_-^{'}(\tau, q') \int_0^{\infty} \dd r \, r^2 j_a(qr) j_{a'}(q'r) j_c(kr). \nonumber
\end{align}
We recognize that the angular integral is simply the Gaunt integral (\ref{eq:Gaunt}). 
These results show that we may write Eq. (\ref{eq:Gen_FT_2}) as 
\begin{align}
\tilde{\text{H}}_{L' \lambda \lambda'}^{M \nu_3 (j)} (\vb{k}, x, x', \tau) = \frac{1}{\pi^3} \sum_{a, b} \sum_{a', b'} \sum_{c,d} (-1)^{a+a'}& i^{a+a'+c} \mathcal{C}_{(1)a L'}^{(j) b M'} \mathcal{C}_{(2) a' 1}^{(j) b' \nu_3} \mathcal{G}_{a a' c}^{b b' d} \label{eq:simpl_FT_2}  \\ 
&\times Y_c^{d}(\vu k) Z_{\lambda \lambda' a a' c}(k, x, x', \tau)\nonumber
\end{align}
where
\begin{align}\label{Z}
Z_{\lambda \lambda' a a' c}(k, x, x', \tau) &\equiv  \int_0^{\infty} \dd q \ q^3  A_-(\tau, q) j_{\lambda}(q x) j_{\lambda'}(q x') \int_0^{\infty} \dd q' \ q'^2  A_-^{'}(\tau, q') \\
&\times \int_0^{\infty} \dd r \, r^2 j_a(qr) j_{a'}(q'r) j_c(kr). \nonumber
\end{align}
Similarly, for Eq. (\ref{eq:Gen_FT_1}), we have
\begin{align}
\text{H}_{\Lambda \lambda \lambda'}^{M (j)} (\vb{k}, x, x', \tau) = \frac{1}{\pi^3} \sum_{a, b} \sum_{a', b'} \sum_{c,d} (-1)^{a+a'}& i^{a+a'+c} \mathcal{C}_{(1)a L'}^{(j) b M'} C_{(2) a'}^{(j) b' } \mathcal{G}_{a a' c}^{b b' d} \label{eq:simpl_FT_1} \\
&\times Y_c^{d}(\vu k) Z_{\lambda \lambda' a a' c}(k, x, x', \tau). \nonumber   
\end{align}
We may now use Eqs. (\ref{eq:simpl_FT_1}) and (\ref{eq:simpl_FT_2}) to rewrite Eq. (\ref{I_int_simpl}) as
\begin{align}    
& \text{I}_{\Lambda \lambda \lambda'}^{M}(\vb{k}, x, x', \tau) = \frac{1}{\pi^3} \sum_{a, b} \sum_{a', b'} \sum_{c,d} (-1)^{a+a'} i^{a+a'+c}\mathcal{G}_{a a' c}^{b b' d} \Bigg \{ \frac{1}{2} \sum_{j=1}^5 (-1)^{\kron_{j 3} + \kron_{j 4}} \mathcal{C}_{(1) a \Lambda}^{(j) b M } C_{(2)a'}^{(j)b'} \nonumber\\
&+ \frac{4\pi}{\sqrt{6}} \sum_{\nu_3}  C_{0 -\nu_3 \nu_3}^{1 1 1} \sum_{L', M'}^\infty (-1)^{M'} \mathcal{G}_{\Lambda 1 L'}^{M -\nu_3 -M'}\sum_{j=6}^7 (-1)^{j} \mathcal{C}_{(1) a L'}^{(j) b M} \mathcal{C}_{(2) a' 1}^{(j) b' \nu_3 } \Bigg \} \nonumber\\ 
&\times Y_c^{d}(\vu k) Z_{\lambda \lambda' a a' c}(k,x,x',\tau).
\end{align}
We thus find that 
\begin{align} 
&\text{I}_{\Lambda \lambda \lambda'}^{M}(\vb k, x, x', \tau) = \sum_{a, b} \sum_{a', b'} \sum_{c,d} B_{\Lambda a a' c}^{M b b' d} Y_c^{d}(\vu k) Z_{\lambda \lambda' a a' c}(k,x,x',\tau) \label{eq:Int_simpl}
\end{align}
where 
\begin{align}
&B_{\Lambda a a' c}^{M b b' d} \equiv \frac{(-1)^{a+a'}}{\pi^3} i^{a+a'+c}\mathcal{G}_{a a' c}^{b b'd} \Bigg \{ \frac{1}{2} \sum_{j=1}^5 (-1)^{\kron_{j 3} + \kron_{j 4}} \mathcal{C}_{(1) a \Lambda}^{(j) b M}  C_{(2) a'}^{(j) b'} \\
&+ \frac{4\pi}{\sqrt{6}} \sum_{\nu_3}  C_{0 -\nu_3 \nu_3}^{1 1 1} \sum_{L', M'}^\infty (-1)^{M'} \mathcal{G}_{\Lambda 1 L'}^{M -\nu_3 -M'}\sum_{j=6}^7 (-1)^{j} \mathcal{C}_{(1) a L'}^{(j) b M} \mathcal{C}_{(2) a' 1}^{(j) b' \nu_3}  \Bigg \}. \nonumber 
\end{align}
With Eq. (\ref{eq:Int_simpl}), we have found an expression that separates the angular and radial parts of Eq. (\ref{eq:I_int}). The angular parts of the Fourier integrals have been computed analytically and are included via the coefficients (\ref{eq:ang_coeffs}), and the radial parts of the Fourier integrals are contained within Eq. (\ref{Z}).

\section{Reduction to Low-Dimensional Radial Integrals}
\label{Section7}
We found in \S\ref{Section5} that each contribution to the trispectrum involves integrals of the form 
\begin{align}\label{FancyI1}
\mathcal{I}_{\Lambda \lambda \lambda'}^{M}(\vb k, x, x', \tau) \equiv \int_{-\infty}^0 \dd \tau' \; G(\tau, \tau', k) \text{I}_{\Lambda \lambda \lambda'}^{M}(\vb k, x, x', \tau').  
\end{align}
We may now rewrite this integral using Eq. (\ref{eq:Int_simpl}), so that we have
\begin{align}
\mathcal{I}_{\Lambda \lambda \lambda'}^{M}(\vb k, x, x', \tau) &= \sum_{a, b} \sum_{a', b'} \sum_{c,d} B_{\Lambda a a' c}^{M b b' d} Y_c^{d}(\vu k) \int_{-\infty}^0 \dd \tau' \; G(\tau, \tau', k) Z_{\lambda \lambda' a a' c}(k,x,x',\tau') \\
& \label{FancyI}\equiv \sum_{a, b} \sum_{a', b'} \sum_{c,d} B_{\Lambda a a' c}^{M b b' d} Y_c^{d}(\vu k) \mathcal{Z}_{\lambda \lambda' a a' c}(k,x,x',\tau), \nonumber
\end{align}
where
\begin{align} 
&\mathcal{Z}_{\lambda \lambda' a a' c}(k,x,x',\tau) \equiv \int_{-\infty}^0 \dd \tau' \; G(\tau, \tau', k) Z_{\lambda \lambda' a a' c}(k,x,x',\tau') \\
&= \int_{-\infty}^0 \dd \tau' \;\frac{a(\tau) H}{k^3\tau'}(-k\tau)^{\frac{n_s - 1}{2}} \left (k\tau'\cos k\tau' - \sin k\tau' \right ) \int_0^{\infty} \dd q \ q^3  A_-(\tau', q) j_{\lambda}(q x) j_{\lambda'}(q x') \nonumber \\
&\times \int_0^{\infty} \dd q' \ q'^2  A_-^{'}(\tau', q') \int_0^{\infty} \dd r \, r^2 j_a(qr) j_{a'}(q'r) j_c(kr). \nonumber
\end{align}
We have used the definition of $Z_{\lambda \lambda' a a' c}(k,x,x',\tau')$ given in Eq. (\ref{Z}) and we also inserted Eq. (\ref{Green_function_limit}), the $-k\tau \rightarrow 0$ limit of the Green's function, since we are interested in modes which are stretched outside of the horizon.
We now change the order of integration, finding
\begin{align} \label{FancyZ}
&\mathcal{Z}_{\lambda \lambda' a a' c}(k,x,x',\tau) = \int_0^{\infty} \dd r \; r^2 j_c(kr) \int_0^{\infty} \dd q \ q^3 j_{\lambda}(q x) j_{\lambda'}(q x') j_a(qr)\\
&\times \int_{-\infty}^0 \dd \tau'\frac{a(\tau) H(\tau)}{k^3\tau'} (-k\tau)^{\frac{n_s - 1}{2}} \left (k\tau'\cos k\tau' - \sin k\tau' \right ) A_-(\tau', q) W_{a'}(\tau',r), \nonumber
\end{align}
where
\begin{align}
W_{a'}(\tau',r) \equiv \int_0^{\infty} \dd q' \ q'^2  A_-^{'}(\tau', q') j_{a'}(q'r). 
\end{align}
Our goal is to reduce the computational complexity of the integrals in (\ref{FancyZ}) as much as possible. In order to accomplish this, we will perform several changes of variable to remove the factors of $k$ from the functions in the inner integrals. This will be possible because we can factor out a $1/\sqrt{2k}$ from Eq. (\ref{A:approx_1}) so that we have a function that depends only on $k\tau$. The ability to perform this transformation also holds for the exact solution to Eq. (\ref{eq: A_mf_eom}). Motivated by this, we now write the mode function as
\begin{align}
&A_-(\tau, k) = \frac{1}{\sqrt{2k}} \mathcal{A}(k\tau),  
\end{align}
where 
\begin{align} \label{FancyA}
\mathcal{A}(k\tau) \equiv \left(\frac{-k\tau}{2\xi}\right)^{1/4} e^{\pi \xi - 2\sqrt{-2\xi k\tau}}.
\end{align}
We have used the fact that $\tau \approx -1/(aH)$ for $\epsilon \ll 1$. For the derivatives of the mode function, we follow \citep{Barnaby_2011_2} and employ the approximation
\begin{align}
&A^{'}_{-}(\tau, k) \approx \sqrt{\frac{2k\xi}{-\tau}}A_{-}(\tau, k) = \sqrt{\frac{\xi}{-\tau}} \mathcal{A}(k\tau), \label{A_approx}
\end{align} 
which is valid for $1/(8\xi) \lesssim -k\tau \lesssim 2\xi$. This approximation yields
\begin{align}
&W_{a'}(\tau',r) = \int_0^{\infty} \dd q' \ q'^2 \sqrt{\frac{\xi}{-\tau'}} \, \mathcal{A}(q'\tau') j_{a'}(q'r). 
\end{align}
Making the change of variables $z = q'\tau'$, we find
\begin{align}\label{W}
&W_{a'}(\tau',r) = \int_{-\infty}^0 \dd z \, \frac{z^2}{\tau'^3} \sqrt{\frac{\xi}{-\tau'}}\mathcal{A}(z) j_{a'}(zr/\tau') \equiv \frac{1}{\tau'^3} \sqrt{\frac{\xi}{-\tau'}} M_{a'}(r/\tau'), 
\end{align}
where 
\begin{align} \label{M_integral}
&M_{a'}(r/\tau') \equiv \int_{-\infty}^0 \dd z \, z^2 \mathcal{A}(z) j_{a'}(zr/\tau').  
\end{align}
We now insert Eq. (\ref{W}) into Eq. (\ref{FancyZ}) and find
\begin{align}
&\mathcal{Z}_{\lambda \lambda' a a' c}(k,x,x',\tau) = (-k\tau)^{\frac{n_s - 1}{2}} \int_0^{\infty} \dd r \; r^2 j_c(kr) \int_0^{\infty} \dd q \ q^3 j_{\lambda}(q x) j_{\lambda'}(q x') j_a(qr)\\
&\times \int_{-\infty}^0 \dd \tau'\frac{a(\tau) H(\tau)}{k^3\tau'^4} \sqrt{\frac{\xi}{-\tau'}} \frac{1}{\sqrt{2q}}\left (k\tau'\cos k\tau' - \sin k\tau' \right ) \mathcal{A}(q\tau') M_{a'}(r/\tau'). \nonumber
\end{align}
Making the change of variables $w = k\tau'$, $u = q/k$ and defining $v = kr$, we obtain
\begin{align}\label{FancyZ2}
&\mathcal{Z}_{\lambda \lambda' a a' c}(k,x,x',\tau) = (-k\tau)^{\frac{n_s - 1}{2}} k a(\tau) H(\tau) \sqrt{\xi} \int_0^{\infty} \dd v \; v^2 j_c(v) \nonumber \\ &\times \int_0^{\infty} \dd u \; u^3 j_{\lambda}(kux) j_{\lambda'}(kux') j_a(uv) 
\int_{-\infty}^0 \dd w \; \mathcal{S}(w) \frac{1}{\sqrt{2u}} \, \mathcal{A}(uw) M_{a'}(v/w), 
\end{align}
where 
\begin{align}\mathcal{S}(w) \equiv \left(w\cos w - \sin w\right)/(w^4\sqrt{-w}). \label{S_def}
\end{align} 
We once again change the order of integration to find
\begin{align} \label{FancyZ_Final}
\mathcal{Z}_{\lambda \lambda' a a' c}(k,x,x',\tau) = (-k\tau)^{\frac{n_s - 1}{2}} k a(\tau) H(\tau) \sqrt{\xi}\int_{0}^{\infty} \dd u \; u^3 j_{\lambda}(kux) j_{\lambda'}(kux') \mathcal{M}_{aa'c}(u),   
\end{align}
where
\begin{align} \label{Curly_M}
\mathcal{M}_{aa'c}(u) \equiv \int_{0}^{\infty} \dd v \; v^2 j_c(v) j_a(uv) \int_{-\infty}^0 \dd w \; \mathcal{S}(w) \frac{1}{\sqrt{2u}} \, \mathcal{A}(uw) M_{a'}(v/w).
\end{align}
Finally, we may use Eq. (\ref{FancyZ_Final}) to rewrite Eq. (\ref{FancyI}) as 
\begin{align}\label{FancyI_Final}
\mathcal{I}_{\Lambda \lambda \lambda'}^{M}(\vb k, x, x', \tau) &= (-k\tau)^{\frac{n_s - 1}{2}} k a(\tau) H(\tau) \sqrt{\xi} \sum_{a, b} \sum_{a', b'} \sum_{c,d} B_{\Lambda a a' c}^{M b b' d} Y_c^{d}(\vu k) \nonumber\\
&\times \int_{0}^{\infty} \dd u \; u^3 j_{\lambda}(kux) j_{\lambda'}(kux') \mathcal{M}_{aa'c}(u).  
\end{align}
We have thus shown that Eq. (\ref{FancyI1}) can be written in terms of low-dimensional radial integrals that will be efficient to compute numerically. We will return to an exact definition of what we mean by dimension (in terms of how it impacts numerical integration) in \S \ref{Section8.1}. We note that this simplification was possible because we were able to manipulate the mode functions so that they were dependent only on the combination $k\tau$ in their argument. We will return to the question of how to compute these integrals numerically in \S \ref{Section9}. First, however, in  the following section we will use this result to write down a final expression for each diagram of \S \ref{Section5}.    

\section{The Final Inverse Decay Trispectrum} \label{Section8}
In \S \ref{Section6} and \S \ref{Section7}, we assembled all the ingredients necessary to compute the trispectrum. We now return to the simplified expressions that we found in \S\ref{Section5} for the four Reference Diagrams.

\subsection{Final Result for \hyperref[Diagrams1-4]{Reference Diagram 1}} \label{Section8.1}
After replacing the Green's function integrals in Eq. (\ref{eq:Trispectrum_1}) with Eq. (\ref{FancyI1}) we find that the contribution to the trispectrum for \hyperref[Diagrams1-4]{Reference Diagram 1} is 
\begin{align}
&\expval{Q^{\text{inv}}(\tau, \vb k_1) Q^{\text{inv}}(\tau, \vb k_2)  Q^{\text{inv}}(\tau, \vb k_3) Q^{\text{inv}}(\tau, \vb k_4)}^{(1)} = \\  
&\frac{4{,}096}{\Lambda^4 a^4(\tau)}\sum_{\vb \Bell, \vb m} \sum_{\bm{\lambda}, \bm{\mu}} \sum_{\bm \lambda', \bm \mu'} \Bigg [\prod_{j=1}^{4}  i^{\ell_j + \lambda_j + \lambda'_j} (-1)^{\lambda'_j + \mu_j + \mu'_j}  Y_{\ell_j}^{m_j *}(\vu k_j) \sum_{\Lambda_j M_j} (-1)^{M_j} \mathcal{G}_{\lambda'_j \lambda_j \Lambda_j}^{\mu'_j \mu_j M_j} \Bigg ] \nonumber  \\
&\times \mathcal{G}_{\lambda_2 \ell_4 \lambda'_4}^{\mu_2 m_4 \mu'_4} \mathcal{G}_{\lambda_1 \ell_3 \lambda'_3}^{\mu_1 m_3 \mu'_3} \mathcal{G}_{\ell_1 \lambda'_1 \lambda_4}^{m_1 \mu'_1 \mu_4} \mathcal{G}_{\ell_2 \lambda'_2 \lambda_3}^{m_2 \mu'_2 \mu_3} \mathcal{T}_{\bm \ell \vb \Lambda \bm \lambda \bm \lambda'}^{(1) \bm M}(\vb k_1, \vb k_2, \vb k_3, \vb k_4, \tau).\nonumber 
\end{align}
Bold subscripts represent all the individual indices, i.e. $$\mathcal{T}_{\bm \ell} = \mathcal{T}_{\ell_1 \ell_2 \ell_3 \ell_4}, $$ and we have defined
\begin{align} \label{T1}
&\mathcal{T}_{\bm \ell \vb \Lambda \bm \lambda \bm \lambda'}^{(1) \bm M}(\vb k_1, \vb k_2, \vb k_3, \vb k_4, \tau) \equiv \int_0^{\infty} \int_0^{\infty} \dd x_1 \, \dd x_2 \, x_1^2 x_2^2 \, j_{\ell_4}(k_4 x_1) j_{\ell_3}(k_3 x_2) \\
&\times \int_0^{\infty} \dd x_3 \, x_3^2 \, j_{\ell_1}(k_1 x_3) \mathcal{I}_{\Lambda_1 \lambda_1 \lambda_1'}^{M_1}(\vb k_1, x_2, x_3, \tau) \mathcal{I}_{\Lambda_4 \lambda_4 \lambda_4'}^{M_4}(\vb k_4, x_3, x_1, \tau)  \nonumber\\
&\times \int_0^{\infty} \dd x_4 \, x_4^2 \, j_{\ell_2}(k_2 x_4) \mathcal{I}_{\Lambda_2 \lambda_2 \lambda_2'}^{M_2}(\vb k_2, x_1, x_4, \tau) \mathcal{I}_{\Lambda_3 \lambda_3 \lambda_3'}^{M_3}(\vb k_3, x_4, x_2, \tau) \nonumber.
\end{align}
These four integrals are then to be computed numerically; we discuss various approaches in \S\ref{Section9}. Here, we simply briefly comment on the dimension of each integral, as regards the grid of values that must be explored. 

First we note that the dependence of $\mathcal{I}_{\Lambda_i \lambda_i \lambda_i'}^{M_i}$ on $\vu{k}_i$ is entirely contained in the spherical harmonic in Eq. \eqref{FancyI_Final} and thus can be pulled out of the integral. This is why we only need to consider the magnitude of the $\vb k$ vectors when integrating. For a given integral, we must generate a grid for all of the variables that are not being integrated over, and perform the integration at each point in that grid. For example, the last integral in Eq. (\ref{T1}) is with respect to the $x_4$. We must perform the numerical integration with respect to  $x_4$ at every point of a grid in $k_2, x_1, k_3,$ and $x_2$ ($\tau$ will be evaluated at a single value). Consequently, we will call this a ``five-dimensional integral''.  If the integration is done in the order shown in (\ref{T1}), then the most computationally intensive integration is that with respect to $x_2$, which is six-dimensional. A similar analysis for the other three reference diagram integrals shows that they also contain at most six-dimensional integrals, and the integral with the highest dimensionality depends on the order of integration.

\subsection{Final Result for \hyperref[Diagrams1-4]{Reference Diagram 2}} 
The contribution to the trispectrum for \hyperref[Diagrams1-4]{Reference Diagram 2} is 
\begin{align}
&\expval{Q^{\text{inv}}(\tau, \vb k_1) Q^{\text{inv}}(\tau, \vb k_2)  Q^{\text{inv}}(\tau, \vb k_3) Q^{\text{inv}}(\tau, \vb k_4)}^{(2)} = \\
&\frac{4{,}096}{\Lambda^4 a^4(\tau)} \sum_{\vb \Bell, \vb m} \sum_{\bm{\lambda}, \bm{\mu}} \sum_{\bm \lambda', \bm \mu'} \sum_{A,B} \Bigg [\prod_{j=1}^{4} i^{\ell_j + \lambda_j + \lambda'_j} (-1)^{\lambda'_j + \mu_j + \mu'_j} Y_{\ell_j}^{m_j *}(\vu k_j) \sum_{\Lambda_j} \sum_{M_j} (-1)^{M_j} \Bigg ](-1)^{\mu_4} \nonumber \\
&\times \mathcal{K}_{\ell_1 \lambda'_1 \ell_2 \lambda'_2 A}^{m_1 \mu'_1 m_2 \mu'_2 B} \mathcal{G}_{\lambda_1 \ell_4 \lambda'_4}^{\mu_1 m_4 \mu'_4} \mathcal{G}_{\ell_3 \lambda'_3 \lambda_2}^{m_3 \mu'_3 \mu_2} \kron_{\lambda_3 \lambda_4} \kron_{\mu_3 \mu_4} \nonumber \mathcal{G}_{\lambda'_4 \lambda_4 \Lambda_4}^{-\mu'_4 \mu_4 -M_4}\mathcal{G}_{\lambda'_3 \lambda_3 \Lambda_3}^{\mu'_3 \mu_3 M_3} \mathcal{G}_{\lambda'_2 \lambda_2 \Lambda_2}^{\mu'_2 \mu_2 M_2} \mathcal{G}_{\lambda'_1 \lambda_1 \Lambda_1}^{\mu'_1 \mu_1 M_1}\\
&\times \mathcal{T}_{\bm \ell \vb \Lambda \bm \lambda \bm \lambda'}^{(2) \bm M}(\vb k_1, \vb k_2, \vb k_3, \vb k_4, \tau),\nonumber \nonumber
\end{align}
where
\begin{align}\label{T2}
&\mathcal{T}_{\bm \ell \vb \Lambda \bm \lambda \bm \lambda'}^{(2) \bm M}(\vb k_1, \vb k_2, \vb k_3, \vb k_4, \tau) \equiv \\
&\int_0^{\infty} \int_0^{\infty} \dd x_1 \, \dd x_2 \, x_1^2 x_2^2 \, j_{\ell_1}(k_1 x_1) j_{\ell_4}(k_4 x_2) j_{\ell_2}(k_2 x_1) \mathcal{I}_{\Lambda_1 \lambda_1 \lambda_1'}^{M_1}(\vb k_1, x_2, x_1, \tau) \nonumber \\
&\times \int_0^{\infty} \dd x_3 \, x_3^2 \, j_{\ell_3}(k_3 x_3) \mathcal{I}_{\Lambda_2 \lambda_2 \lambda_2'}^{M_2}(\vb k_2, x_3, x_1, \tau) \nonumber\\
&\times \int_0^{\infty} \dd x_4 \, x_4^2 \, \mathcal{I}_{\Lambda_3 \lambda_3 \lambda_3'}^{M_3}(\vb k_3, x_4, x_3, \tau) \mathcal{I}_{\Lambda_4 \lambda_4 \lambda_4'}^{M_4}(\vb k_4, x_4, x_2, \tau). \nonumber
\end{align}

\subsection{Final Result for \hyperref[Diagrams1-4]{Reference Diagram 3}}
\begin{align}
&\expval{Q^{\text{inv}}(\tau, \vb k_1) Q^{\text{inv}}(\tau, \vb k_2)  Q^{\text{inv}}(\tau, \vb k_3) Q^{\text{inv}}(\tau, \vb k_4)}^{(3)} \equiv \\
&\frac{4{,}096}{\Lambda^4 a^4(\tau)} \sum_{\vb \Bell, \vb m} \sum_{\bm{\lambda}, \bm{\mu}} \sum_{\bm \lambda', \bm \mu'} \sum_{A,B} \sum_{A',B'} \Bigg [\prod_{j=1}^{4} i^{\ell_j + \lambda_j + \lambda'_j} (-1)^{\lambda'_j + \mu_j + \mu'_j} Y_{\ell_j}^{m_j *}(\vu k_j) \sum_{\Lambda_j} \sum_{M_j} (-1)^{M_j} \Bigg ](-1)^{\mu_3 + \mu_4} \nonumber \\
&\times \mathcal{K}_{\ell_1 \lambda'_1 \ell_2 \lambda'_2 A}^{m_1 \mu'_1 m_2 \mu'_2 B} \mathcal{K}_{\ell_3 \lambda'_3 \ell_4 \lambda'_4 A'}^{m_3 \mu'_3 m_4 \mu'_4 B'} \kron_{\lambda_1 \lambda_3} \kron_{\mu_1 \mu_3} \kron_{\lambda_2 \lambda_4} \kron_{\mu_2 \mu_4} \nonumber \mathcal{G}_{\lambda'_4 \lambda_4 \Lambda_4}^{-\mu'_4 \mu_4 -M_4}\mathcal{G}_{\lambda'_3 \lambda_3 \Lambda_3}^{-\mu'_3 \mu_3 -M_3} \mathcal{G}_{\lambda'_2 \lambda_2 \Lambda_2}^{\mu'_2 \mu_2 M_2} \\ 
&\times \mathcal{G}_{\lambda'_1 \lambda_1 \Lambda_1}^{\mu'_1 \mu_1 M_1} \mathcal{T}_{\bm \ell \vb \Lambda \bm \lambda \bm \lambda'}^{(3) \bm M}(\vb k_1, \vb k_2, \vb k_3, \vb k_4, \tau),\nonumber 
\end{align}
where
\begin{align}\label{T3}
&\mathcal{T}_{\bm \ell \vb \Lambda \bm \lambda \bm \lambda'}^{(3) \bm M}(\vb k_1, \vb k_2, \vb k_3, \vb k_4, \tau) \equiv \\
&\int_0^{\infty} \int_0^{\infty} \dd x_1 \, \dd x_2 \, x_1^2 x_2^2 \, j_{\ell_1}(k_1 x_1) j_{\ell_2}(k_2 x_1) j_{\ell_4}(k_4 x_2) j_{\ell_3}(k_3 x_2) \nonumber \\
&\times \int_0^{\infty} \dd x_3 \, x_3^2 \, \mathcal{I}_{\Lambda_3 \lambda_3 \lambda_3'}^{M_3}(\vb k_3, x_3, x_2, \tau) \mathcal{I}_{\Lambda_1 \lambda_1 \lambda_1'}^{M_1}(\vb k_1, x_3, x_1, \tau) \nonumber\\
&\times \int_0^{\infty} \dd x_4 \, x_4^2 \, \mathcal{I}_{\Lambda_2 \lambda_2 \lambda_2'}^{M_2}(\vb k_2, x_4, x_1, \tau)  \mathcal{I}_{\Lambda_4 \lambda_4 \lambda_4'}^{M_4}(\vb k_4, x_4, x_2, \tau). \nonumber 
\end{align}

\subsection{Final Result for \hyperref[Diagrams1-4]{Reference Diagram 4}}
\begin{align}
&\expval{Q^{\text{inv}}(\tau, \vb k_1) Q^{\text{inv}}(\tau, \vb k_2)  Q^{\text{inv}}(\tau, \vb k_3) Q^{\text{inv}}(\tau, \vb k_4)}^{(4)} = \\
&\frac{4{,}096}{\Lambda^4 a^4(\tau)} \sum_{\vb \Bell, \vb m} \sum_{\bm{\lambda}, \bm{\mu}} \sum_{\bm \lambda', \bm \mu'} \sum_{A,B} \Bigg [\prod_{j=1}^{4} i^{\ell_j + \lambda_j + \lambda'_j} (-1)^{\lambda'_j + \mu_j + \mu'_j} Y_{\ell_j}^{m_j *}(\vu k_j) \sum_{\Lambda_j} \sum_{M_j} (-1)^{M_j} \Bigg ](-1)^{\mu_3} \nonumber \\
&\times \mathcal{K}_{\ell_1 \lambda'_1 \ell_4 \lambda'_4 A}^{m_1 \mu'_1 m_4 \mu'_4 B} \mathcal{G}_{\lambda_4 \ell_2 \lambda'_2}^{\mu_4 m_2 \mu'_2} \mathcal{G}_{\ell_3 \lambda'_3 \lambda_2}^{m_3 \mu'_3 \mu_2} \kron_{\lambda_1 \lambda_3} \kron_{\mu_1 \mu_3} \nonumber \mathcal{G}_{\lambda'_4 \lambda_4 \Lambda_4}^{\mu'_4 \mu_4 M_4}\mathcal{G}_{\lambda'_3 \lambda_3 \Lambda_3}^{-\mu'_3  \mu_3 -M_3} \mathcal{G}_{\lambda'_2 \lambda_2 \Lambda_2}^{\mu'_2 \mu_2 M_2} \mathcal{G}_{\lambda'_1 \lambda_1 \Lambda_1}^{\mu'_1 \mu_1 M_1}\\
&\times \mathcal{T}_{\bm \ell \vb \Lambda \bm \lambda \bm \lambda'}^{(4) \bm M}(\vb k_1, \vb k_2, \vb k_3, \vb k_4, \tau),\nonumber
\end{align}
where
\begin{align}\label{T4}
&\mathcal{T}_{\bm \ell \vb \Lambda \bm \lambda \bm \lambda'}^{(4) \bm M}(\vb k_1, \vb k_2, \vb k_3, \vb k_4, \tau) \equiv \\
&\int_0^{\infty} \int_0^{\infty} \dd x_1 \, \dd x_2 \, x_1^2 x_2^2 \, j_{\ell_1}(k_1 x_1) j_{\ell_4}(k_4 x_1) j_{\ell_2}(k_2 x_2) \mathcal{I}_{\Lambda_1 \lambda_1 \lambda_1'}^{M_1}(\vb k_1, x_2, x_1, \tau) \nonumber\\
&\times \int_0^{\infty} \dd x_3 \, x_3^2 \, j_{\ell_3}(k_3 x_3) \mathcal{I}_{\Lambda_2 \lambda_2 \lambda_2'}^{M_2}(\vb k_2, x_3, x_1, \tau) \nonumber\\
&\times \int_0^{\infty} \dd x_4 \, x_4^2 \,   \mathcal{I}_{\Lambda_3 \lambda_3 \lambda_3'}^{M_3}(\vb k_3, x_4, x_3, \tau) \mathcal{I}_{\Lambda_4 \lambda_4 \lambda_4'}^{M_4}(\vb k_4, x_4, x_2, \tau). \nonumber 
\end{align}

\subsection{Final Result for the Inverse Decay Trispectrum} \label{Final_Trispectrum}
In the previous four subsections, we wrote down the final result for the trispectrum contribution from each of the four Reference Diagrams in \S \ref{Section5}. These diagrams were arbitrarily chosen to serve as a representative for each connection structure. The numbering of each Reference Diagram corresponds to its connection structure. 

We can see from Figure \ref{fig:diagram_structure} that there are a number of diagrams for each connection structure. Each of these diagrams can be related to their corresponding Reference Diagram by a relabeling of the external momenta. We may therefore include the contribution to the trispectrum for any one of the 48 one-loop diagrams by identifying its connection structure, writing down the final results for the corresponding Reference Diagram, and relabeling the external momenta in the $\mathcal{T}_{ \bm \ell \vb \Lambda \bm \lambda \bm \lambda'}^{(i) \bm M}(\vb k_1, \vb k_2, \vb k_3, \vb k_4, \tau)$.  

With this in mind, we may write Eq. (\ref{eq:inv_4PCF}) as
\begin{align}
&\expval{Q^{\text{inv}}(\tau, \vb k_1) Q^{\text{inv}}(\tau, \vb k_2)  Q^{\text{inv}}(\tau, \vb k_3) Q^{\text{inv}}(\tau, \vb k_4)} = \\
&\frac{4{,}096}{\Lambda^4 a^4(\tau)}\sum_{\vb \Bell, \vb m} \sum_{\bm{\lambda}, \bm{\mu}} \sum_{\bm \lambda', \bm \mu'} \Bigg [\prod_{j=1}^{4}  i^{\ell_j + \lambda_j + \lambda'_j} (-1)^{\lambda'_j + \mu_j + \mu'_j}  Y_{\ell_j}^{m_j *}(\vu k_j) \sum_{\Lambda_j M_j} (-1)^{M_j} \mathcal{G}_{\lambda'_j \lambda_j \Lambda_j}^{\mu'_j \mu_j M_j} \Bigg ] \nonumber  \\
&\times \mathcal{G}_{\lambda_2 \ell_4 \lambda'_4}^{\mu_2 m_4 \mu'_4} \mathcal{G}_{\lambda_1 \ell_3 \lambda'_3}^{\mu_1 m_3 \mu'_3} \mathcal{G}_{\ell_1 \lambda'_1 \lambda_4}^{m_1 \mu'_1 \mu_4} \mathcal{G}_{\ell_2 \lambda'_2 \lambda_3}^{m_2 \mu'_2 \mu_3} \tilde{\mathcal{T}}_{\bm \ell \vb \Lambda \bm \lambda \bm \lambda'}^{(1) \bm M}(\vb k_1, \vb k_2, \vb k_3, \vb k_4, \tau) \nonumber \\
&+ \frac{4{,}096}{\Lambda^4 a^4(\tau)} \sum_{\vb \Bell, \vb m} \sum_{\bm{\lambda}, \bm{\mu}} \sum_{\bm \lambda', \bm \mu'} \sum_{A,B} \Bigg [\prod_{j=1}^{4} i^{\ell_j + \lambda_j + \lambda'_j} (-1)^{\lambda'_j + \mu_j + \mu'_j} Y_{\ell_j}^{m_j *}(\vu k_j) \sum_{\Lambda_j} \sum_{M_j} (-1)^{M_j} \Bigg ](-1)^{\mu_4} \nonumber \\
&\times \mathcal{K}_{\ell_1 \lambda'_1 \ell_2 \lambda'_2 A}^{m_1 \mu'_1 m_2 \mu'_2 B} \mathcal{G}_{\lambda_1 \ell_4 \lambda'_4}^{\mu_1 m_4 \mu'_4} \mathcal{G}_{\ell_3 \lambda'_3 \lambda_2}^{m_3 \mu'_3 \mu_2} \kron_{\lambda_3 \lambda_4} \kron_{\mu_3 \mu_4} \nonumber \mathcal{G}_{\lambda'_4 \lambda_4 \Lambda_4}^{-\mu'_4 \mu_4 -M_4}\mathcal{G}_{\lambda'_3 \lambda_3 \Lambda_3}^{\mu'_3 \mu_3 M_3} \mathcal{G}_{\lambda'_2 \lambda_2 \Lambda_2}^{\mu'_2 \mu_2 M_2} \mathcal{G}_{\lambda'_1 \lambda_1 \Lambda_1}^{\mu'_1 \mu_1 M_1}\\
&\times \tilde{\mathcal{T}}_{\bm \ell \vb \Lambda \bm \lambda \bm \lambda'}^{(2) \bm M}(\vb k_1, \vb k_2, \vb k_3, \vb k_4, \tau) \nonumber \\
& + \frac{4{,}096}{\Lambda^4 a^4(\tau)} \sum_{\vb \Bell, \vb m} \sum_{\bm{\lambda}, \bm{\mu}} \sum_{\bm \lambda', \bm \mu'} \sum_{A,B} \sum_{A',B'} \Bigg [\prod_{j=1}^{4} i^{\ell_j + \lambda_j + \lambda'_j} (-1)^{\lambda'_j + \mu_j + \mu'_j} Y_{\ell_j}^{m_j *}(\vu k_j) \sum_{\Lambda_j} \sum_{M_j} (-1)^{M_j} \Bigg ](-1)^{\mu_3 + \mu_4} \nonumber \\
&\times \mathcal{K}_{\ell_1 \lambda'_1 \ell_2 \lambda'_2 A}^{m_1 \mu'_1 m_2 \mu'_2 B} \mathcal{K}_{\ell_3 \lambda'_3 \ell_4 \lambda'_4 A'}^{m_3 \mu'_3 m_4 \mu'_4 B'} \kron_{\lambda_1 \lambda_3} \kron_{\mu_1 \mu_3} \kron_{\lambda_2 \lambda_4} \kron_{\mu_2 \mu_4} \nonumber \mathcal{G}_{\lambda'_4 \lambda_4 \Lambda_4}^{-\mu'_4 \mu_4 -M_4}\mathcal{G}_{\lambda'_3 \lambda_3 \Lambda_3}^{-\mu'_3 \mu_3 -M_3} \mathcal{G}_{\lambda'_2 \lambda_2 \Lambda_2}^{\mu'_2 \mu_2 M_2} \\ 
&\times \mathcal{G}_{\lambda'_1 \lambda_1 \Lambda_1}^{\mu'_1 \mu_1 M_1} \tilde{\mathcal{T}}_{\bm \ell \vb \Lambda \bm \lambda \bm \lambda'}^{(3) \bm M}(\vb k_1, \vb k_2, \vb k_3, \vb k_4, \tau) \nonumber \\ 
& + \frac{4{,}096}{\Lambda^4 a^4(\tau)} \sum_{\vb \Bell, \vb m} \sum_{\bm{\lambda}, \bm{\mu}} \sum_{\bm \lambda', \bm \mu'} \sum_{A,B} \Bigg [\prod_{j=1}^{4} i^{\ell_j + \lambda_j + \lambda'_j} (-1)^{\lambda'_j + \mu_j + \mu'_j} Y_{\ell_j}^{m_j *}(\vu k_j) \sum_{\Lambda_j} \sum_{M_j} (-1)^{M_j} \Bigg ](-1)^{\mu_3} \nonumber \\
&\times \mathcal{K}_{\ell_1 \lambda'_1 \ell_4 \lambda'_4 A}^{m_1 \mu'_1 m_4 \mu'_4 B} \mathcal{G}_{\lambda_4 \ell_2 \lambda'_2}^{\mu_4 m_2 \mu'_2} \mathcal{G}_{\ell_3 \lambda'_3 \lambda_2}^{m_3 \mu'_3 \mu_2} \kron_{\lambda_1 \lambda_3} \kron_{\mu_1 \mu_3} \nonumber \mathcal{G}_{\lambda'_4 \lambda_4 \Lambda_4}^{\mu'_4 \mu_4 M_4}\mathcal{G}_{\lambda'_3 \lambda_3 \Lambda_3}^{-\mu'_3  \mu_3 -M_3} \mathcal{G}_{\lambda'_2 \lambda_2 \Lambda_2}^{\mu'_2 \mu_2 M_2} \mathcal{G}_{\lambda'_1 \lambda_1 \Lambda_1}^{\mu'_1 \mu_1 M_1}\\
&\times \tilde{\mathcal{T}}_{\bm \ell \vb \Lambda \bm \lambda \bm \lambda'}^{(4) \bm M}(\vb k_1, \vb k_2, \vb k_3, \vb k_4, \tau).\nonumber
\end{align}
Here, 
\begin{align}
\tilde{\mathcal{T}}_{\bm \ell \vb \Lambda \bm \lambda \bm \lambda'}^{(1) \bm M}(\vb k_1, \vb k_2, \vb k_3, \vb k_4, \tau)& \equiv  \mathcal{T}_{\bm \ell \vb \Lambda \bm \lambda \bm \lambda'}^{(1) \bm M}(\vb k_1, \vb k_2, \vb k_3, \vb k_4, \tau) + \mathcal{T}_{\bm \ell \vb \Lambda \bm \lambda \bm \lambda'}^{(1) \bm M}(\vb k_1, \vb k_2, \vb k_4, \vb k_3, \tau)  \nonumber \\
&+ \mathcal{T}_{\bm \ell \vb \Lambda \bm \lambda \bm \lambda'}^{(1) \bm M}(\vb k_3, \vb k_2, \vb k_1, \vb k_4, \tau) + \mathcal{T}_{\bm \ell \vb \Lambda \bm \lambda \bm \lambda'}^{(1) \bm M}(\vb k_1, \vb k_3, \vb k_2, \vb k_4, \tau) \nonumber \\
&+ \mathcal{T}_{\bm \ell \vb \Lambda \bm \lambda \bm \lambda'}^{(1) \bm M}(\vb k_4, \vb k_2, \vb k_3, \vb k_1, \tau) + \mathcal{T}_{\bm \ell \vb \Lambda \bm \lambda \bm \lambda'}^{(1) \bm M}(\vb k_1, \vb k_4, \vb k_3, \vb k_2, \tau),
\end{align}
\begin{align}
\tilde{\mathcal{T}}_{\bm \ell \vb \Lambda \bm \lambda \bm \lambda'}^{(2) \bm M}(\vb k_1, \vb k_2, \vb k_3, \vb k_4, \tau)& \equiv  \mathcal{T}_{\bm \ell \vb \Lambda \bm \lambda \bm \lambda'}^{(2) \bm M}(\vb k_1, \vb k_2, \vb k_3, \vb k_4, \tau) + \mathcal{T}_{\bm \ell \vb \Lambda \bm \lambda \bm \lambda'}^{(2) \bm M}(\vb k_4, \vb k_2, \vb k_3, \vb k_1, \tau) \nonumber \\
&+ \mathcal{T}_{\bm \ell \vb \Lambda \bm \lambda \bm \lambda'}^{(2) \bm M}(\vb k_4, \vb k_1, \vb k_3, \vb k_2, \tau) + \mathcal{T}_{\bm \ell \vb \Lambda \bm \lambda \bm \lambda'}^{(2) \bm M}(\vb k_2, \vb k_3, \vb k_1, \vb k_4, \tau) \nonumber \\
&+ \mathcal{T}_{\bm \ell \vb \Lambda \bm \lambda \bm \lambda'}^{(2) \bm M}(\vb k_1, \vb k_3, \vb k_2, \vb k_4, \tau) + \mathcal{T}_{\bm \ell \vb \Lambda \bm \lambda \bm \lambda'}^{(2) \bm M}(\vb k_1, \vb k_4, \vb k_3, \vb k_2, \tau) \nonumber \\
&+ \mathcal{T}_{\bm \ell \vb \Lambda \bm \lambda \bm \lambda'}^{(2) \bm M}(\vb k_3, \vb k_2, \vb k_1, \vb k_4, \tau) + \mathcal{T}_{\bm \ell \vb \Lambda \bm \lambda \bm \lambda'}^{(2) \bm M}(\vb k_1, \vb k_3, \vb k_4, \vb k_2, \tau) \nonumber \\
&+ \mathcal{T}_{\bm \ell \vb \Lambda \bm \lambda \bm \lambda'}^{(2) \bm M}(\vb k_2, \vb k_4, \vb k_3, \vb k_1, \tau) + \mathcal{T}_{\bm \ell \vb \Lambda \bm \lambda \bm \lambda'}^{(2) \bm M}(\vb k_2, \vb k_1, \vb k_3, \vb k_4, \tau) \nonumber \\
&+ \mathcal{T}_{\bm \ell \vb \Lambda \bm \lambda \bm \lambda'}^{(2) \bm M}(\vb k_3, \vb k_4, \vb k_2, \vb k_1, \tau) + \mathcal{T}_{\bm \ell \vb \Lambda \bm \lambda \bm \lambda'}^{(2) \bm M}(\vb k_3, \vb k_4, \vb k_1, \vb k_2, \tau),
\end{align}
\begin{align}
\tilde{\mathcal{T}}_{\bm \ell \vb \Lambda \bm \lambda \bm \lambda'}^{(3) \bm M}(\vb k_1, \vb k_2, \vb k_3, \vb k_4, \tau)& \equiv  \mathcal{T}_{\bm \ell \vb \Lambda \bm \lambda \bm \lambda'}^{(3) \bm M}(\vb k_1, \vb k_2, \vb k_3, \vb k_4, \tau) + \mathcal{T}_{\bm \ell \vb \Lambda \bm \lambda \bm \lambda'}^{(3) \bm M}(\vb k_1, \vb k_4, \vb k_3, \vb k_2, \tau) \nonumber \\
&+ \mathcal{T}_{\bm \ell \vb \Lambda \bm \lambda \bm \lambda'}^{(3) \bm M}(\vb k_1, \vb k_2, \vb k_4, \vb k_3, \tau) + \mathcal{T}_{\bm \ell \vb \Lambda \bm \lambda \bm \lambda'}^{(3) \bm M}(\vb k_3, \vb k_1, \vb k_2, \vb k_4, \tau) \nonumber \\
&+ \mathcal{T}_{\bm \ell \vb \Lambda \bm \lambda \bm \lambda'}^{(3) \bm M}(\vb k_1, \vb k_3, \vb k_2, \vb k_4, \tau) + \mathcal{T}_{\bm \ell \vb \Lambda \bm \lambda \bm \lambda'}^{(3) \bm M}(\vb k_1, \vb k_4, \vb k_2, \vb k_3, \tau), 
\end{align}
and
\begin{align}
\tilde{\mathcal{T}}_{\bm \ell \vb \Lambda \bm \lambda \bm \lambda'}^{(4) \bm M}(\vb k_1, \vb k_2, \vb k_3, \vb k_4, \tau)& \equiv  \mathcal{T}_{\bm \ell \vb \Lambda \bm \lambda \bm \lambda'}^{(4) \bm M}(\vb k_1, \vb k_2, \vb k_3, \vb k_4, \tau) + \mathcal{T}_{\bm \ell \vb \Lambda \bm \lambda \bm \lambda'}^{(4) \bm M}(\vb k_3, \vb k_4, \vb k_1, \vb k_2, \tau) \nonumber \\
&+ \mathcal{T}_{\bm \ell \vb \Lambda \bm \lambda \bm \lambda'}^{(4) \bm M}(\vb k_4, \vb k_3, \vb k_2, \vb k_1, \tau)+ \mathcal{T}_{\bm \ell \vb \Lambda \bm \lambda \bm \lambda'}^{(4) \bm M}(\vb k_2, \vb k_1, \vb k_3, \vb k_4, \tau) \nonumber \\
&+ \mathcal{T}_{\bm \ell \vb \Lambda \bm \lambda \bm \lambda'}^{(4) \bm M}(\vb k_4, \vb k_3, \vb k_1, \vb k_2, \tau) + \mathcal{T}_{\bm \ell \vb \Lambda \bm \lambda \bm \lambda'}^{(4) \bm M}(\vb k_3, \vb k_2, \vb k_1, \vb k_4, \tau) \nonumber \\
&+ \mathcal{T}_{\bm \ell \vb \Lambda \bm \lambda \bm \lambda'}^{(4) \bm M}(\vb k_3, \vb k_4, \vb k_2, \vb k_1, \tau) + \mathcal{T}_{\bm \ell \vb \Lambda \bm \lambda \bm \lambda'}^{(4) \bm M}(\vb k_2, \vb k_1, \vb k_4, \vb k_3, \tau) \nonumber \\
&+ \mathcal{T}_{\bm \ell \vb \Lambda \bm \lambda \bm \lambda'}^{(4) \bm M}(\vb k_1, \vb k_4, \vb k_3, \vb k_2, \tau) + \mathcal{T}_{\bm \ell \vb \Lambda \bm \lambda \bm \lambda'}^{(4) \bm M}(\vb k_4, \vb k_1, \vb k_2, \vb k_3, \tau) \nonumber \\
&+ \mathcal{T}_{\bm \ell \vb \Lambda \bm \lambda \bm \lambda'}^{(4) \bm M}(\vb k_3, \vb k_1, \vb k_2, \vb k_4, \tau) + \mathcal{T}_{\bm \ell \vb \Lambda \bm \lambda \bm \lambda'}^{(4) \bm M}(\vb k_2, \vb k_4, \vb k_3, \vb k_1, \tau) \nonumber \\
&+ \mathcal{T}_{\bm \ell \vb \Lambda \bm \lambda \bm \lambda'}^{(4) \bm M}(\vb k_4, \vb k_2, \vb k_3, \vb k_1, \tau) + \mathcal{T}_{\bm \ell \vb \Lambda \bm \lambda \bm \lambda'}^{(4) \bm M}(\vb k_1, \vb k_4, \vb k_2, \vb k_3, \tau) \nonumber \\
&+ \mathcal{T}_{\bm \ell \vb \Lambda \bm \lambda \bm \lambda'}^{(4) \bm M}(\vb k_2, \vb k_4, \vb k_1, \vb k_3, \tau) + \mathcal{T}_{\bm \ell \vb \Lambda \bm \lambda \bm \lambda'}^{(4) \bm M}(\vb k_4, \vb k_1, \vb k_3, \vb k_2, \tau) \nonumber \\
&+ \mathcal{T}_{\bm \ell \vb \Lambda \bm \lambda \bm \lambda'}^{(4) \bm M}(\vb k_3, \vb k_1, \vb k_4, \vb k_2, \tau) + \mathcal{T}_{\bm \ell \vb \Lambda \bm \lambda \bm \lambda'}^{(4) \bm M}(\vb k_1, \vb k_2, \vb k_4, \vb k_3, \tau) \nonumber \\
&+ \mathcal{T}_{\bm \ell \vb \Lambda \bm \lambda \bm \lambda'}^{(4) \bm M}(\vb k_3, \vb k_2, \vb k_4, \vb k_1, \tau)  + \mathcal{T}_{\bm \ell \vb \Lambda \bm \lambda \bm \lambda'}^{(4) \bm M}(\vb k_2, \vb k_3, \vb k_4, \vb k_1, \tau) \nonumber \\
&+ \mathcal{T}_{\bm \ell \vb \Lambda \bm \lambda \bm \lambda'}^{(4) \bm M}(\vb k_1, \vb k_3, \vb k_4, \vb k_2, \tau) + \mathcal{T}_{\bm \ell \vb \Lambda \bm \lambda \bm \lambda'}^{(4) \bm M}(\vb k_2, \vb k_3, \vb k_1, \vb k_4, \tau) \nonumber \\
&+ \mathcal{T}_{\bm \ell \vb \Lambda \bm \lambda \bm \lambda'}^{(4) \bm M}(\vb k_1, \vb k_3, \vb k_2, \vb k_4, \tau) + \mathcal{T}_{\bm \ell \vb \Lambda \bm \lambda \bm \lambda'}^{(4) \bm M}(\vb k_4, \vb k_2, \vb k_1, \vb k_3, \tau). 
\end{align}
We note that the $\mathcal{T}_{\bm \ell \vb \Lambda \bm \lambda \bm \lambda'}^{(i) \bm M}(\vb k_1, \vb k_2, \vb k_3, \vb k_4, \tau), i = 1,2,3,4$ were defined in Eqs. (\ref{T1}), (\ref{T2}), (\ref{T3}), and (\ref{T4}).  
Each $\tilde{\mathcal{T}}_{\bm \ell \vb \Lambda \bm \lambda \bm \lambda'}^{(i) \bm M}$ contains contributions from all diagrams with connection structure $i$. We have used Figure \ref{fig:diagram_structure} to determine the relabeling of the external momenta that transforms Reference Diagram $i$ into each of the other diagrams with connection structure $i$. The requisite functions for constructing the final expression for the trispectrum are listed in Table \ref{Trispectrum_Table}.

\begin{table}[h] 
    \centering
    \renewcommand{\arraystretch}{1.5} 
    \small
    \begin{tabular}{l l l l}
        \toprule
        \textbf{Symbol} & \textbf{Equation} & \textbf{Mass Dimension} & \textbf{Definition} \\
        \midrule
        $\mathcal{T}_{\bm \ell \vb \Lambda \bm \lambda \bm \lambda'}^{(1) \bm M}(\vb k_1, \vb k_2, \vb k_3, \vb k_4, \tau)$ & \ref{T1} & $-4$ & Integral for \hyperref[Diagrams1-4]{Reference Diagram 1} \\
        $\mathcal{T}_{\bm \ell \vb \Lambda \bm \lambda \bm \lambda'}^{(2) \bm M}(\vb k_1, \vb k_2, \vb k_3, \vb k_4, \tau)$ & \ref{T2} & $-4$ & Integral for \hyperref[Diagrams1-4]{Reference Diagram 2} \\
        $\mathcal{T}_{\bm \ell \vb \Lambda \bm \lambda \bm \lambda'}^{(3) \bm M}(\vb k_1, \vb k_2, \vb k_3, \vb k_4, \tau)$ & \ref{T3} & $-4$ & Integral for \hyperref[Diagrams1-4]{Reference Diagram 3} \\
        $\mathcal{T}_{\bm \ell \vb \Lambda \bm \lambda \bm \lambda'}^{(4) \bm M}(\vb k_1, \vb k_2, \vb k_3, \vb k_4, \tau)$ & \ref{T4} & $-4$ & Integral for \hyperref[Diagrams1-4]{Reference Diagram 4} \\
        $\mathcal{I}_{\Lambda \lambda \lambda'}^{M}(\vb k, x, x', \tau)$ & \ref{FancyI_Final} & \hspace{1.5mm} $2$ & Integral of gauge field and Green function \\
        $\mathcal{M}_{aa'c}(u)$ & \ref{Curly_M} & $-2$ & Nested radial and conformal time integrals\\
        $\mathcal{S}(w)$ & \ref{S_def} & \hspace{1.5mm} $0$ & Conformal time integral \\
        $\mathcal{A}(k\tau)$ & \ref{FancyA} & $-1$ & Modified gauge field mode function\\
        $M_{a'}(r/\tau')$ & \ref{M_integral} & $-1$ & Integral of gauge field against a Bessel function\\
        $\text{I}_{\Lambda \lambda \lambda'}^{M}(\vb{k}, x, x', \tau)$ & \ref{eq:I_int} & \hspace{1.5mm} $4$ & Loop integral for the gauge field\\
        $G(\tau, \tau', k)$ & \ref{eq:Greens_Function} & $-1$ & Inflaton retarded Green function \\
        $J(\tau, \vb k)$ & \ref{J_def} & \hspace{1.5mm} $0$ & Source term for the inflaton fluctuations \\
        $Q(\tau, \vb k)$ & \ref{Q_MF} & $-2$ & Inflaton mode function \\
        $A_{-}(\tau, k)$ & \ref{A:approx_1} & $-\frac{1}{2}$ & Gauge field mode function \\
        $\zeta(\tau, \vb k)$ & \ref{zeta} & $-3$ & Curvature perturbation \\
        
        \bottomrule
    \end{tabular}
    \caption{Table of functions used to construct the final expression for the trispectrum given in \S \ref{Final_Trispectrum}. A column for the symbol, location, mass dimension, and a short description are shown. The lower part of the table contains more fundamental functions from which the more complicated functions in the upper part of the table are constructed.}
    \label{Trispectrum_Table}
\end{table}

\section{Numerical Integration Scheme for the Radial Integrals}
\label{Section9}

The integrals we encountered in \S\ref{Section7} and \S\ref{Section8}  cannot be computed analytically; we will need to turn to numerical methods. There is extensive literature on numerical integration of spherical Bessel functions \cite{LUCAS1995269, Ogata}; we refer the reader to \cite{slepian_rot_method_19} for a recent fast method that exploits rotation in the space of free frequencies to reduce dimensionality, as well as \cite{Fang_2020, Grasshorn_Gebhardt_2018, Umeh_2021, Simonovi_2018, Schoneberg_2018, Assassi_2017} using the FFTLog algorithm. 

We defer a full numerical implementation of the expressions of this work to the second paper in this series. Here we restrict ourselves to a brief discussion to make it clear that the expressions we have obtained carry with them minimal computational cost, especially in comparison to the naive 16-dimensional integrations required when the loop integrals are first written down. 

The integrals $\mathcal{I}_{\Lambda_i \lambda_i \lambda'_i}^{M}(\vb k_i, x_j, x_k, \tau)$ given in Eq. (\ref{FancyI_Final}) can be computed using the 2D FFTLog algorithm \cite{Hamilton_2000}, after which $\mathcal{I}_{\Lambda_i \lambda_i \lambda'_i}^{M}(\vb k_i, x_j, x_k, \tau)$ will be represented by an $n \times m \times m$ tensor ($\tau$ assumes a single value, corresponding to the end of inflation). One must then integrate these tensors against spherical Bessel functions as shown in \S\ref{Section8}. In principle these latter integrals could also be done in this way except that code to do FFTLog in more than two arguments does not yet exist. So instead, we return to the simple method of Riemann sums for the subsequent integrations. This enables numerical computation in \textsc{python} using the \texttt{numpy} library's \texttt{einsum} function. A method using these functions was first presented in \cite{williams_paper} and we briefly review it here.

We may represent each $k_i \in \{k_1, k_2, k_3, k_4\}$ as an $n$-dimensional numerical vector, and $x_i \in \{x_1, x_2, x_3, x_4\}$ as an $m$-dimensional numerical vector. Then each spherical Bessel function $j_{\ell_i}(k_ix_j)$ will be an $n \times m$ matrix for a fixed value of $\ell_i$. In \S\ref{Section8}, these functions must then be integrated over. Let us consider as an example the integral over $x_4$ in Eq. (\ref{T1}):
\begin{align}
\int_0^{\infty} \dd x_4 \, x_4^2 \, j_{\ell_2}(k_2 x_4) \mathcal{I}_{\Lambda_2 \lambda_2 \lambda_2'}^{M_2}(\vb k_2, x_1, x_4, \tau) \mathcal{I}_{\Lambda_3 \lambda_3 \lambda_3'}^{M_3}(\vb k_3, x_4, x_2, \tau) \nonumber.
\end{align}
The integrand contains the $m$-dimensional numerical vector $x_4$, an $n \times m$ matrix $j_{\ell_2}(k_2x_4)$ and two $n \times m \times m$ tensors. We may numerically integrate these functions simply by contracting the arrays along the dimension corresponding to $x_4$ (and multiplying the result by the discretized differential $dx_4$). This operation may be done conveniently with the \texttt{einsum} function of \textsc{python}'s \texttt{numpy} library. The result is a 4D array, with the indices corresponding to the values of $k_2$, $k_3$, $x_1$, and $x_2$. This same method can be used to compute all of the integrals of \S\ref{Section8}.
The most complicated integrals that we must compute numerically are 6D, such as the integral over $x_2$ in Eq. (\ref{T1}).

\section{Discussion \& Conclusions}
\label{Section10}
Recent measurements of the galaxy 4PCF have found evidence for parity violation in the 3D large-scale structure \cite{4PCF, Philcox_2022}. If true, this finding indicates that there was very likely a parity-violating interaction during inflation. An inflationary model with an axion coupled to a $U(1)$ gauge field is a leading candidate for such a parity-violating theory. The primordial trispectrum only has a parity-odd component at one-loop and beyond; the loop integrals that appear in this calculation are extremely difficult to calculate analytically, and are also very expensive if one sought to do them fully numerically. We have demonstrated a new method to simplify this computation, which relies on rewriting the Dirac delta functions using the plane wave expansion and making use of the Convolution Theorem. We have drastically reduced the dimensionality of the trispectrum integrals.

One previous work computed the trispectrum for the model of axion inflation we explore here; however they were restricted to just two wave-vector configurations due to the high cost of numerically evaluating the loop integrals \cite{Niu_2023}. Another work that computed the trispectrum for a similar model, but with an axion as a spectator field, also explored only a small portion of the parameter space \cite{Fujita_2024}. 

With the method presented in this paper, it is now feasible to compute these integrals numerically and thus calculate the primordial trispectrum of axion inflation for all configurations. One could use this method to compare with recent lattice simulations of the axion inflation trispectrum \cite{Caravano_2023, Angelo_Thesis, Figueroa_2023}. Our analytic treatment assumes $\xi$ is constant in time; a lattice simulation would not need to make this assumption.

The ultimate aim of our work is constraining axion inflation through fitting a model of the galaxy 4PCF from it to the measured 4PCF. In future work, we will develop code that implements the methods outlined in this paper, and numerically compute the primordial trispectrum for a large range of the parameter space. Next we will apply the linear transfer function to evolve the primordial trispectrum in time. After accounting for galaxy biasing, applying redshift space distortions, and taking an inverse FT, we will obtain the galaxy 4PCF. With the galaxy 4PCF in hand, one can determine the tetrahedra that  have the highest parity-odd signal to noise ratio. Considering only these tetrahedra will greatly reduce the number of degrees of freedom in our covariance matrix, significantly reducing the uncertainty as compared to previous galaxy 4PCF measurements (\cite{PhysRevLett.130.201002}, page 4). This will allow us to place the first ever constraints on parity violation from axion inflation using higher-order galaxy correlations.

\appendix
\section{Expansions of the Multi-Argument Dirac Delta Function} \label{AppendixA}
We will rewrite the Dirac delta functions of multiple arguments that occur in the main text in a form more suitable for our trispectrum calculations.
\subsection{Definitions}
 Our expansions decompose the Dirac delta functions into a sum of angular coefficients multiplied by radial integrals; we define these expressions now. The Gaunt integral is
\begin{align}\label{eq:Gaunt}
&\mathcal{G}_{\ell_1 \ell_2 \ell_3}^{m_1 m_2 m_3} \equiv \int Y_{\ell_1}^{m_1}(\vu r ) Y_{\ell_2}^{m_2}(\vu r) Y_{\ell_3}^{m_3}(\vu r) \dd \Omega_r \\
&\qquad \qquad = \sqrt{\frac{(2\ell_1 + 1) (2\ell_2 + 1) (2\ell_3 + 1)}{4\pi}} \mqty( \ell_1 & \ell_2 & \ell_3 \\ 0 & 0 & 0) \mqty( \ell_1 & \ell_2 & \ell_3 \\ m_1 & m_2 & m_3), \nonumber 
\end{align}
where the $2 \times 3$ matrices are Wigner 3-$j$ symbols\footnote{https://dlmf.nist.gov/34.3} \cite{Varshalovich}, representing the addition of angular momenta. Each $z$-component is bounded by its corresponding total angular momentum. Non-zero entries in the bottom row of a 3-$j$ symbol describe the $z$-component of the angular momentum. The Gaunt integral vanishes unless the angular momenta satisfy the triangle inequality $|\ell_1 - \ell_2| \leq \ell_3 \leq \ell_1 + \ell_2 $ and the $z$-component of the sum of the angular momenta vanishes, $m_1 + m_2 + m_3 = 0$. A 3-$j$ with an identically zero bottom row requires that $\ell_1 + \ell_2 + \ell_3$ be even.

In the expansion of the four-argument Dirac delta function, we will  have the coefficient
\begin{multline}
\mathcal{K}_{\ell_1 \ell_2 \ell_3 \ell_4 L}^{m_1 m_2 m_3 m_4 M} \equiv (-1)^M(2L+1)\sqrt{\frac{(2\ell_1 + 1) (2\ell_2 + 1) (2\ell_3 + 1) (2\ell_4 + 1)}{4\pi}} \\ \times
\mqty( \ell_3 & \ell_4 & L \\ m_3 & m_4 & M) \mqty( \ell_3 & \ell_4 & L \\ 0 & 0 & 0) \mqty( \ell_1 & \ell_2 & L \\ 0 & 0 & 0) \mqty( \ell_1 & \ell_2 & L \\ m_1 & m_2 & -M). 
\end{multline}
Additionally, we will have the three radial integrals:
\begin{equation}
\mathcal{R}_{\ell_1 \ell_2}(k, p) \equiv \int \dd x \ x^2 j_{\ell_1}(kx) j_{\ell_2}(px) ,
\label{eq:R_2}
\end{equation}

\begin{equation}
\mathcal{R}_{\ell_1 \ell_2 \ell_3}(k, p, q) \equiv \int \dd x \ x^2 j_{\ell_1}(kx) j_{\ell_2}(px) j_{\ell_3}(qx) ,
\label{eq:R_3}
\end{equation}
and 
\begin{equation}
\mathcal{R}_{\ell_1 \ell_2 \ell_3 \ell_4}(k, p, q, s) \equiv \int \dd x \ x^2 j_{\ell_1}(kx) j_{\ell_2}(px) j_{\ell_3}(qx) j_{\ell_4}(sx). 
\label{eq:R_4}
\end{equation}
We note that Eq. (\ref{eq:R_3}) can be obtained by taking limit as $s \to 0$ of Eq. (\ref{eq:R_4}), and (\ref{eq:R_2}) can be obtained by taking the limit as $q \to 0$ of (\ref{eq:R_3}).
\subsection{Two-Argument Dirac Delta Expansion}
The Dirac delta function in Fourier space can be written as an FT of unity:
\begin{equation}\label{2-Dird}
(2\pi)^3\delta_{\rm D}^{[3]} (\vb q + \vb p) = \int \dd [3] \vb x  \ e^{i\vb q \vdot \vb x} e^{i\vb p \vdot \vb x}.  
\end{equation}
We proceed by using the plane wave expansion:
\begin{equation}
e^{i \vb q \vdot \vb x} = 4\pi \sum_{\ell = 0}^\infty \sum_{m = -\ell}^\ell i^\ell j_\ell(qx) Y_\ell^m(\vu x) Y_\ell^{m *}(\vu q).   \label{eq:plane_exp}
\end{equation}
We note that we can exchange the complex conjugate between either spherical harmonic in (\ref{eq:plane_exp}) due to commutativity of the dot product.
Spherical harmonics have several properties we will use, which we list here:
\begin{equation}
Y_{\ell}^{m *}(\vu q) = (-1)^mY_{\ell}^{-m}(\vu q), \hspace{5mm} Y_{\ell}^{m}(- \vu q) = (-1)^{\ell}Y_l^{m}(\vu q),
\label{eq: Y_props}
\end{equation}

\begin{equation}
Y_{\ell}^{m}(\vu q) Y_{\ell'}^{m'}(\vu q)
= \sum_{L=0}^\infty \sum_{M = -L}^L (-1)^M \mathcal{G}_{\ell \ell' L}^{m m' -M} Y_L^{M}(\vu q),
\label{eq:Y_contr}
\end{equation}
where $\mathcal{G}$ is defined in Eq. (\ref{eq:Gaunt}). We will call the two properties in Eq. (\ref{eq: Y_props}) conjugation and parity, respectively, and the property (\ref{eq:Y_contr}) linearization. From Eq. (\ref{eq:Y_contr}) one may also derive the analogous expression for linearization of products of associated Legendre polynomials, $P_L^M$:
\begin{align}
&P_{\ell}^m(\cos\theta) P_{\ell'}^{m'}(\cos\theta) \label{P:contr} \\
&= \sum_{L=0}^{\infty} \sum_{M=-L}^L (-1)^M \sqrt{\frac{4\pi(2L+1)}{(2\ell+1)(2\ell'+1)}\frac{(\ell+m)!(\ell'+m')!(L - M)!}{(\ell-m)!(\ell'-m')!(L+M)!}}\;
\mathcal{G}_{\ell \ell' L}^{m m' -M} P_{L}^{M}(\cos\theta).  \nonumber
\end{align}

Applying the  expansion (\ref{eq:plane_exp}) to Eq. (\ref{2-Dird}) yields
\begin{multline}
(2\pi)^3\delta_{\rm D}^{[3]} (\vb q + \vb p) =  (4\pi)^2 \sum_{\ell_1 , \ell_2} \sum_{m_1 , m_2} i^{\ell_1 + \ell_2} Y_{\ell_1}^{m_1 *}(\vu q)Y_{\ell_2}^{m_2}(\vu p) \int \dd \Omega_x Y_{\ell_1}^{m_1}(\vu x) Y_{\ell_2}^{m_2 *}(\vu x) \\ 
\times \int \dd x \ x^2 j_{\ell_1}(qx) j_{\ell_2}(px).    
\end{multline}
Applying the orthogonality of the spherical harmonics,
\begin{equation}
\int \dd \Omega_x Y_{\ell_1}^{m_1}(\vu x) Y_{\ell_2}^{m_2 *}(\vu x) = \kron_{\ell_1 \ell_2} \kron_{m_1 m_2},    
\end{equation}
we have
\begin{equation}
\delta_{\rm D}^{[3]} (\vb q + \vb p) = \frac{2}{\pi} \sum_{\ell = 0}^\infty (-1)^\ell \mathcal{R}_{\ell \ell}(q, p) \sum_{m = - \ell}^\ell Y_{\ell}^{m *}(\vu q)Y_{\ell}^{m}(\vu p).  
\label{eq:2D_Dird}
\end{equation}

\subsection{Three-Argument Dirac Delta Expansion}

We begin by writing a Dirac delta function in Fourier space with three arguments as a forward FT:
\begin{equation}
(2\pi)^3\delta_{\rm D}^{[3]} (\vb k + \vb q + \vb p) = \int \dd [3] \vb x  \ e^{i\vb k \vdot \vb x} e^{i\vb q \vdot \vb x} e^{i\vb p \vdot \vb x};     
\end{equation}
we then proceed as before. Making use of the plane wave expansion (\ref{eq:plane_exp}) we have 
\begin{multline}
(2\pi)^3\delta_{\rm D}^{[3]} (\vb k + \vb q + \vb p) = (4\pi)^3 \sum_{\ell_1 \ell_2 \ell_3} \sum_{m_1 m_2 m_3} i^{\ell_1 + \ell_2 + \ell_3} Y_{\ell_1}^{m_1 *}(\vu k) Y_{\ell_2}^{m_2 *}(\vu q) Y_{\ell_3}^{m_3 *}(\vu p) \\ \times \int \dd \Omega_x Y_{\ell_1}^{m_1}(\vu x) Y_{\ell_2}^{m_2}(\vu x) Y_{\ell_3}^{m_3}(\vu x) \int \dd x \ x^2 j_{\ell_1}(kx) j_{\ell_2}(qx) j_{\ell_3}(px). \nonumber 
\end{multline}
We may rewrite this using Eqs. (\ref{eq:Gaunt}) and (\ref{eq:R_3}) to obtain
\begin{equation}
\label{eq:3D_Dird}
\delta_{\rm D}^{[3]} (\vb k + \vb q + \vb p) = 8 \sum_{\ell_1 \ell_2 \ell_3} \sum_{m_1 m_2 m_3} i^{\ell_1 + \ell_2 + \ell_3} \mathcal{G}_{\ell_1 \ell_2 \ell_3}^{m_1 m_2 m_3} \mathcal{R}_{\ell_1 \ell_2 \ell_3}(k, q, p) Y_{\ell_1}^{m_1 *}(\vu k) Y_{\ell_2}^{m_2 *}(\vu q) Y_{\ell_3}^{m_3 *}(\vu p). 
\end{equation}

\subsection{Four-Argument Dirac Delta Expansion}
We proceed as in the previous two subsections. Writing the four-argument Dirac delta function in Fourier space as a forward FT, we have
\begin{equation}
(2\pi)^3\delta_{\rm D}^{[3]} (\vb k + \vb q + \vb p + \vb s) = \int \dd [3] \vb x  \ e^{i\vb k \vdot \vb x} e^{i\vb q \vdot \vb x} e^{i\vb p \vdot \vb x}  e^{i\vb s \vdot \vb x}   
\end{equation}

\begin{multline}
 = (4\pi)^4 \sum_{\ell_1 \ell_2 \ell_3 \ell_4} \; \sum_{m_1 m_2 m_3 m_4} i^{\ell_1 + \ell_2 + \ell_3 + \ell_4} Y_{\ell_1}^{m_1 *}(\vu k) Y_{\ell_2}^{m_2 *}(\vu q) Y_{\ell_3}^{m_3 *}(\vu p) Y_{\ell_4}^{m_4 *}(\vu s) 
\\ \times  \int \dd \Omega_x \; Y_{\ell_1}^{m_1}(\vu x) Y_{\ell_2}^{m_2}(\vu x) Y_{\ell_3}^{m_3}(\vu x) Y_{\ell_4}^{m_4}(\vu x) \int \dd x \ x^2 j_{\ell_1}(kx) j_{\ell_2}(qx) j_{\ell_3}(px) j_{\ell_4}(sx) \nonumber      
\end{multline}

\begin{multline}
= (4\pi)^4 \sum_{\ell_1 \ell_2 \ell_3 \ell_4} \; \sum_{m_1 m_2 m_3 m_4} \; \sum_{L, M} i^{\ell_1 + \ell_2 + \ell_3 + \ell_4} \mathcal{K}_{\ell_1 \ell_2 \ell_3 \ell_4 L}^{m_1 m_2 m_3 m_4 M} \mathcal{R}_{\ell_1 \ell_2 \ell_3 \ell_4}(k, q, p, s) \nonumber\\ 
\times Y_{\ell_1}^{m_1 *}(\vu k) Y_{\ell_2}^{m_2 *}(\vu q) Y_{\ell_3}^{m_3 *}(\vu p) Y_{\ell_4}^{m_4 *}(\vu s), \nonumber
\label{eq:4D_Dird}
\end{multline}
where we used the linearization rule (\ref{eq:Y_contr}) to combine two of the spherical harmonics into a sum of single spherical harmonics. We find
\begin{align}
&\delta_{\rm D}^{[3]} (\vb k + \vb q + \vb p + \vb s) =  32\pi \sum_{\ell_1 \ell_2 \ell_3 \ell_4} \; \sum_{m_1 m_2 m_3 m_4} \; \sum_{L M} i^{\ell_1 + \ell_2 + \ell_3 + \ell_4} \mathcal{K}_{\ell_1 \ell_2 \ell_3 \ell_4 L}^{m_1 m_2 m_3 m_4 M} \mathcal{R}_{\ell_1 \ell_2 \ell_3 \ell_4}(k, q, p, s) \nonumber \\ 
&\hspace{78mm} \times Y_{\ell_1}^{m_1 *}(\vu k) Y_{\ell_2}^{m_2 *}(\vu q) Y_{\ell_3}^{m_3 *}(\vu p) Y_{\ell_4}^{m_4 *}(\vu s). 
\end{align}
These Dirac delta function expansions are equivalent to those given in \cite{slepian_2024}; we also note this expansion for three arguments was used in \cite{3pt}, with the original idea going back to \cite{Mehrem_1991}.

\section{Convolutions}
\label{AppendixB}
The convolution of two functions $f(\vb q)$ and $g(\vb q)$, evaluated at a 3-vector $\vb k$, is defined as

\begin{align}\label{convolution}
[f \star g](\vb k) = \int \dd [3] \vb q \  f(\vb q) g(\vb k - \vb q).    
\end{align}
The Convolution Theorem states that 
\begin{align}
&[f \star g](\vb k) = \mathcal{F} \left \{\mathcal{F}^{-1}\{f(\vb q)\} \mathcal{F}^{-1}\{g(\vb q)\} \right \}(\vb k).
\label{eq:conv_thm}
\end{align}
Here $\mathcal{F}$ represents an FT and $\mathcal{F}^{-1}$ an inverse FT.
Eq. (\ref{eq:conv_thm}) allows us to convert between convolutions and FTs. We make repeated use of this theorem in \S\ref{Section6}. Here we list all of the convolutional integrals that appear in the computation of the inverse decay trispectrum. Specifically, the angular parts of these integrals come from the expansion of the polarization vector dot product in Eq. (\ref{dot_prod}). Analytical simplification of these angular integrals is essential in order to numerically compute the trispectrum for a large portion of the parameter space.
Here we rewrite each term as a sequence of FTs using the Convolution Theorem.
\begin{align}
&\text{I}_{\Lambda \lambda \lambda'}^{M(1)}(\vb{k}, x, x', \tau) \equiv \label{eqn:I1} \frac{1}{2}  \int \dd[3] \vb{q} \ Y_{\Lambda}^M(\vu q) \cos\alpha \cos\theta \cos\phi \cos\beta \, q A_-^{'}(\tau, |\vb{k - q}|)A_-(\tau, q) j_{\lambda}(q x) j_{\lambda'}(q x') \nonumber \\   
&  \hspace{27mm} = \frac{1}{2} \mathcal{F} \Bigg \{ \mathcal{F}^{-1} \bigg \{\cos\theta \cos\phi Y_{\Lambda}^{M} (\vu q) q A_-(\tau, q) j_{\lambda}(q x) j_{\lambda'}(q x') \bigg \}(\vb r) \\
& \hspace{33 mm}\times \mathcal{F}^{-1} \bigg \{ \cos\theta \cos\phi A_-^{'}(\tau, q) \bigg \}(\vb r) \Bigg \}(\vb k), \nonumber \\ \nonumber\\
&\text{I}_{\Lambda \lambda \lambda'}^{M(2)}(\vb{k}, x, x', \tau) \equiv \label{eq:I2} \frac{1}{2} \int \dd[3] \vb{q} \ Y_{\Lambda}^M(\vu q) \cos\alpha \cos\theta \sin\phi \sin\beta \, q A_-^{'}(\tau, |\vb{k - q}|)A_-(\tau, q) j_{\lambda}(q x) j_{\lambda'}(q x') \nonumber \\
&\hspace{27mm}=\frac{1}{2}\mathcal{F} \Bigg \{ \mathcal{F}^{-1} \bigg \{ \cos\theta \sin\phi \, Y_{\Lambda}^{M} (\vu q) q A_-(\tau, q) j_{\lambda}(q x) j_{\lambda'}(q x') \bigg \}(\vb r) \\
&\hspace{33mm}\times \mathcal{F}^{-1} \bigg \{ \cos\theta \sin\phi \, A_-^{'}(\tau, q) \bigg \}(\vb r) \Bigg \}(\vb k), \nonumber \\ \nonumber \\
&\text{I}_{\Lambda \lambda \lambda'}^{M(3)}(\vb{k}, x, x', \tau) \equiv \label{eq:I3} -\frac{1}{2} \int \dd[3] \vb{q} \ Y_{\Lambda}^M(\vu q) \cos\phi \cos\beta \, q A_-^{'}(\tau, |\vb{k - q}|)A_-(\tau, q) j_{\lambda}(q x) j_{\lambda'}(q x') \nonumber \\ 
&\hspace{27mm}=-\frac{1}{2}\mathcal{F} \Bigg \{ \mathcal{F}^{-1} \bigg \{ \cos\phi \, Y_{\Lambda}^{M} (\vu q) q A_-(\tau, q) j_{\lambda}(q x) j_{\lambda'}(q x') \bigg \}(\vb r) \\
&\hspace{33mm}\times \mathcal{F}^{-1} \bigg \{\cos\phi \, A_-^{'}(\tau, q) \bigg \}(\vb r) \Bigg \}(\vb k), \nonumber \\ \nonumber \\
&\text{I}_{\Lambda \lambda \lambda'}^{M(4)}(\vb{k}, x, x', \tau) \equiv \label{eq:I4} -\frac{1}{2} \int \dd[3] \vb{q} \ Y_{\Lambda}^M(\vu q) \sin\phi \sin\beta \, q A_-^{'}(\tau, |\vb{k - q}|)A_-(\tau, q) j_{\lambda}(q x) j_{\lambda'}(q x') \nonumber \\ 
&\hspace{27mm}=-\frac{1}{2}\mathcal{F} \Bigg \{ \mathcal{F}^{-1} \bigg \{ \sin\phi \, Y_{\Lambda}^{M} (\vu q) q A_-(\tau, q) j_{\lambda}(q x) j_{\lambda'}(q x') \bigg \}(\vb r) \\
&\hspace{33mm}\times \mathcal{F}^{-1} \bigg \{\sin\phi \, A_-^{'}(\tau, q) \bigg \}(\vb r) \Bigg \}(\vb k), \nonumber \\ \nonumber \\
&\text{I}_{\Lambda \lambda \lambda'}^{M(5)}(\vb{k}, x, x', \tau) \equiv \label{eq:I5} \frac{1}{2} \int \dd[3] \vb{q} \ Y_{\Lambda}^M(\vu q) \sin\alpha \sin\theta \, q A_-^{'}(\tau, |\vb{k - q}|)A_-(\tau, q) j_{\lambda}(q x) j_{\lambda'}(q x') \nonumber \\ 
&\hspace{27mm}= \frac{1}{2} \mathcal{F} \Bigg \{ \mathcal{F}^{-1} \bigg \{ \sin\theta \, Y_{\Lambda}^{M} (\vu q) q A_-(\tau, q) j_{\lambda}(q x) j_{\lambda'}(q x') \bigg \}(\vb r) \\
&\hspace{33mm}\times \mathcal{F}^{-1} \bigg \{\sin\theta \, A_-^{'}(\tau, q) \bigg \}(\vb r) \Bigg \}(\vb k), \nonumber \\ \nonumber \\
&\text{I}_{\Lambda \lambda \lambda'}^{M (6)} (\vb{k}, x, x', \tau) \equiv \label{eqn:I6} -\frac{i}{2} \int \dd[3] \vb{q} \; \bigg(\vu q \cross \frac{\vb{k-q}}{|\vb{k-q}|} \vdot \vu z \bigg ) \frac{q\cos\alpha}{\sin\alpha \sin\theta} Y_\Lambda^M(\vu q) A_-^{'}(\tau, |\vb{k - q}|)A_-(\tau, q) \nonumber \\ 
&\hspace{118 mm}\times j_{\lambda}(q x) j_{\lambda'}(q x') \nonumber \\
& \hspace{27mm}= \frac{\sqrt{2}(4\pi)^{3/2}}{6} \sum_{\nu_1 \nu_2 \nu_3}  C_{\nu_1 \nu_2 \nu_3}^{1 1 1} Y_1^{\nu_1}(\vu z) \sum_{L', M'}^\infty (-1)^{M'} \mathcal{G}_{\Lambda 1 L'}^{M \nu_2 -M'} \\ 
&\times \mathcal{F} \Bigg \{\mathcal{F}^{-1} \bigg \{\frac{1}{\sin\theta} Y_{L'}^{M'} (\vu q) q A_-(\tau, q) j_{\lambda}(q x) j_{\lambda'}(q x') \bigg \}(\vb r) \, \mathcal{F}^{-1} \left \{ \frac{\cos\theta}{\sin\theta} Y_1^{m_3}(\vu q)A_-^{'}(\tau, q) \right \}(\vb r) \Bigg \}(\vb k), \nonumber \\ \nonumber \\
&\text{I}_{\Lambda \lambda \lambda'}^{M (7)} (\vb{k}, x, x', \tau) \equiv \label{eqn:I7} \frac{i}{2} \int \dd[3] \vb{q} \; \bigg(\vu q \cross \frac{\vb{k-q}}{|\vb{k-q}|} \vdot \vu z \bigg ) \frac{q\cos\theta}{\sin\alpha \sin\theta} Y_\Lambda^M(\vu q) A_-^{'}(\tau, |\vb{k - q}|)A_-(\tau, q) \nonumber \\ 
&\hspace{118 mm}\times j_{\lambda}(q x) j_{\lambda'}(q x') \nonumber \\
& \hspace{27mm}= -\frac{\sqrt{2}(4\pi)^{3/2}}{6} \sum_{\nu_1 \nu_2 \nu_3}  C_{\nu_1 \nu_2 \nu_3}^{1 1 1} Y_1^{\nu_1}(\vu z) \sum_{L', M'}^\infty (-1)^{M'} \mathcal{G}_{\Lambda 1 L'}^{M \nu_2 -M'} \\ 
&\times \mathcal{F} \Bigg \{\mathcal{F}^{-1} \bigg \{\frac{\cos\theta}{\sin\theta} Y_{L'}^{M'} (\vu q) q A_-(\tau, q) j_{\lambda}(q x) j_{\lambda'}(q x') \bigg \}(\vb r) \, \mathcal{F}^{-1} \left \{ \frac{1}{\sin\theta} Y_1^{m_3}(\vu q)A_-^{'}(\tau, q) \right \}(\vb r) \Bigg \}(\vb k). \nonumber
\end{align}
\section{Angular Integrals}
\label{AppendixC}
\subsection{General Result}
In \S\ref{FTs}, we introduced the coefficients (\ref{eq:ang_coeffs}), corresponding to the integral of a spherical harmonic against another function:
\begin{align}\label{C_def}
&C_{(i) \alpha}^{(j) \beta} \equiv \int \dd \Omega \ Y_\alpha^\beta(\vu q) \vb f_i^{(j)}(\vu q).
\end{align}
where $\Omega$ is the solid angle explored by the $\vu q$ vector, and $\vb f_i^{(j)}(\vu q)$ is defined in Eq. (\ref{f}).

Here we show explicitly how to compute these integrals.
Each can be reduced to the form 
\begin{align}
\label{eq:LP_Int}
&\int_{-1}^{1} \dd x \ (1-x^2)^{\lambda - 1}P_{\nu}^{\mu}(x) \\
&= \frac{\pi 2^{\mu} \Gamma(\lambda + \mu/2)\Gamma(\lambda - \mu/2)}{\Gamma(\lambda + \nu/2 + 1/2)\Gamma(\lambda - \nu/2)\Gamma(-\mu/2 + \nu/2 + 1)\Gamma(-\mu/2 -\nu/2 + 1/2)}; \nonumber 
\end{align}
this result is given in \cite{Gradshteyn:1702455}. It is valid when $2\,\text{Re}\;\lambda > |\text{Re}\;\mu|$, which will always be the case for the integrals we require in this paper.

To cast the angular integrals in the form of Eq. (\ref{eq:LP_Int}), in some cases we use the following linearization rule for Legendre polynomials $P_L$, first given in \cite{Adams}, and written in a more succinct form in the Appendix of \citep{Slepian_2015}:
\begin{align}
\label{eq:LP_Contr}
P_k(x)P_l(x) = \sum_{m = |k-l|}^{k+l}\mqty(k & l & m \\ 0 & 0 & 0)^2 (2m+1) P_m(x).
\end{align}

\subsection{Specific Results}
Here we present explicit results for each coefficient $C$ as defined by Eq. (\ref{C_def}). In the next subsection we will demonstrate how these results were calculated. 
\begin{align}
&C_{(1)\alpha}^{(1)\beta} = C_{(2)\alpha}^{(1)\beta} = \int \dd \Omega_q \ Y_\alpha^\beta(\vu q) \cos\theta \cos\phi \label{C1} \\
&=  -\pi \sum_{\gamma = |\alpha - 1|}^{\alpha + 1} 
\Bigg(\sqrt{\frac{(2\gamma + 1)}{3} \frac{(\gamma+1)!}{(\gamma-1)!}} \mathcal{G}_{\alpha 1 \gamma}^{-1 0 1} \kron_{\beta -1} \frac{\pi \Gamma(1/2)\Gamma(3/2)}{2\Gamma(3/2+\gamma/2)\Gamma(1-\gamma/2)\Gamma(3/2 +\gamma/2 )\Gamma(1-\gamma/2)}  \nonumber \\
&+ \sqrt{\frac{(2\gamma + 1)}{3} \frac{(\gamma-1)!}{(\gamma+1)!}} \mathcal{G}_{\alpha 1 \gamma}^{1 0 -1} \kron_{\beta 1} \frac{2\pi \Gamma(3/2)\Gamma(1/2)}{\Gamma(3/2+\gamma/2)\Gamma(1-\gamma/2)\Gamma(1/2 +\gamma/2)\Gamma(-\gamma/2)}\Bigg), \nonumber \\ \nonumber \\
&C_{(1)\alpha}^{(2)\beta} = C_{(2)\alpha}^{(2)\beta} = \int \dd \Omega_q \ Y_\alpha^\beta(\vu q) \cos\theta \sin\phi \label{C2}\\
&= -i\pi \sum_{\gamma = |\alpha - 1|}^{\alpha + 1} 
\Bigg(\sqrt{\frac{(2\gamma + 1)}{3} \frac{(\gamma+1)!}{(\gamma-1)!}} \mathcal{G}_{\alpha 1 \gamma}^{-1 0 1} \kron_{\beta -1} \frac{-\pi \Gamma(1/2)\Gamma(3/2)}{2\Gamma(3/2+\gamma/2)\Gamma(1-\gamma/2)\Gamma(3/2 +\gamma/2 )\Gamma(1-\gamma/2)} \nonumber \\
&+ \sqrt{\frac{(2\gamma + 1)}{3} \frac{(\gamma-1)!}{(\gamma+1)!}} \mathcal{G}_{\alpha 1 \gamma}^{1 0 -1} \kron_{\beta 1} \frac{2\pi \Gamma(3/2)\Gamma(1/2)}{\Gamma(3/2+\gamma/2)\Gamma(1-\gamma/2)\Gamma(1/2 +\gamma/2)\Gamma(-\gamma/2)} \Bigg ), \nonumber \\ \nonumber\\
& C_{(1)\alpha}^{(3)\beta} = C_{(2)\alpha}^{(3)\beta} = \int \dd \Omega_q \ Y_\alpha^\beta(\vu q) \cos\phi \\
&= \pi \left(\sqrt{\frac{(2\alpha + 1)}{4\pi}\frac{(\alpha+1)!}{(\alpha-1)!}} \kron_{\beta -1} \frac{\pi \Gamma(1/2)\Gamma(3/2)}{2\Gamma(3/2+\alpha/2)\Gamma(1-\alpha/2)\Gamma(3/2 +\alpha/2 )\Gamma(1-\alpha/2)} \right. \nonumber \\
&\left.+ \sqrt{\frac{(2\alpha + 1)}{4\pi}\frac{(\alpha-1)!}{(\alpha+1)!}} \kron_{\beta 1} \frac{2\pi \Gamma(3/2)\Gamma(1/2)}{\Gamma(3/2+\alpha/2)\Gamma(1-\alpha/2)\Gamma(1/2 +\alpha/2)\Gamma(-\alpha/2)}\right), \nonumber \\ \nonumber\\
& C_{(1)\alpha}^{(4)\beta} = C_{(2)\alpha}^{(4)\beta} = \int \dd \Omega_q \ Y_\alpha^\beta(\vu q) \sin\phi \\
&= -i\pi \left(\sqrt{\frac{(2\alpha + 1)}{4\pi}\frac{(\alpha+1)!}{(\alpha-1)!}} \kron_{\beta -1} \frac{\pi \Gamma(1/2)\Gamma(3/2)}{2\Gamma(3/2+\alpha/2)\Gamma(1-\alpha/2)\Gamma(3/2 +\alpha/2 )\Gamma(1-\alpha/2)} \right. \nonumber \\
&\left.- \sqrt{\frac{(2\alpha + 1)}{4\pi}\frac{(\alpha-1)!}{(\alpha+1)!}} \kron_{\beta 1} \frac{2\pi \Gamma(3/2)\Gamma(1/2)}{\Gamma(3/2+\alpha/2)\Gamma(1-\alpha/2)\Gamma(1/2 +\alpha/2)\Gamma(-\alpha/2)}\right), \nonumber \\ \nonumber \\
&C_{(1)\alpha}^{(5) \beta} = C_{(2)\alpha}^{(5) \beta} = \int \dd \Omega_q \ Y_\alpha^\beta(\vu q) \sin\theta \\
&= 2\pi \kron_{\beta 0} \left(\sqrt{\frac{(2\alpha + 1)}{4\pi}}  \frac{\pi \Gamma(3/2)\Gamma(3/2)}{\Gamma(2+\alpha/2)\Gamma(1+\alpha/2)\Gamma(3/2 -\alpha/2 )\Gamma(1/2-\alpha/2)} \right), \nonumber \\ \nonumber\\
& C_{(1)\alpha}^{(6) \beta} = C_{(2)\alpha}^{(7) \beta} = \int \dd \Omega_q \ Y_\alpha^\beta(\vu q) \; \frac{1}{\sin\theta} \label{C6} \\
&= 2\pi^2 \kron_{\beta 0} \sqrt{\frac{2\alpha + 1}{4\pi}} \frac{\Gamma(3/2)\Gamma(3/2)}{\Gamma(2 + \alpha/2)\Gamma(3/2-\alpha/2)\Gamma(\alpha/2 + 1)\Gamma(1/2-\alpha/2)},\nonumber \\ \nonumber \\
& C_{(1)\alpha}^{(7) \beta} = C_{(2)\alpha}^{(6) \beta} = \int \dd \Omega_q \ Y_\alpha^\beta(\vu q) \; \frac{\cos\theta}{\sin\theta} \label{C7} \\
&= 2\pi \sqrt{\frac{2\alpha + 1}{4\pi}} \kron_{\beta 0} \sum_{\gamma = |\alpha - 1|}^{\alpha + 1} \mqty(\alpha & 1 & \gamma \\ 0 & 0 & 0)^2 (2\gamma + 1) \frac{\pi \Gamma(1/2)\Gamma(1/2)}{\Gamma(1 + \gamma/2)\Gamma(1/2-\gamma/2)\Gamma(1+\gamma/2)\Gamma(1/2-\gamma/2)}.\nonumber
\end{align}

\subsection{Inverse Fourier Transform Angular Integrals}
We now demonstrate how to perform the angular integrals associated with the inverse FTs of \S \ref{Section6}. We will show explicitly how to get the results (\ref{C1}) and (\ref{C7}). We have chosen to explicitly calculate only these two results as (\ref{C1}) contains a trigonometric function in $\theta$ and $\phi$ in the numerator and (\ref{C7}) contains trigonometric function in the denominator. The other integrals can be computed in a very similar fashion as these two examples. We begin by computing integral (\ref{C1}):

\begin{align}
&C_{(1)\alpha}^{(1)\beta} = C_{(2)\alpha}^{(1)\beta} = \int \dd \Omega_q \ Y_\alpha^\beta(\vu q) \cos\theta \cos\phi.
\end{align}
Inserting the definition of the spherical harmonics and rewriting $\cos\phi$ in terms of complex exponentials, we have
\begin{align}
&C_{(1)\alpha}^{(1)\beta} = -\sqrt{\frac{(2\alpha+1)(\alpha - \beta)!}{4\pi(\alpha+\beta)!}} \int_{1}^{-1} \dd (\cos\theta) P_{\alpha}^{\beta}(\cos\theta) P_{1}^{0}(\cos\theta) \int_0^{2\pi} \dd \phi \ \frac{e^{i \beta \phi}(e^{i \phi} + e^{-i \phi})}{2} \\
&= \sqrt{\frac{(2\alpha+1)(\alpha - \beta)!}{4\pi(\alpha+\beta)!}} \sum_{\gamma = 0}^\infty \sum_{\delta = -\gamma}^{\gamma} (-1)^{\delta} \sqrt{\frac{4\pi(2\gamma + 1)}{(2\alpha + 1)(2+1)}\frac{(\alpha+\beta)!}{(\alpha-\beta)!}\frac{(\gamma-\delta)!}{(\gamma+\delta)!}} \mathcal{G}_{\alpha 1 \gamma}^{\beta 0 -\delta} \pi (\kron_{\beta1} + \kron_{\beta -1}) \nonumber \\
&\times \int_{-1}^{1} \dd x \; P_{\gamma}^{\delta} (x) \nonumber \\
&= \pi (\kron_{\beta1} + \kron_{\beta -1}) \sum_{\gamma = 0}^\infty \sum_{\delta = -\gamma}^{\gamma} (-1)^{\delta} \sqrt{\frac{(2\gamma + 1)}{3} \frac{(\gamma-\delta)!}{(\gamma+\delta)!}} \mathcal{G}_{\alpha 1 \gamma}^{\beta 0 -\delta} \nonumber \\
& \times \frac{\pi 2^{\delta} \Gamma(1+\delta/2)\Gamma(1-\delta/2)}{\Gamma(1+ \gamma/2 + 1/2)\Gamma(1-\gamma/2)\Gamma(-\delta/2 +\gamma/2 +1)\Gamma(-\delta/2 - \gamma/2 + 1/2)}\nonumber \\
& = -\pi \sum_{\gamma = |\alpha - 1|}^{\alpha + 1} 
\Bigg(\sqrt{\frac{(2\gamma + 1)}{3} \frac{(\gamma+1)!}{(\gamma-1)!}} \mathcal{G}_{\alpha 1 \gamma}^{-1 0 1} \kron_{\beta -1} \frac{\pi \Gamma(1/2)\Gamma(3/2)}{2\Gamma(3/2+\gamma/2)\Gamma(1-\gamma/2)\Gamma(3/2 +\gamma/2 )\Gamma(1-\gamma/2)}  \nonumber \\
&+ \sqrt{\frac{(2\gamma + 1)}{3} \frac{(\gamma-1)!}{(\gamma+1)!}} \mathcal{G}_{\alpha 1 \gamma}^{1 0 -1} \kron_{\beta 1} \frac{2\pi \Gamma(3/2)\Gamma(1/2)}{\Gamma(3/2+\gamma/2)\Gamma(1-\gamma/2)\Gamma(1/2 +\gamma/2)\Gamma(-\gamma/2)}\Bigg). \nonumber 
\end{align}
To obtain the second equality we linearized the two associated Legendre polynomials via Eq. (\ref{P:contr}), and to obtain the third equality, we used Eq. (\ref{eq:LP_Int}). 

We now turn to computing the integral (\ref{C7}). We have
\begin{align}
&C_{\alpha (1)}^{\beta (7)} = C_{\alpha (2)}^{\beta (6)} = \int \dd \Omega \ Y_\alpha^\beta(\vu q) \frac{\cos\theta}{\sin\theta} \\ 
&= -\sqrt{\frac{(2\alpha+1)(\alpha - \beta)!}{4\pi(\alpha+\beta)!}} \int_{1}^{-1} \dd (\cos\theta) P_{\alpha}^{\beta}(\cos\theta) \frac{P_{1}^{0}(\cos\theta)}{\sqrt{1-\cos^2\theta}} \int_0^{2\pi} \dd \varphi \ e^{i \beta \varphi} \nonumber \\ 
&= -2\pi \sqrt{\frac{2\alpha + 1}{4\pi}} \kron_{\beta 0} \int_{1}^{-1} \dd \mu \ P_{\alpha}(\mu) \frac{P_{1}(\mu)}{\sqrt{1-\mu^2}} \nonumber \\ &= 2\pi \sqrt{\frac{2\alpha + 1}{4\pi}} \kron_{\beta 0} \sum_{\gamma = |\alpha - 1|}^{\alpha + 1} \mqty(\alpha & 1 & \gamma \\ 0 & 0 & 0)^2 (2\gamma + 1)\int_{-1}^{1} \dd x \ \frac{P_{\gamma}(x)}{\sqrt{1-x^2}} \nonumber \\ 
&= 2\pi \sqrt{\frac{2\alpha + 1}{4\pi}} \kron_{\beta 0} \sum_{\gamma = |\alpha - 1|}^{\alpha + 1} \mqty(\alpha & 1 & \gamma \\ 0 & 0 & 0)^2 (2\gamma + 1) \frac{\pi \Gamma(1/2)\Gamma(1/2)}{\Gamma(1 + \gamma/2)\Gamma(1/2-\gamma/2)\Gamma(1+\gamma/2)\Gamma(1/2-\gamma/2)} ,\nonumber
\end{align}
where we have used the linearization rule for a product of Legendre polynomials given in Eq. (\ref{eq:LP_Contr}), and we have again used Eq. (\ref{eq:LP_Int}) to obtain the last line above.

\section{Validity of the Mode Function Approximation} \label{AppendixD}
Throughout this work, we use Eq. (\ref{A:approx_1}) to approximate the gauge field mode functions $A_-(\tau, k)$, which is valid in the region $1/(8\xi) \leq -k\tau \leq 2\xi$. In the method we here present, one must frequently integrate these mode functions with respect to momentum and conformal time, during which the quantity $k\tau$ falls out of the region of good approximation. In this section we check the accuracy of this approximation by computing the integrals of \S\ref{Section7} numerically. We must ensure that widening the bounds of integration of these integrals outside of the region of good approximation does not significantly alter their result. We begin with the integral given in Eq. (\ref{M_integral}),

\begin{align}\label{M_int}
&M_{a'}(\eta) \equiv \int_{z_{\rm min}}^{z_{\rm max}} \dd z \, z^2 \mathcal{A}(z) j_{a'}(z\eta), \nonumber 
\end{align}
where $\eta \equiv r/\tau' \leq 0$. We will have two versions of this function, one for the restricted integral $M^R_{a'}(\eta)$ with $z_{\rm min} = -2\xi$ and $z_{\rm max} = -1/(8\xi)$, and one for the full integration region of $z_{\rm min} = -2\xi$ and $z_{\rm max} = -1/(8\xi)$. We experimented with different bounds for the full integral, which we denote $M_{a'}(\eta)$, and found that the difference in results is negligible if the bounds are widened any further than $z_{\rm min} = -10$ and $z_{\rm max} = -0.01$.

These integrals suffer from extreme oscillations due to the increasing frequency of the Bessel functions for smaller values of $\eta$. We apply smoothing 

\begin{align}
&M^S_{a'}(\eta) \equiv e^{-(\eta/\eta_0)^2} M_{a'}(\eta)
\end{align}
and 
\begin{align}
&M^{R, S}_{a'}(\eta) \equiv e^{-(\eta/\eta_0)^2} M^R_{a'}(\eta).
\end{align}

\begin{figure}[H]
\vspace{-2mm}
    \centering
    \includegraphics[width=0.8\textwidth]{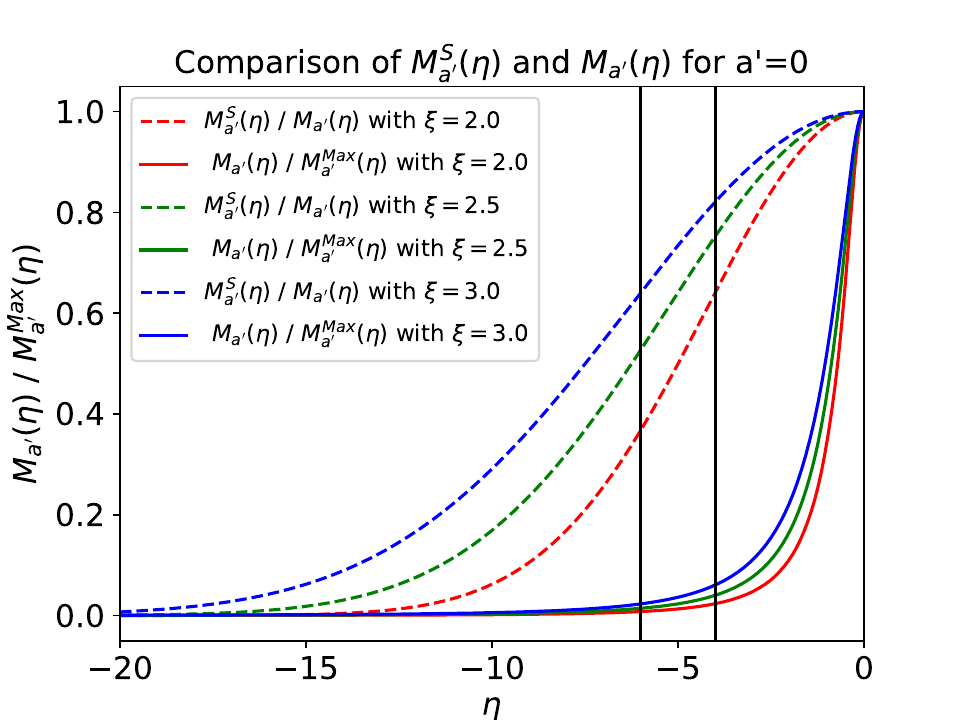}
    \vspace{-2mm}
    \caption{\footnotesize{This figure shows the ratio of the smoothed to unsmoothed function, $M^{S}_{a'}(\eta)$ / $M_{a'}(\eta)$, as well as the normalized integral $M_{a'}(\eta)$, for spherical Bessel order $a^\prime = 0$. Since we are integrating Eq. (\ref{M_int}) numerically, we must truncate the infinite integration region at some point. This truncation removes high-frequency information coming from the extreme oscillations of the spherical Bessel function at large negative values of $\eta$. This leads to a ringing behavior known as the Gibbs phenomenon. We remove the effects of the Gibbs phenomenon by applying a smoothing to the function $M_{a'}(\eta)$. The vertical black lines are at $\eta = -4$ and $\eta = -6$. These locations correspond respectively to $\sim$20\% and $\sim$40\% error from applying the smoothing for the $\xi=3$ case. We can see that the smoothing only begins to significantly alter the function for values of eta where $M_{a'}(\eta)$ is a small fraction of its maximum value. This gives us confidence that the smoothing is not significantly affecting our result. }} 
    \label{M_ratio_1}
\end{figure}

\begin{figure}[H]
\vspace{-2mm}
    \centering
    \includegraphics[width=0.8\textwidth]{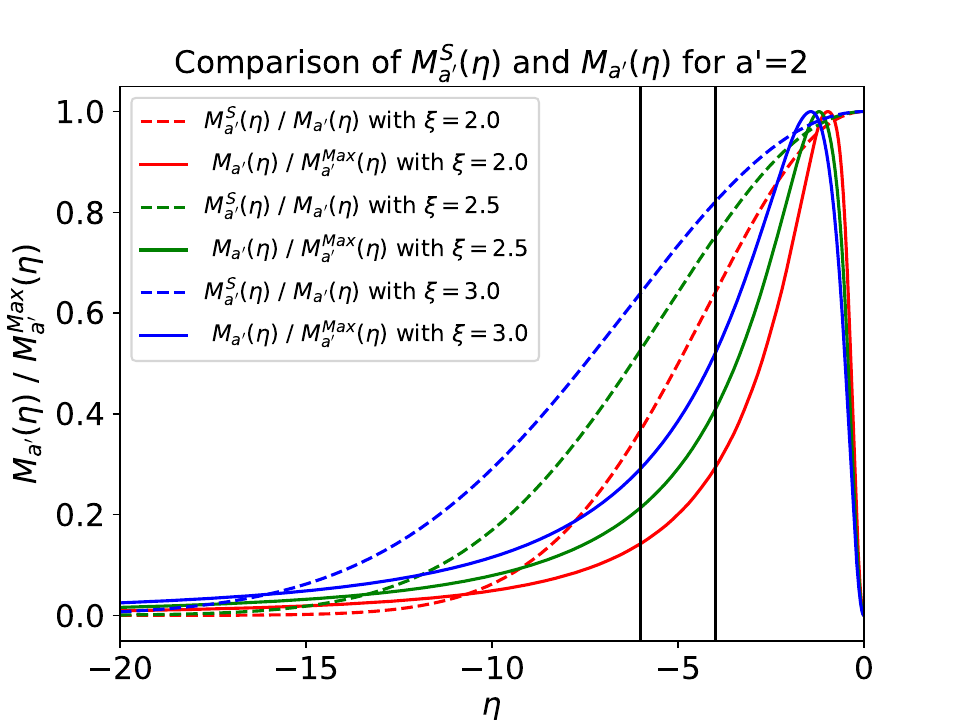}
    \vspace{-2mm}
    \caption{\footnotesize{Similarly to Figure \ref{M_ratio_1}, this figure shows the ratio of the smoothed to unsmoothed function $M^{S}_{a'}(\eta)$ / $M_{a'}(\eta)$ as well as the normalized integral $M_{a'}(\eta)$ for spherical Bessel order $a^\prime = 2$. The smoothing only begins to significantly alter the function for values of eta where $M_{a'}(\eta)$ is a small fraction of its maximum value. We can therefore conclude that the smoothing is not significantly affecting our result.}} 
    \label{M_ratio_3}
\end{figure}

We have experimented with different values of the smoothing cutoff scale $\eta_0$ and found $\eta_0 = 3\xi$ to be a good heuristic. In Figures \ref{M_ratio_1}\,-\ref{M_ratio_3}, we show that this choice of cutoff does not significantly alter either $M^R_{a'}(\eta)$ or $M_{a'}(\eta)$. 

The next integral in Eq. (\ref{FancyZ2}), also contains a gauge field mode function: 
\begin{align}
d_{a'}(u,v) = \int_{-\infty}^0 \dd w \; \mathcal{S}(w) \mathcal{A}(uw) M_{a'}(v/w). \nonumber
\end{align}

We again want to compute the restricted and full integrals. We use the following change of variables to facilitate efficient numerical integration: $h=uv$, $\eta'=1/u$, and $y=uw$. Thus we have 
\begin{align}\label{d_int}
d_{a'}(\eta',h) = \int_{-\infty}^0 \dd y \; \mathcal{S}(y\eta') \mathcal{A}(y) M^S_{a'}(h/y) \nonumber
\end{align}
and
\begin{align}
d_{a'}^R(\eta',h) = \int_{-2\xi}^{(-8\xi)^{-1}} \dd y \; \mathcal{S}(y\eta') \mathcal{A}(y) M^{R, S}_{a'}(h/y). \nonumber
\end{align}

\begin{figure}[H]
\vspace{-2mm}
    \centering
    \includegraphics[width=1\textwidth]{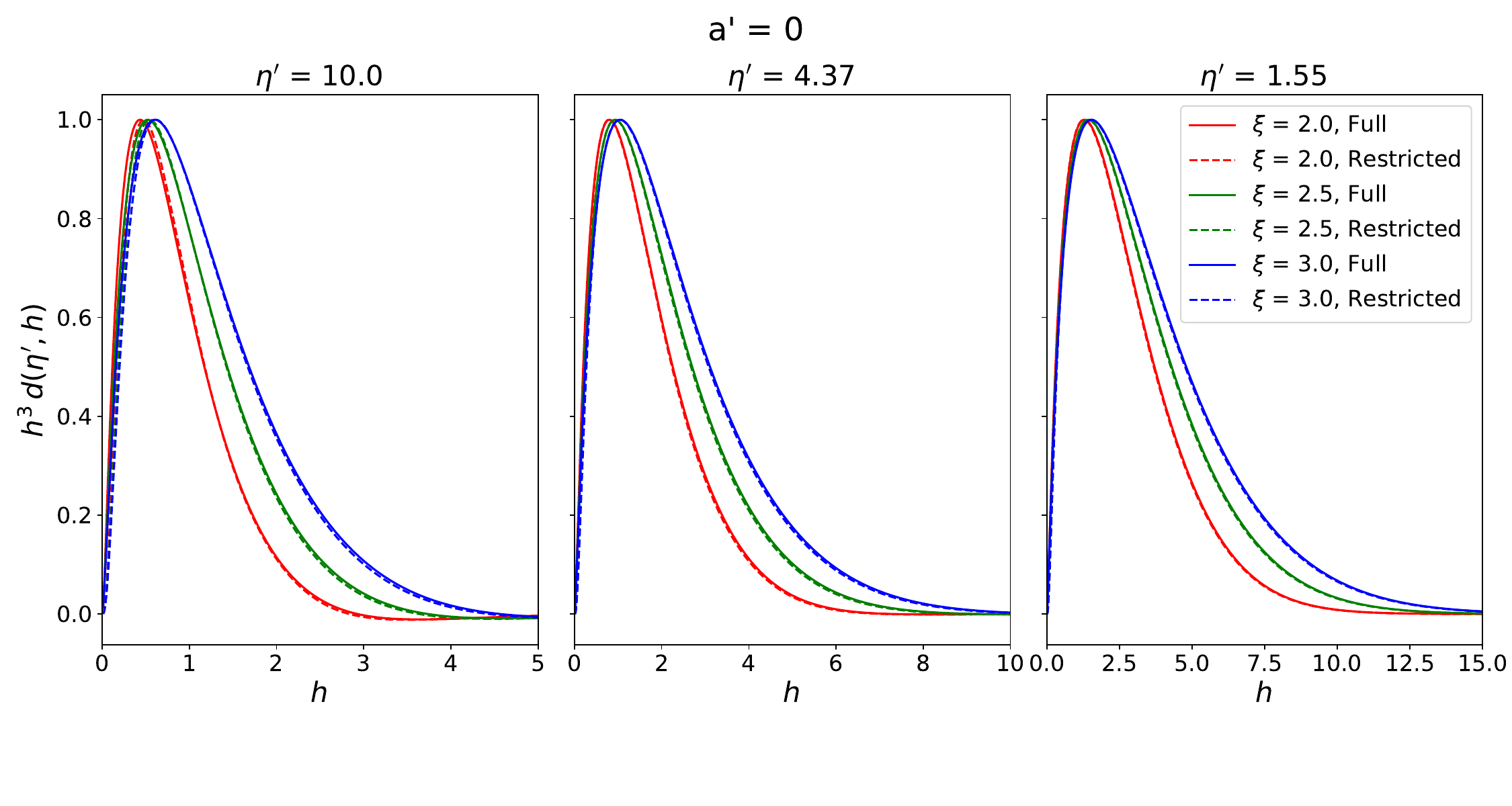}
    \vspace{-2mm}
    \caption{\footnotesize{In these panels we compare the full and restricted functions $d(\eta', h)$ from Eq. (\ref{d_int}) at three different values of $\eta'$. We have re-scaled the function by a factor of $h^3$ to improve visualization of it. We can see that the full and restricted forms of the integral yield results that are in good agreement. Thus we can have confidence that integrating our function over a region containing points outside of which Eq. (\ref{A:approx_1}) is a good approximation for the mode functions will not significantly change the result.}} 
    \label{d_panels}
\end{figure}

\begin{figure}[H]
\vspace{-2mm}
    \centering
    \includegraphics[width=0.8\textwidth]{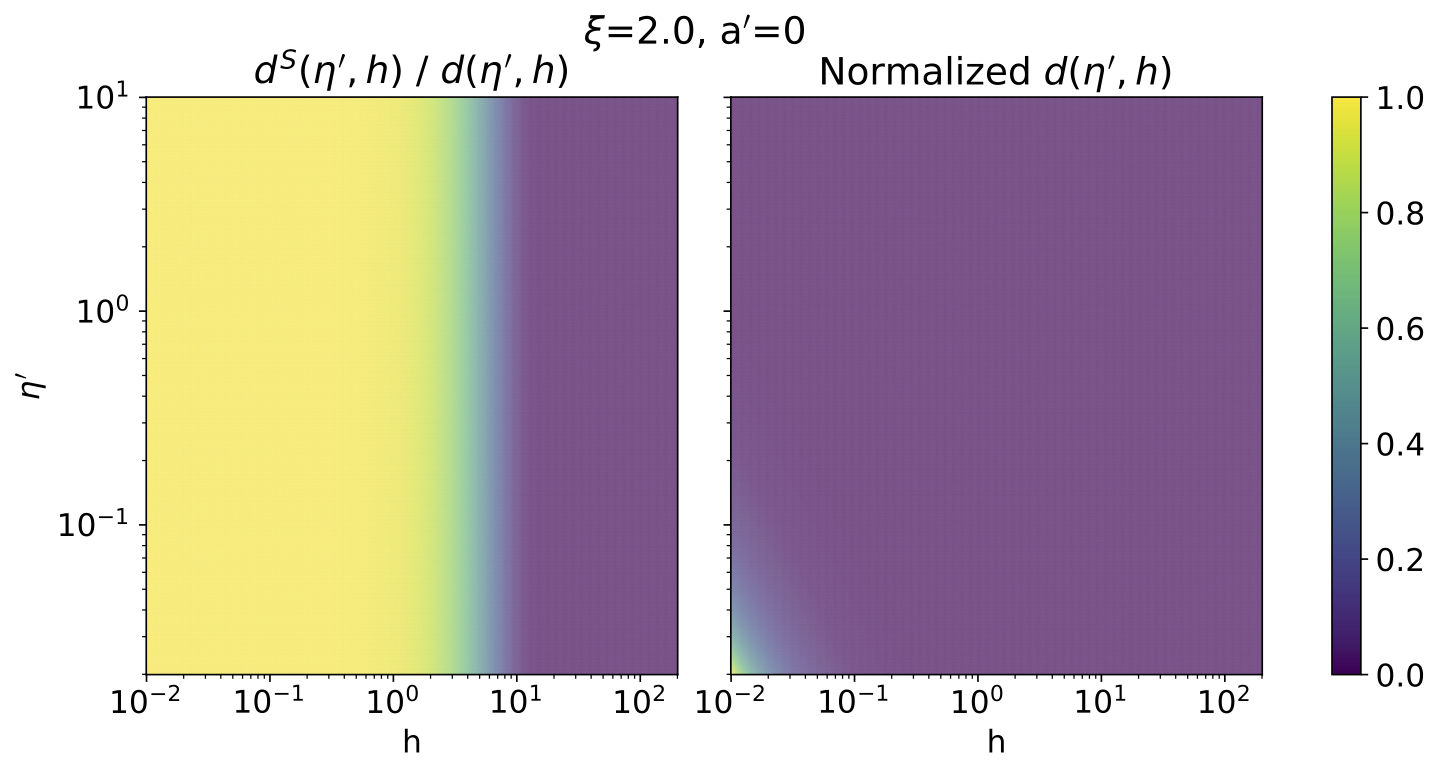}
    \vspace{-2mm}
    \caption{\footnotesize{This figure shows the ratio of the smoothed, to the unsmoothed function $d^S(\eta', h)$ / $d(\eta', h)$,  as well as a normalized plot of the function $d(\eta', h)$ for $\xi=2$ and spherical Bessel order $a'=0$. As was the case with the function $M_{a^\prime}(\eta)$, truncating the $d(\eta', h)$ integral in Eq. (\ref{d_int}) at a finite value, leads to a ringing effect known as Gibbs Phenomenon. To remove this effect, we again apply a smoothing to our result. The plot shows that the function $d(\eta', h)$ rapidly falls to zero outside of the points $h=0$, $\eta'=0$, and the smoothing only significantly begins to alter the function far from this point. Thus we can have confidence that our results is unchanged by applying the smoothing.}} 
    \label{d_ratio}
\end{figure}

\begin{figure}[H]
\vspace{-2mm}
    \centering
    \includegraphics[width=0.8\textwidth]{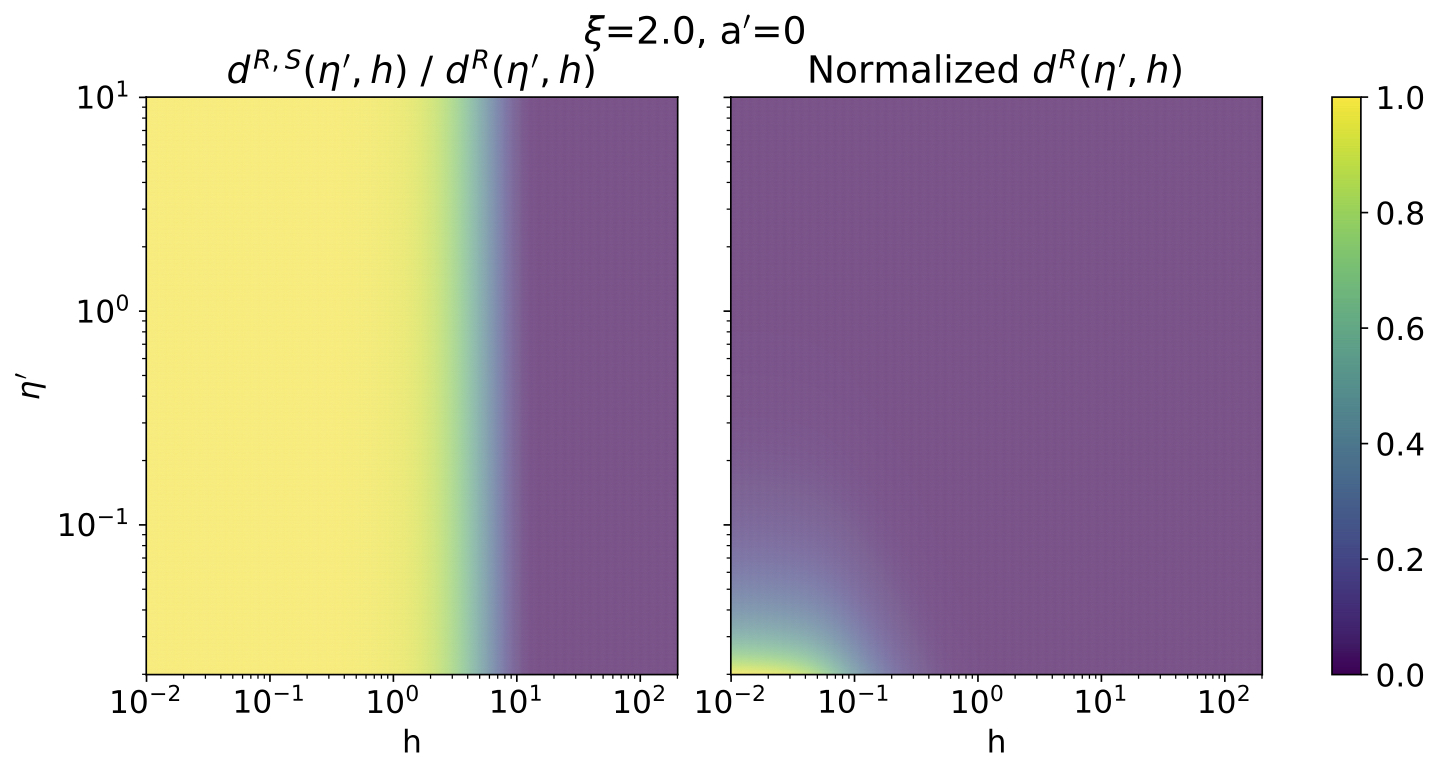}
    \vspace{-2mm}
    \caption{\footnotesize{Similarly to Figure \ref{d_ratio}, this figure shows the ratio of the smoothed, to the unsmoothed restricted function $d^{R,S}(\eta', h)$ / $d^{R}(\eta', h)$,  as well as a normalized plot of the function $d^R(\eta', h)$ for $\xi=2$ and spherical Bessel order $a'=0$. The fact that the function $d(\eta', h)$ rapidly falls to zero outside of the points $h=0$, $\eta'=0$, and the smoothing only significantly begins to alter the function far from this point, gives us confidence that our result is not significantly altered by the smoothing.}} 
    \label{dr_ratio}
\end{figure}

We find that the result of $d(\eta',h)$ is unchanged when the bounds are widened further than $y_{\rm min} = -10$ and $y_{\rm max} = -0.001$. We again apply a smoothing to these integrals with the same cutoff value of $h_0 = 3\xi$:
\begin{align}
d_{a'}^S(\eta',h)\equiv e^{-(h/h_0)^2} d_{a'}(\eta',h),
\end{align}
and
\begin{align}
d_{a'}^{R, S}(\eta',h)\equiv e^{-(h/h_0)^2} d^R_{a'}(\eta',h).
\end{align}
In Figures \ref{d_ratio} and \ref{dr_ratio}, we see that this choice of cutoff does not significantly alter our results.
Finally, we compute both cases of the leftmost integral in Eq. (\ref{Curly_M}),

\begin{align}
\label{M1}
\mathcal{M}_{c a a'}(u) = \int_{0}^{\infty} \dd h \; h^2 j_c(h\eta') j_a(h) d^S_{a'}(\eta',h),
\end{align}
\begin{align}
\label{M2}
\mathcal{M}^R_{c a a'}(u) = \int_{0}^{\infty} \dd h \; h^2 j_c(h\eta') j_a(h) d^{R, S}_{a'}(\eta',h),
\end{align}
where we have used our change of variables, and removed a factor of $u^{-4}$ that can be pulled out of the integral. We can see from Figure \ref{d_panels} that $h^3d^S_{a'}(\eta',h)$ and $h^3d^{R, S}_{a'}(\eta',h)$ are in excellent agreement (the extra factor of $h$ appears in the integrand of Eqs. (\ref{M2}) and (\ref{M2}) if one integrates over $\ln h$). Thus we have found that Eq. (\ref{M1}) will be a very good approximation for Eq. (\ref{M2}) . This shows that extending our bounds of integration outside of the region $1/(8\xi) \leq -k\tau \leq 2\xi$ will not significantly affect our result. 
\begin{figure}[H]
\vspace{-2mm}
    \centering
    \includegraphics[width=0.8\textwidth]{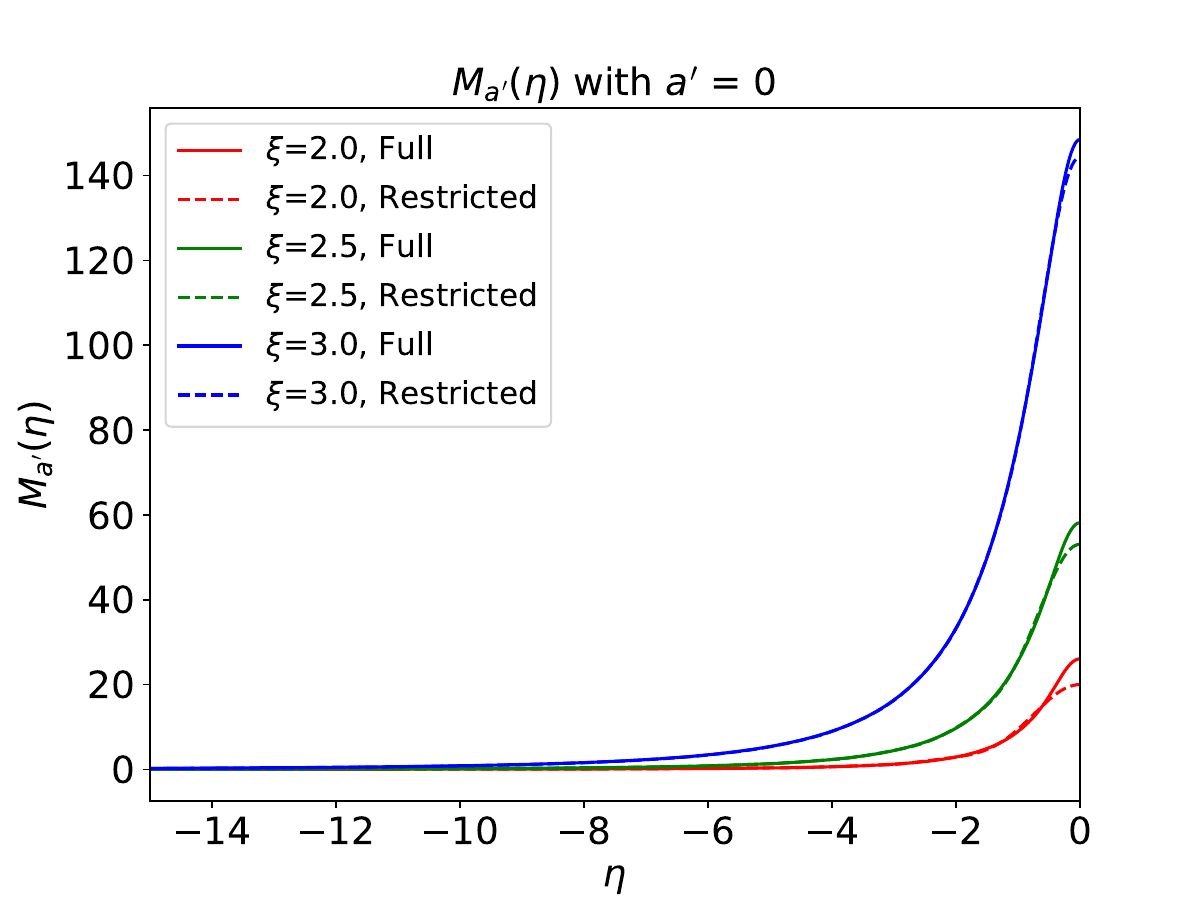}
    \vspace{-2mm}
    \caption{\footnotesize{This plot shows how the function $M_{a'}(\eta)$ of Eq. (\ref{M_int}) varies with parameter $\eta$ for both the full and restricted integrals with spherical Bessel function order $a^\prime = 0$. In our trispectrum computation, we must integrate the gauge field mode functions from $z_{\rm min} = -\infty$ to $z_{\rm max}=0$; however we must truncate the integration range to some finite value to do the integrals numerically. For the full integral we integrate from $z_{\rm min}=-10$ to $z_{\rm max}=-0.01$, and we have verified that widening these bounds further does not affect the integration result. For the restricted case, we do the same integration from $z_{\rm min}=-2\xi$ to $z_{\rm max}=-(8\xi)^{-1}$. Throughout this paper, we use the approximation for the mode functions (\ref{A:approx_1}), which is valid only in the restricted integration range. From the figure we verify that widening our integration bounds outside of the region of good approximation does not significantly change the result.}}
    \label{M_1}
\end{figure}

\begin{figure}[H]
\vspace{-2mm}
    \centering
    \includegraphics[width=0.8\textwidth]{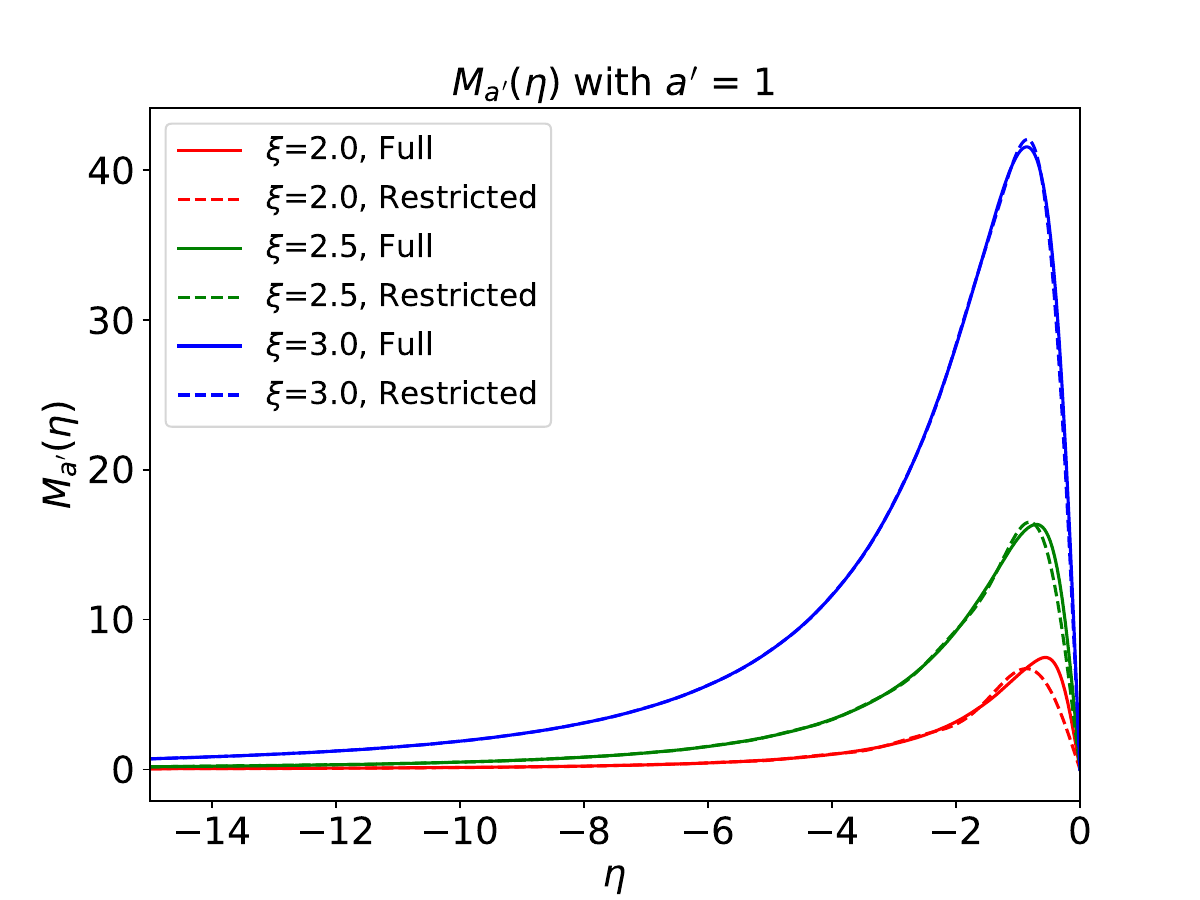}
    \vspace{-2mm}
    \caption{\footnotesize{Similarly to Figure \ref{M_1}, this plot verifies that the function $M_{a'}(\eta)$, with spherical Bessel order $a^\prime = 1$, does not change significantly when we integrate Eq. (\ref{M_int}) over the full range ($z_{\rm min}=-10$ to $z_{\rm max}=-0.01$) or the restricted range ($z_{\rm min}=-2\xi$ to $z_{\rm max}=-(8\xi)^{-1}$). }} 
    \label{M_2}
\end{figure}

\begin{figure}[H]
\vspace{-2mm}
    \centering
    \includegraphics[width=0.8\textwidth]{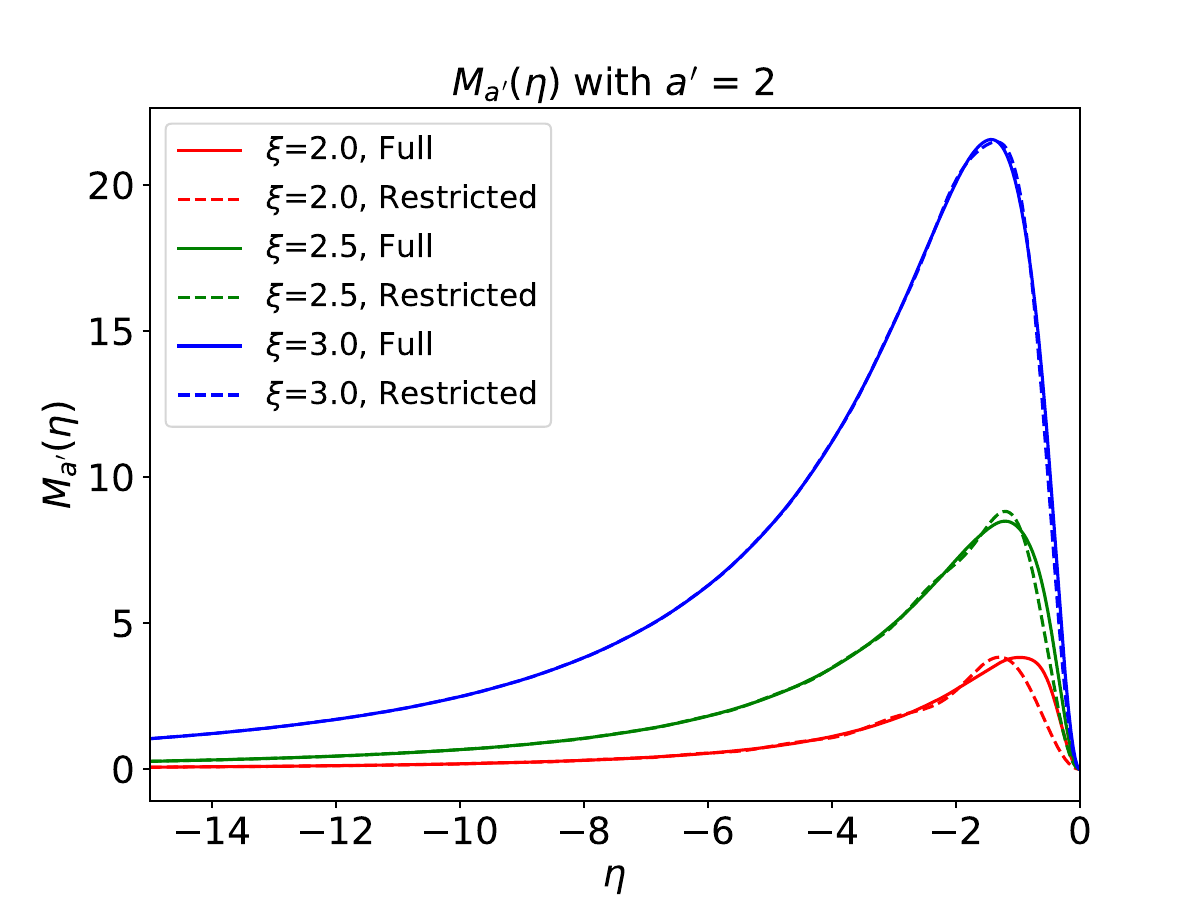}
    \vspace{-2mm}
    \caption{\footnotesize{Similarly to Figure \ref{M_1}, this plot verifies that the function $M_{a'}(\eta)$, with spherical Bessel order $a^\prime = 2$, does not change significantly when we integrate Eq. (\ref{M_int}) over the full range ($z_{\rm min}=-10$ to $z_{\rm max}=-0.01$) or the restricted range ($z_{\rm min}=-2\xi$ to $z_{\rm max}=-(8\xi)^{-1}$). }} 
    \label{M_3}
\end{figure}

\begin{figure}[H]
\vspace{-2mm}
    \centering
    \includegraphics[width=0.8\textwidth]{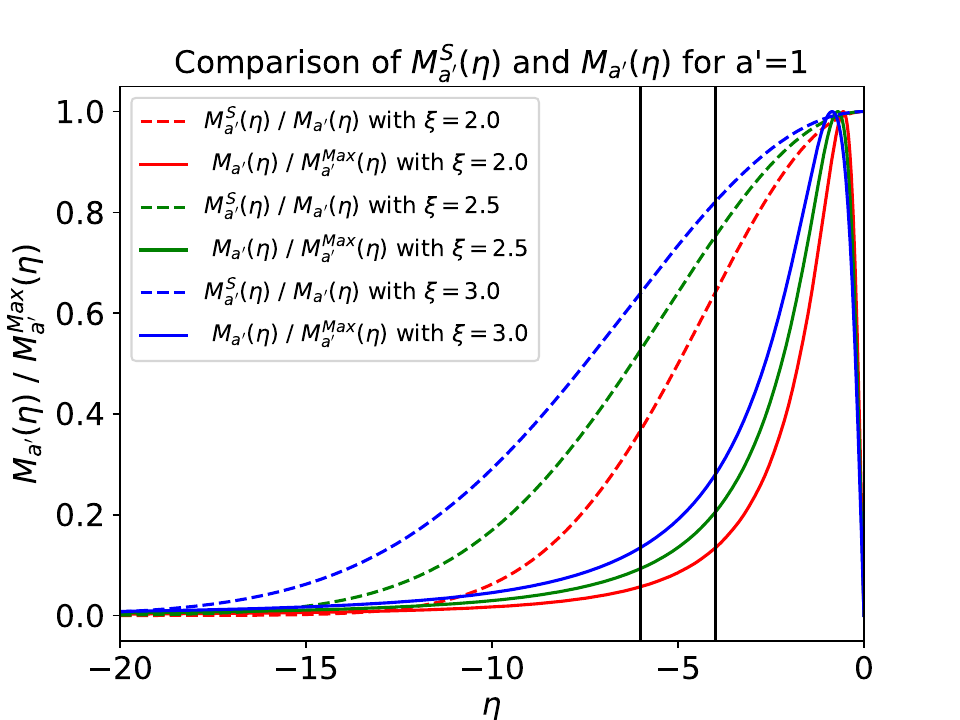}
    \vspace{-2mm}
    \caption{\footnotesize{Similarly to Figure \ref{M_ratio_1}, this figure shows the ratio of the smoothed to unsmoothed function $M^{S}_{a'}(\eta)$ / $M_{a'}(\eta)$ as well as the normalized integral $M_{a'}(\eta)$ for spherical Bessel order $a^\prime = 1$. The smoothing only begins to significantly alter the function for values of eta where $M_{a'}(\eta)$ is a small fraction of its maximum value. We can therefore conclude that the smoothing is not significantly affecting our result.}} 
    \label{M_ratio_2}
\end{figure}

\section{Proof that Trispectrum Cross Terms are Parity-Even} \label{AppendixE}

In \S \ref{sec:corrfun}, we state that the trispectrum cross terms 
\begin{align}
\expval{Q^{\text{vac}}(\tau, \vb k_i) Q^{\text{vac}}(\tau, \vb k_j) Q^{\text{inv}}(\tau, \vb k_l) Q^{\text{inv}}(\tau, \vb k_m)},
\end{align}
are parity-even, where $i,j,l,m \in \{1,2,3,4\}$. Here we prove this statement. The imaginary part of the trispectrum is its parity-odd component, so it will suffice to show that these cross terms are real. We consider the term
\begin{align}\label{2pcf_Q}
\expval{Q^{\text{vac}}(\tau, \vb k_1) Q^{\text{vac}}(\tau, \vb k_2) Q^{\text{inv}}(\tau, \vb k_3) Q^{\text{inv}}(\tau, \vb k_4)}
\end{align}
without loss of generality as the proof will be the same for any other arrangement of the $\vb k$ vectors. Since the creation and annihilation operators of the gauge field and inflaton field commute, we can separate this term into two expectation values
\begin{align}
&\expval{Q^{\text{vac}}(\tau, \vb k_1) Q^{\text{vac}}(\tau, \vb k_2) Q^{\text{inv}}(\tau, \vb k_3) Q^{\text{inv}}(\tau, \vb k_4)} = \\
&\expval{Q^{\text{vac}}(\tau, \vb k_1) Q^{\text{vac}}(\tau, \vb k_2)} \expval{Q^{\text{inv}}(\tau, \vb k_3) Q^{\text{inv}}(\tau, \vb k_4)}. \nonumber
\end{align}

The correlation function of the $Q^{\text{vac}}$ mode functions is the power spectrum of standard inflation, which is known to be parity even, so we will focus on the power spectrum of the $Q^{\text{inv}}$ mode functions:
\begin{align} \label{Q34}
\expval{Q^{\text{inv}}(\tau, \vb k_3) Q^{\text{inv}}(\tau, \vb k_4)} = \int_{-\infty}^0 \dd \tau_3 \int_{-\infty}^0 \dd \tau_4 \; G(\tau, \tau_3, k_3) \, G(\tau, \tau_4, k_4) \expval{J(\tau_3, \vb k_3) \, J(\tau_4, \vb k_4)}. 
\end{align}
Inserting Eq. (\ref{eq:J}) into the 2-point function of the source terms $J$ in Eq. (\ref{Q34}), we find 
\begin{equation}
\label{2pcf_J}
\begin{split} 
&\expval{J(\tau_3, \vb k_3) \, J(\tau_4, \vb k_4)} =  \\
&\frac{1}{(\Lambda a(\tau))^2} \int \int \frac{\dd [3] \vb q_3}{(2\pi)^{3}}\, \frac{\dd [3] \vb q_4}{(2\pi)^{3}} \, q_3 \, q_4 \, [ \bm{\hat \epsilon}_- (\widehat{\vb k_3 - \vb q_3}) \vdot \bm{\hat \epsilon}_- (\vu q_3)] [ \bm{\hat \epsilon}_- (\widehat{\vb k_4 - \vb q_4}) \vdot \bm{\hat \epsilon}_- (\vu q_4)]\\
&\hspace{40mm} \times A_-'(\tau_3, |\vb k_3 - \vb q_3|) A_-(\tau, q_3) A_-'(\tau_4, |\vb q_4 - \vb k_4|)A_-(\tau, q_4)  \\
&\hspace{40mm} \times \expval{a_-(\vb k_3 - \vb q_3) a_-(\vb q_3) a_-^\dag(\vb q_4 - \vb k_4) a_-^\dag(-\vb q_4)},
\end{split}
\end{equation}
where we have excluded the creation and annihilation operator combinations that have a vanishing vacuum expectation value, and we used the approximation that the mode functions are real as discussed in the beginning of \S \ref{sec:corrfun}.

As in \S\ref{Section5}, we proceed by considering all possible Wick contractions:
\begin{equation}
\label{2pcf_a}
\begin{split}
\expval{a_-(\vb k_3 - \vb q_3) a_-(\vb q_3) a_-^\dag(\vb q_4 - \vb k_4) a_-^\dag(-\vb q_4)} = & (2\pi)^6  \Dird (\vb k_3 - \vb q_3 + \vb k_4 - \vb q_4) \Dird (\vb q_3 + \vb q_4)  \\
+ &(2\pi)^6  \Dird (\vb q_4 + \vb k_3 - \vb q_3) \Dird (\vb q_3 + \vb k_4 - \vb q_4).
\end{split}
\end{equation}

After inserting Eq. \eqref{2pcf_a} back into Eq. \eqref{2pcf_J}, we must integrate over both pairs of delta functions. Consider the integral over the first pair
\begin{equation}
\label{2pcf_J1}
\begin{split}
&\expval{J(\tau_3, \vb k_3) \, J(\tau_4, \vb k_4)}^{(1)} \equiv   \\
&\frac{1}{(\Lambda a(\tau))^2} \int \int \dd [3] \vb q_3 \, \dd [3] \vb q_4 \, q_3 \, q_4 \, [ \bm{\hat \epsilon}_- (\widehat{\vb k_3 - \vb q_3}) \vdot \bm{\hat \epsilon}_- (\vu q_3)] [ \bm{\hat \epsilon}_- (\widehat{\vb k_4 - \vb q_4}) \vdot \bm{\hat \epsilon}_- (\vu q_4)]\\
&\hspace{40mm} \times A_-'(\tau_3, |\vb k_3 - \vb q_3|) A_-(\tau_3, q_3) A_-'(\tau_4, |\vb q_4 - \vb k_4|)A_-(\tau_4, q_4)   \\
&\hspace{40mm} \times \Dird (\vb k_3 - \vb q_3 + \vb k_4 - \vb q_4) \Dird (\vb q_3 + \vb q_4).
\end{split}
\end{equation}

From the definition of the polarization vector $\epsilon_-(\vu q) = (\vu \theta - i\vu \phi)/\sqrt{2}$, it can be shown that 
\begin{align}
\Re{\bm{\hat \epsilon}_- (\widehat{\vb k - \vb q}) \vdot \bm{\hat \epsilon}_- (\vu q)} = \Re{\bm{\hat \epsilon}_- (-(\widehat{\vb k - \vb q})) \vdot \bm{\hat \epsilon}_- (-\vu q)}
\end{align}
and
\begin{align}
\Im{\bm{\hat \epsilon}_- (\widehat{\vb k - \vb q}) \vdot \bm{\hat \epsilon}_- (\vu q)} = -\Im{\bm{\hat \epsilon}_- (-(\widehat{\vb k - \vb q})) \vdot \bm{\hat \epsilon}_- (-\vu q)}.
\end{align}
Therefore, when we multiply the two polarization vector dot products in Eq. (\ref{2pcf_J1}), the two imaginary terms cancel and we are left with a real function. For precisely the same reason, the integral over the other pair of delta functions in Eq. (\ref{2pcf_a}) is also a real function. We have thus shown that the power spectrum of the $Q^{\text{inv}}$ mode functions is real and therefore the trispectrum cross term given in Eq. (\ref{2pcf_Q}) is parity-even. 

\acknowledgments
MR gratefully acknowledges many useful conversations with the Slepian research group at UF, especially William Ortol\'a-Leonard, Jess Chellino, Farshad Kamalinejad, and Robert Cahn. MR and ZS thank Richard Woodard and Wei Xue as well for much useful input. MR thanks Xuce Niu, Moinul Rahat, and David Sadek for helpful discussions. JH has received funding from the European Union’s Horizon 2020 research and innovation program under the Marie Sk\l{}odowska-Curie grant agreement No 101025187. AG acknowledges funding from NASA grant number 80NSSC24M0021.






\bibliographystyle{utphys}
\bibliography{bibliography.bib}
\end{document}